\documentclass[12pt]{scrartcl}

\areaset{158mm}{238mm}
\usepackage{setspace}
\usepackage{amsmath}
\usepackage{mathtools}

\usepackage{amssymb}
\usepackage{hyperref}
\usepackage{cite}
\usepackage{bm}
\usepackage{slashed}

\usepackage{tikz,pgf}
\usepackage{xcolor}
  \definecolor{rblue}{RGB}{65,105,225}
  \definecolor{sgreen}{RGB}{46,139,87}

\numberwithin{equation}{section}

\title{\vspace{2cm} Open superstring field theory \\
based on the supermoduli space}
\author{Kantaro Ohmori$^{\spadesuit,\clubsuit}$ and~ Yuji Okawa$^\diamondsuit$
\\[0.5cm]
	\large\slshape $^\spadesuit$ Department of Physics, The University of Tokyo,\\[-0.2cm] 
	\large\slshape Hongo, Bunkyo-ku, Tokyo 133-0022, Japan \\[-0.05cm]
	\large\slshape $^\clubsuit$ School of Natural Sciences, Institute for Advanced Study,\\[-0.2cm] 
	\large\slshape Princeton, NJ 08540, USA \\[-0.05cm]
	\large\slshape $^\diamondsuit$ Institute of Physics, The University of Tokyo, \\[-0.2cm]
	\large\slshape  Komaba, Meguro-ku, Tokyo 153-8902, Japan \\[-0.1cm]
}
\date{}
\begin{document}
\maketitle

\vspace{-13.5cm}


\hfill  UT-17-08\\[-.5cm]

\hfill UT-Komaba/17-1

\vspace{11.5cm}
\paragraph{\hspace{.9cm}\large{Abstract}}
\vspace{-.1cm}
\begin{abstract}
We present a new approach to formulating open superstring field theory
based on the covering of the supermoduli space of super-Riemann surfaces
and explicitly construct a gauge-invariant action
in the Neveu-Schwarz sector up to quartic interactions.
The cubic interaction takes a form of an integral
over an odd modulus of disks with three punctures
and the associated ghost is inserted.
The quartic interaction takes a form of an integral
over one even modulus and two odd moduli,
and it can be interpreted as the integral over the region
of the supermoduli space of disks with four punctures
which is not covered by Feynman diagrams with two cubic vertices
and one propagator.
As our approach is based on the covering of the supermoduli space,
the resulting theory naturally realizes an $A_\infty$ structure,
and the two-string product and the three-string product
used in defining the cubic and quartic interactions
are constructed to satisfy the $A_\infty$ relations to this order.

\end{abstract}
\thispagestyle{empty}

\newpage

\tableofcontents

\section{Introduction}

On-shell scattering amplitudes of the open bosonic string
are calculated by integrating correlation functions
of the conformal field theory (CFT)
associated with the D-brane we consider
over the moduli space of Riemann surfaces with boundary.
The integration over the moduli space is crucial
for consistency of the theory,
and, for example, unphysical states decouple
after the integration over the moduli space. 
In string field theory, on the other hand, gauge invariance plays a crucial role.
Where is the structure of the integration over the moduli space
of Riemann surfaces encoded?

In string field theory, the equation of motion of the free theory
corresponds to the physical state condition of the world-sheet theory
and takes the form
\begin{equation}
Q \Psi = 0 \,,
\end{equation}
where $\Psi$ is the string field and $Q$ is the BRST operator of the world-sheet CFT.
The equivalence relation of the physical states is then realized
as a gauge symmetry in string field theory.
The gauge transformation of the free theory takes the form
\begin{equation}
\delta_\Lambda \Psi = Q \Lambda \,,
\end{equation}
where $\Lambda$ is the gauge parameter.
The action of the free theory which is invariant under this gauge transformation
is written as
\begin{equation}
S = {}-\frac{1}{2} \, \langle \, \Psi, Q \Psi \, \rangle \,,
\end{equation}
where $\langle \, A, B \, \rangle$ is an appropriate inner product
for a pair of string fields $A$ and $B$.

In the interacting theory,
this gauge transformation is nonlinearly extended
by introducing products of string fields.
A famous example of such string products is
the star product $A \ast B$ defined for two string fields $A$ and $B$
in open bosonic string field theory~\cite{Witten:1985cc}.
The gauge transformation of the interacting theory is given by
\begin{equation}
\delta_\Lambda \Psi = Q \Lambda +g \, ( \, \Psi \ast \Lambda -\Lambda \ast \Psi \, ) \,,
\end{equation}
where $g$ is the open string coupling constant,
and the action invariant under this nonlinearly extended transformation
takes the following Chern-Simons form:
\begin{equation}
S = {}-\frac{1}{2} \, \langle \, \Psi, Q \Psi \, \rangle
-\frac{g}{3} \, \langle \, \Psi, \Psi \ast \Psi \, \rangle \,.
\end{equation}

This formulation of the cubic theory is beautiful and practically useful,
and recent developments of the analytic method
in open string field theory,
which were triggered
by the construction of an analytic solution
by Schnabl in~\cite{Schnabl:2005gv} 
and are based on the $KBc$ algebra~\cite{Okawa:2006vm},
crucially depend on properties specific to the star product.
However, gauge-invariant actions can be constructed
in terms of a more general class of string products.
Consider an action of the form
\begin{equation}
\begin{split}
S & = {}-\frac{1}{2} \, \langle \, \Psi, Q \Psi \, \rangle
-\sum_{n=2}^\infty \frac{g^{n-1}}{n+1} \, \langle \, \Psi,
[ \,\, \underbrace{\Psi, \ldots , \Psi}_n \,\, ] \, \rangle \\
& = {}-\frac{1}{2} \, \langle \, \Psi, Q \Psi \, \rangle
-\frac{g}{3} \, \langle \, \Psi, [ \, \Psi, \Psi \, ] \, \rangle
-\frac{g^2}{4} \, \langle \, \Psi, [ \, \Psi, \Psi, \Psi \, ] \, \rangle
+O(g^3) \,,
\end{split}
\end{equation}
where $[ \, A_1, A_2, \ldots, A_n \, ]$ is an $n$-string product
defined for $n$ string fields $A_1$, $A_2$, \ldots, $A_{n-1}$, and $A_n$.
If the $n$-string products have appropriate cyclic properties
and satisfy a set of relations called
$A_\infty$~\cite{Stasheff:I, Stasheff:II, Getzler-Jones, Markl, Penkava:1994mu, Gaberdiel:1997ia},
the action is invariant under a gauge transformation
written in terms of the $n$-string products.
Furthermore, the quantization of the theory
based on the Batalin-Vilkovisky formalism~\cite{Batalin:1981jr, Batalin:1984jr}
is straightforward
when the action has this $A_\infty$ structure.
The cubic theory based on the star product is a special case
where the $A_\infty$ relations are satisfied
without $n$-string products for $n > 2$ because of the associativity of the star product,
and we can think of $A_\infty$
as the structure underlying gauge invariance for open string field theory.

Actually, the $A_\infty$ structure is closely related
to the integration over the moduli space of disks with punctures on the boundary.
We can construct a set of string products
satisfying the $A_\infty$ relations for the open bosonic string
based on the decomposition of the moduli space of disks with punctures on the boundary.
For closed string field theory, the corresponding structure is called
$L_\infty$~\cite{Zwiebach:1992ie, Lada:1992wc, Schlessinger-Stasheff},
and gauge-invariant actions with the $L_\infty$ structure for the closed bosonic string
can be constructed based on the decomposition
of the moduli space of spheres with punctures. 
We could therefore say that the integration of the moduli space of disks or spheres with punctures
is encoded through the $A_\infty$ structure or the $L_\infty$ structure, respectively,
in bosonic string field theory.

In superstring field theory, the structure underlying the gauge invariance
should be related to the decomposition of the supermoduli space of super-Riemann surfaces,
but the understanding in this perspective has not been fully developed.
While we are seeing exciting breakthroughs
in the construction of superstring field theory over the last few years,
we consider that it is important to explore the relation
of the gauge invariance to the supermoduli space for further developments.
Below let us describe the current status of the relation
to the supermoduli space from three aspects.

First, consider the Berkovits formulation of open superstring field theory
in the Neveu-Schwarz (NS) sector~\cite{Berkovits:1995ab}
based on the large Hilbert space
of the superconformal ghost sector~\cite{Friedan:1985ge}.
It is one of the most successful formulations of superstring field theory,
and the action is beautifully written in a Wess-Zumino-Witten-like (WZW-like) form.
However, the relation to the supermoduli space is obscure
because the gauge invariance is based on the large Hilbert space.
The quantization of this theory based on the Batalin-Vilkovisky formalism
has turned out to be formidably
complicated~\cite{Kroyter:2012ni, Torii:2012nj, Torii:2011zz, Berkovits:2012np, Berkovits:201X},
and this might be coming from our insufficient understanding
of the relation to the supermoduli space.
In~\cite{Iimori:2013kha}, the four-point amplitude in the Berkovits formulation
was calculated in a class of gauges where the relation to the world-sheet calculation
can be manifestly seen.
It was found that the assignment of picture-changing operators
is different in the $s$ channel and in the $t$ channel
of Feynman diagrams with two vertices and one propagator,
and this different behavior is adjusted by the contribution from the quartic interaction.
In other words, we found that the quartic interaction of the Berkovits formulation
precisely implements the {\normalfont \itshape vertical integration}~\cite{Sen:2014pia, Sen:2015hia}.
This different behavior persists in the limit
where the picture-changing operator localizes at the open-string midpoint.
This way the difficulty in the attempt of formulating the cubic theory with a local insertion
of a picture-changing operator at the open-string midpoint can be understood
in the context of the covering of the supermoduli space of super-Riemann surfaces.
To summarize, we have obtained preliminary insight into the understanding
of why the Berkovits formulation based on the large Hilbert space
is successful from the context of the supermoduli space of super-Riemann surfaces,
and we consider that it is important to explore this direction further.

Second, consider formulating the NS sector of open superstring filed theory
based on the small Hilbert space.
It had long been thought that a regular formulation
based on the small Hilbert space would be difficult
because of singularities coming from local picture-changing operators.
However, it was demonstrated in~\cite{Iimori:2013kha} that a regular formulation
of open superstring field theory based on the small Hilbert space
can be obtained from the Berkovits formulation by partial gauge fixing.
In the process of the partial gauge fixing,
a line integral of the superconformal ghost $\xi (z)$ of~\cite{Friedan:1985ge}
in the large Hilbert space is used.
Its BRST transformation yields a line integral of the picture-changing operator,
and insertions of local picture-changing operators are avoided in this approach.
While a line integral of $\xi (z)$ does not have simple transformation properties
under conformal transformations, the partial gauge fixing guarantees
that the resulting theory is gauge invariant.
It turned out, however, that the resulting theory does not exhibit
an $A_\infty$ structure.
Once we recognize that a line integral of $\xi (z)$ can be used
in constructing a gauge-invariant action,
we do not have to start from the Berkovits formulation.
Using a line integral of $\xi (z)$ as a new ingredient,
an action with an $A_\infty$ structure based on the small Hilbert space
was constructed in~\cite{Erler:2013xta}.\footnote{
The construction was further generalized to the NS sector
of heterotic string field theory
and the NS-NS sector of type II superstring field theory in~\cite{Erler:2014eba}
and to the equations of motion including the Ramond sector in~\cite{Erler:2015lya}.
}
Furthermore, the equivalence of the theory with an $A_\infty$ structure 
for the open superstring in~\cite{Erler:2013xta}
to the Berkovits formulation
is shown~\cite{Erler:2015rra, Erler:2015uba, Erler:2015uoa}.
We can now extract an $A_\infty$ structure hidden in the Berkovits formulation
by field redefinition.
While our understanding of the NS sector of open superstring field theory
has significantly developed,
the relation to the supermoduli space has not be seen explicitly.
This is partly because 
the large Hilbert space is used in a fairly essential way
during the construction of the theory with an $A_\infty$ structure in~\cite{Erler:2013xta},
although the dynamical string field is in the small Hilbert space.

Third, consider the Ramond sector of open superstring field theory.
While construction of an action including the Ramond sector
had been a major obstacle to formulating superstring field theory
for many years,
gauge-invariant actions for open superstring field theory
including both the NS sector and the Ramond sector
were recently constructed~\cite{Kunitomo:2015usa, Erler:2016ybs, Konopka:2016grr}.
Sen also developed a remarkable approach to covariant actions
including Ramond sectors by allowing spurious free fields~\cite{Sen:2015uaa}.\footnote{
See \cite{deLacroix:2017lif} for a recent review.
}
While Sen described the approach
in the context of heterotic string field theory
and type II superstring field theory, the idea can be applied to open superstring field theory,
and in that case we can explicitly construct interactions based on the star product
as explained in subsection~3.5 of~\cite{Erler:2016ybs}.
The action in~\cite{Kunitomo:2015usa} is based on the WZW-like formulation.
The interactions are written in a closed form to all orders,
but the action does not exhibit an $A_\infty$ structure.
The action in~\cite{Erler:2016ybs, Konopka:2016grr} is based on the approach developed
in~\cite{Erler:2013xta, Erler:2014eba, Erler:2015lya}.
It has a cyclic $A_\infty$ structure, but the form of the interactions
is complicated and is not written in a closed form.
The equivalence of the two actions is also shown in~\cite{Erler:2016ybs}.
In the construction of these actions,
there were two key ingredients.
The first ingredient is to impose a constraint
on the state space of the Ramond sector
and to characterize the constraint using a projector
which is obtained by integrating an odd modulus
of propagator strips in the Ramond sector~\cite{Kunitomo:2015usa}.\footnote{
As mentioned in~\cite{Kunitomo:2015usa}, this important observation
is actually based on the results in this paper.}
This characterization was technically convenient,
but it was conceptually important as well
and the treatment of the Ramond sector
in the construction of one-particle irreducible effective superstring field theory~\cite{Sen:2015hha}
indicated that interactions can be compatible
with the constraint characterized this way.
The second ingredient is an analog in the Ramond sector
of the line integral of $\xi (z)$ used in the NS sector.
Therefore, understanding the relation to the supermoduli space
of super-Riemann surfaces has indeed played a crucial role
in this breakthrough regarding the Ramond sector,
but it is a hybrid between
the ingredient based on the supermoduli space
and the ingredient based on the large Hilbert space.
We expect to acquire more fundamental formulations
when we understand the relation between the supermoduli space and the large Hilbert space.\footnote{
On-shell scattering amplitudes involving fermions were recently calculated in~\cite{Kunitomo:2016bhc},
and the analysis of~\cite{Iimori:2013kha} in the NS sector
is extended to include the Ramond sector.
The construction in~\cite{Kunitomo:2015usa} was generalized to heterotic string field theory,
and gauge-invariant interactions which are up to quartic order in the Ramond string field
but contain the NS string field to all orders were constructed in~\cite{Goto:2016ckh}.
For discussions on supersymmetry
in open superstring field theory,
see~\cite{Erler:2016rxg} for the action constructed in~\cite{Erler:2016ybs, Konopka:2016grr}
and~\cite{Kunitomo:2016kwh} for the action constructed in~\cite{Kunitomo:2015usa}.
}

Since all the formulations described above use the large Hilbert space to some extent,
the first step to improve our understanding on the relation
of gauge invariance to the supermoduli space would be to construct
superstring field theory directly based on the supermoduli space
without using the large Hilbert space at all.
In this paper, we present a new approach to formulating open superstring field theory
based on the covering of the supermoduli space of super-Riemann surfaces.\footnote{
As emphasized in~\cite{Witten:2012bh},
the integration over the supermoduli space of super-Riemann surfaces
should not be understood as ordinary integrals even for Grassmann-even directions,
and it is related to the algebraic treatment of the path integral
of the $\beta \gamma$ system developed in~\cite{Witten:2012bh}.
As a consequence, the covering of the supermoduli space should not be
understood as the covering of integration regions in ordinary integrals.
One way to describe the covering of the supermoduli space
is to use a partition of unity. See section~7 of~\cite{Sen:2015hia} for details.
We will also develop related discussions in subsection~\ref{sec:supermoduli}.
}
We explicitly construct a gauge-invariant action
in the NS sector up to quartic interactions.
The cubic interaction takes a form of an integral
over an odd modulus of disks with three punctures
and the associated ghost is inserted.
The quartic interaction takes a form of an integral
over one even modulus and two odd moduli,
and it can be interpreted as the integral over the region
of the supermoduli space of disks with four punctures
which is not covered by Feynman diagrams with two cubic vertices
and one propagator.
As our approach is based on the covering of the supermoduli space,
the resulting theory naturally realizes an $A_\infty$ structure,
and the two-string product and the three-string product
used in defining the cubic and quartic interactions
are constructed to satisfy the $A_\infty$ relations to this order.

The rest of the paper is organized as follows.
In section~\ref{sec:A_infinity} we briefly summarize the $A_\infty$ structure up to quartic interactions.
Section~\ref{sec:cubic} is for the cubic interaction.
In subsection~\ref{sec:bos4pt} we first discuss the treatment of an even modulus
of disks with four punctures in the open bosonic string.
In subsection~\ref{sec:super3pt} we generalize the discussion
to the treatment of an odd modulus
of disks with three NS punctures in the open superstring.
Then in subsection~\ref{sec:cubic-vertex} we construct the cubic interaction
based on the treatment of an odd modulus
of disks with three NS punctures developed in subsection~\ref{sec:super3pt}.
Section~\ref{sec:quartic} is for the quartic interaction.
In subsection~\ref{sec:non-assoc} we find that the two-string product
constructed from the cubic interaction in section~\ref{sec:cubic}
is not associative and thus we need a three-string product
to realize an $A_\infty$ structure.
A three-string product is also necessary
in open bosonic string field theory
when the two-string product is not associative.
In subsection~\ref{sec:interpolation-bosonic} we explain the essence
of the construction of such a three-string product
for the open bosonic string
in a simplified setting.
In subsection~\ref{sec:interpolation-super} we generalize the discussion
to the open superstring,
and then we construct the quartic interaction in subsection~\ref{sec:quartic-vertex}.
The relation to the covering of the supermoduli space is discussed in subsection~\ref{sec:supermoduli}.
Section~\ref{sec:conclusions-discussion} is devoted to conclusions and discussion.
In our construction of the interactions,
the algebraic treatment of the path integral of the $\beta \gamma$ system
developed in~\cite{Witten:2012bh} plays a crucial role.
We review this treatment of the $\beta \gamma$ system in appendix~\ref{sec:beta-gamma} .

\section{$A_\infty$ structure up to quartic interactions}
\label{sec:A_infinity}

Our goal is to construct a gauge-invariant action of open superstring field theory
with an $A_\infty$ structure
based on the supermoduli space of super-Riemann surfaces,
and in this paper we construct a gauge-invariant action 
up to quartic interactions in the NS sector.
In this section we present the $A_\infty$ structure up to this order.

The open superstring field $\Psi$ is a state in the Hilbert space
of the boundary CFT describing the open string background we consider,
which consists of the matter sector, the $bc$ ghost sector,
and the $\beta \gamma$ ghost sector.
As we aim at constructing open superstring field theory
based on the supermoduli space of super-Riemann surfaces,
we choose the open superstring field $\Psi$ to be a Grassmann-odd state
carrying ghost number $1$ in the $-1$ picture.
The action of the free theory is given by
\begin{equation}
S = {}-\frac{1}{2} \, \langle \, \Psi, Q \Psi \, \rangle \,,
\end{equation}
where $Q$ is the BRST operator
and $\langle \, A, B \, \rangle$ is the BPZ inner product
of states $A$ and $B$.
Three basic properties of the BPZ inner product
and the BRST operator $Q$ are
\begin{equation}
\langle \, A_1, A_2 \, \rangle
= (-1)^{A_1 A_2} \langle \, A_2, A_1 \, \rangle \,, \quad
Q^2 = 0 \,, \quad
\langle \, Q A_1, A_2 \, \rangle
= {}-(-1)^{A_1} \langle \, A_1, QA_2 \, \rangle \,.
\end{equation}
Here and in what follows,
a state in the exponent of $-1$ represents
its Grassmann parity: it is 0 mod 2 for a Grassmann-even state
and 1 mod 2 for a Grassmann-odd state.
We then consider an action of the interacting theory
in the following form:
\begin{equation}
S = {}-\frac{1}{2} \, \langle \, \Psi, Q \Psi \, \rangle
-\frac{g}{3} \, \langle \, \Psi, [ \, \Psi, \Psi \, ] \, \rangle
-\frac{g^2}{4} \, \langle \, \Psi, [ \, \Psi, \Psi, \Psi \, ] \, \rangle +O(g^3) \,,
\label{action}
\end{equation}
where $g$ is the open string coupling constant,
$[ \, A_1, A_2 \, ]$ is a two-string product
defined for a pair of string fields $A_1$ and $A_2$,
and $[ \, A_1, A_2, A_3 \, ]$ is a three-string product
defined for three string fields $A_1$, $A_2$, and $A_3$.
The Grassmann parity of $[ \, A_1, A_2 \, ]$ is $\epsilon( A_1 ) +\epsilon( A_2 )$ mod $2$
and the Grassmann parity of $[ \, A_1, A_2, A_3 \, ]$
is $\epsilon( A_1 ) +\epsilon( A_2 ) +\epsilon( A_3 ) +1$ mod $2$,
where $\epsilon ( A_i )$ is the Grassmann parity of $A_i$ mod $2$ for $i = 1, 2, 3$.
These string products are assumed to have the following cyclic properties:
\begin{equation}
\begin{split}
\langle \, A_1, [ \, A_2, A_3 \, ] \, \rangle
& = (-1)^{A_1 (A_2+A_3)} \langle \, A_2, [ \, A_3, A_1 \, ] \, \rangle \,, \\
\langle \, A_1, [ \, A_2, A_3, A_4 \, ] \, \rangle
& = {}-(-1)^{A_1+A_2+A_1 (A_2+A_3+A_4)} \langle \, A_2, [ \, A_3, A_4, A_1 \, ] \, \rangle \,,
\label{eq:cyclicity}
\end{split}
\end{equation}
or equivalently
\begin{align}
\langle \, A_1, [ \, A_2, A_3 \, ] \, \rangle
& = \langle \, [ \, A_1, A_2 \, ], A_3 \, \rangle \,,
\label{two-string-cyclicity}
\\
\langle \, A_1, [ \, A_2, A_3, A_4 \, ] \, \rangle
& = {}-(-1)^{A_1} \langle \, [ \, A_1, A_2, A_3 \, ], A_4 \, \rangle \,.
\label{three-string-cyclicity}
\end{align}
The variation of the action~\eqref{action} is then given by
\begin{equation}
\delta S = {}-\, \langle \, \delta \Psi, Q \Psi \, \rangle
-g \, \langle \, \delta \Psi, [ \, \Psi, \Psi \, ] \, \rangle
-g^2 \, \langle \, \delta \Psi, [ \, \Psi, \Psi, \Psi \, ] \, \rangle +O(g^3) \,.
\end{equation}
The $A_\infty$ relations up to this order are
\begin{align}
& Q^2 = 0 \,, \\
& Q \, [ \, A_1, A_2 \, ] -[ \, Q A_1, A_2 \, ] -(-1)^{A_1} [ \, A_1, Q A_2 \, ] = 0 \,,
\label{derivation} \\
& Q \, [ \, A_1, A_2, A_3 \, ]
+[ \, Q A_1, A_2, A_3 \, ]
+(-1)^{A_1} [ \, A_1, Q A_2, A_3 \, ]
+(-1)^{A_1+A_2} [ \, A_1, A_2, Q A_3 \, ] \nonumber \\
& -[ \, [ \, A_1, A_2 \, ], A_3 \, ]
+[ \, A_1, [ \, A_2, A_3 \, ] \, ] = 0 \,.
\label{A_infinity}
\end{align}
We can show using these relations that the action~\eqref{action} is invariant
under the following gauge transformation:
\begin{equation}
\delta_\Lambda \Psi = Q \Lambda
+g \, ( \, [ \, \Psi, \Lambda \, ] -[ \, \Lambda, \Psi \, ] \, )
+g^2 \, ( \, [ \, \Psi, \Psi, \Lambda \, ]
-[ \, \Psi, \Lambda, \Psi \, ]
+[ \, \Lambda, \Psi, \Psi \, ] \, ) +O(g^3) \,,
\end{equation}
where the gauge parameter $\Lambda$ is a Grassmann-even state of ghost number $0$
in the $-1$ picture.

The second relation~\eqref{derivation} states that the BRST operator
is a deviation of the two-string product.
In open bosonic string field theory~\cite{Witten:1985cc}, this relation is satisfied
with $[ \, A_1, A_2 \, ] = A_1 \ast A_2$
because the BRST operator is a derivation of the star product:
\begin{equation}
Q \, ( A_1 \ast A_2 ) = Q A_1 \ast A_2 +(-1)^{A_1} A_1 \ast Q A_2 \,.
\end{equation}
The third relation~\eqref{A_infinity} is satisfied
without introducing the three-string product
if the two-string product is associative.
This is the case with $[ \, A_1, A_2 \, ] = A_1 \ast A_2$
because the star product is associative:
\begin{equation}
( A_1 \ast A_2 ) \ast A_3 = A_1 \ast ( A_2 \ast A_3 ) \,.
\end{equation}

In open superstring field theory, we construct
the two-string product and the three-string product
based on the star product.
For the two-string product $[ \, \Psi, \Psi \, ]$,
the ghost number should be $2$ and the picture number should be $-1$
in order for the  BPZ inner product $\langle \, \Psi, [ \, \Psi, \Psi \, ] \, \rangle$
to be well defined.
These conditions can be satisfied if we construct
the two-string product~$[ \, \Psi, \Psi \, ]$
by inserting an operator $X$
carrying ghost number $0$ and picture number $1$
to the star product~$\Psi \ast \Psi$.
The relation~\eqref{derivation} is also satisfied
if $X$ is Grassmann even and commutes with the BRST operator.
To be compatible with the cyclic property~\eqref{two-string-cyclicity}
of the two-string product,
we insert $X$ as follows:
\begin{equation}
[ \, A_1, A_2 \, ]
= \frac{1}{3} \, \Bigl[ \,
X^\star ( A_1 \ast A_2 ) +X A_1 \ast A_2 +A_1 \ast X A_2 \, \Bigr] \,,
\label{eq:cubvertex}
\end{equation}
where $X^\star$ is the BPZ conjugate of $X$.
Let us verify that the cyclic property~\eqref{two-string-cyclicity} is satisfied
with this definition.
From the definition of the BPZ conjugation we have
\begin{equation}
\langle \, A_1, X^\star (A_2\ast A_3) \,\rangle = \langle\, X A_1, A_2\ast A_3\, \rangle \,.
\end{equation}
The relation~\eqref{two-string-cyclicity} then follows from the cyclic property
of the star product
\begin{equation}
\langle \, A_1, A_2 \ast A_3 \, \rangle = \langle \, A_1 \ast A_2, A_3 \, \rangle
\end{equation}
as follows:
\begin{equation}
	\begin{split}
	\langle\, A_1, [ \, A_2, A_3 \, ] \, \rangle
	& = \frac{1}{3} \, \Bigl( \langle \, X A_1, A_2 \ast A_3 \, \rangle 
	+ \langle\, A_1, X A_2 \ast A_3 \, \rangle 
	+ \langle\, A_1, A_2 \ast X A_3 \, \rangle \Bigr) \\
	& = 
	\frac{1}{3} \, \Bigl( \langle \, X A_1 \ast A_2 , A_3 \, \rangle 
	+ \langle \, A_1 \ast X A_2 , A_3 \, \rangle 
	+ \langle \, A_1 \ast A_2, X A_3 \, \rangle \Bigr) \\
	& = \langle\, [ \, A_1, A_2 \, ] , A_3 \, \rangle \,.
	\end{split}
\end{equation}
In~\cite{Erler:2013xta}, a line integral of the picture-changing operator
is used as $X$.
In the next section, we construct the operator $X$
in the context of the integration over the supermoduli space
of disks with three NS punctures.

After the insertion of $X$ to the star product,
the resulting two-string product is no longer associative.
We then need to construct a three-string product
which has the cyclic property~\eqref{three-string-cyclicity}
and satisfies the relation~\eqref{A_infinity}.
We construct such a three-string product in section~\ref{sec:quartic}
based on the covering of the supermoduli space
of disks with four NS punctures.

\section{Cubic interaction}
\label{sec:cubic}

Scattering amplitudes at the tree level
in the world-sheet theory of the open bosonic string
can be calculated from correlation functions on the disk
of the boundary CFT.
Since three points on the boundary of the disk
can be mapped to arbitrary three points by conformal transformations
which are defined globally,
we say that disks with $n$ punctures have $n-3$ moduli.

For the world-sheet theory of the open superstring,
scattering amplitudes in the NS sector at the tree level
can be calculated from correlation functions on the disk
of the boundary superconformal field theory.
If we use the superspace,
the location of a point on the boundary is described
by one Grassmann-even coordinate and one Grassmann-odd coordinate.
We can fix three even coordinates and two odd coordinates
by superconformal transformations which are defined globally,
and the remaining coordinates parameterize
the supermoduli space of disks with $n$ NS punctures.
We say that disks with $n$ NS punctures have
$n-3$ even moduli and $n-2$ odd moduli.

In this section we construct the cubic interaction
of open superstring field theory
based on the integral of the supermoduli space
of disks with three NS punctures.
Disks with three NS punctures do not have even moduli
but have one odd modulus,
so we construct the cubic interaction
in a form of the integration over the odd modulus.
As we need to find an appropriate ghost insertion
associated with the integration over the odd modulus,
we first review the construction of the corresponding ghost insertion
for four-point functions of the bosonic string
in such a way that it can be generalized
to NS three-point functions of the superstring.

\subsection{Four-point functions in the open bosonic string}
\label{sec:bos4pt}

On-shell four-point amplitudes of the open bosonic string at the tree level
are given by integrating CFT four-point functions on the disk over the moduli space.
Disks with four punctures have one modulus,
and the integral over the moduli space consists of terms of the following form:
\begin{equation}
\int_{t_1}^{t_2} \mathrm{d}t \, \langle \, cV_A (t_1) \, V_B (t) \, cV_C (t_2) \, cV_D (t_3) \, \rangle \,,
\end{equation}
where $t_1$, $t_2$, and $t_3$ are arbitrary fixed values
and $t$ corresponds to the modulus.
The operators $V_A$, $V_B$, $V_C$, and $V_D$
are primary fields of weight $1$ in the matter sector,
and the $c$ ghost is inserted for each unintegrated vertex operator.
This can be rewritten as
\begin{equation}
	\begin{split}
		& \int_{t_1}^{t_2} \mathrm{d}t \,
		\langle \, cV_A (t_1) \, V_B (t) \, cV_C (t_2) \, cV_D (t_3) \, \rangle \\
		& = \int_{t_1}^{t_2} \mathrm{d}t \,
		\langle \, cV_A (t_1) \, b_{-1} \cdot cV_B (t) \, cV_C (t_2) \, cV_D (t_3) \, \rangle \\
		& = \int_{t_1}^{t_2} \mathrm{d}t \,
		\langle \, cV_A (t_1) \, b_{-1} \, \mathrm{e}^{t \, L_{-1}} \cdot cV_B (0) \, cV_C (t_2) \, cV_D (t_3) \, \rangle \,,
	\end{split}
	\label{eq:bos4pt}
\end{equation}
where $b_{-1}$ and $L_{-1}$ are modes of the $b$ ghost $b(z)$
and the energy-moment tensor $T(z)$, respectively, given by\footnote{
We use the doubling trick throughout the paper.
}
\begin{equation}
b_{-1} = \oint \frac{\mathrm{d}z}{2 \pi \mathrm{i}} \, b(z) \,, \qquad
L_{-1} = \oint \frac{\mathrm{d}z}{2 \pi \mathrm{i}} \, T(z) \,.
\end{equation}
Here we have used the fact that the operator $L_{-1}$
acts as the derivative $\partial_t$ on a local operator at $t$:
\begin{equation}
L_{-1} \cdot V(t) = \partial_t V(t) \,,
\end{equation}
where $V(t)$ is the local operator at $t$.

As we mentioned in the introduction, the integration over the moduli space
is crucial for the decoupling of BRST-exact states.
Consider the case where the vertex operator $cV_D$ is BRST exact
and is written in terms of an operator $\Lambda$ of ghost number $0$ as
\begin{equation}
cV_D (t_3) = Q \cdot \Lambda (t_3) \,.
\end{equation}
Since the BRST operator acts on $b_{-1}$ to give
\begin{equation}
\{ \, Q, b_{-1} \, \mathrm{e}^{t \, L_{-1}} \, \}
= L_{-1} \, \mathrm{e}^{t \, L_{-1}} = \partial_t \, \mathrm{e}^{t \, L_{-1}} \,,
\end{equation}
we find
\begin{equation}
	\begin{split}
		& \int_{t_1}^{t_2} \mathrm{d}t \,
		\langle \, cV_A (t_1) \, b_{-1} \, \mathrm{e}^{t \, L_{-1}} \cdot cV_B (0) \, cV_C (t_2) \,
		Q \cdot \Lambda (t_3) \, \rangle \\
		& = \int_{t_1}^{t_2} \mathrm{d}t \, \Bigl( \,
		\langle \, Q \cdot \big( \, cV_A (t_1) \, b_{-1} \, \mathrm{e}^{t \, L_{-1}} \cdot cV_B (0) \, cV_C (t_2) \,
		\Lambda (t_3) \, \big) \, \rangle \\
		&\hspace{60pt} +
		\langle \, cV_A (t_1) \, \{Q,b_{-1} \, \mathrm{e}^{t \, L_{-1}}\} \cdot cV_B (0) \, cV_C (t_2) \,
		 \Lambda (t_3) \, \rangle \,
		\Bigr) \\
		& = \int_{t_1}^{t_2} \mathrm{d}t \, \partial_t \,
		\langle \, cV_A (t_1) \, cV_B (t) \, cV_C (t_2) \, \Lambda (t_3) \, \rangle \,,
	\end{split}
\label{decoupling}
\end{equation}
where we used the BRST invariance of the form
\begin{equation}
\langle \, Q \cdot \big( \, \mathcal{O}_1 (t_1) \, \mathcal{O}_2 (t_2) \, \ldots \, \mathcal{O}_n (t_n) \, \big) \, \rangle = 0
\end{equation}
which holds for any operators $\mathcal{O}_1$, \ldots, $\mathcal{O}_{n-1}$, and $\mathcal{O}_n$.
The final form is an integral of a total derivative
and vanishes up to potential contributions from surface terms.
Thus the BRST-exact state decouples after the integration over the modulus $t$.

For generalization to NS three-point functions in the superstring,
it is convenient to use the {\normalfont \itshape extended BRST transformation} introduced in~\cite{Witten:2012bh}.
We introduce a Grassmann-odd parameter~$\tilde{t}$ and write the four-point function as follows:\footnote{
Our convention for the integral over Grassmann-odd variables is
\begin{equation*}
\int \mathrm{d} \tilde{t} \, \tilde{t} = 1 \,.
\end{equation*}
}
\begin{equation}
	\begin{split}
		& \int_{t_1}^{t_2} \mathrm{d}t \,
		\langle \, cV_A (t_1) \, b_{-1} \, \mathrm{e}^{t \, L_{-1}} \cdot cV_B (0) \, cV_C (t_2) \, cV_D (t_3) \, \rangle \\
		& = -\int_{t_1}^{t_2} \mathrm{d}t \int \mathrm{d} \tilde{t} \,
		\langle \, cV_A (t_1) \,
		\mathrm{e}^{\tilde{t} \, b_{-1}} \, \mathrm{e}^{t \, L_{-1}} \cdot cV_B (0) \, cV_C (t_2) \, cV_D (t_3) \, \rangle \\
		& = -\int_{t_1}^{t_2} \mathrm{d}t \int \mathrm{d} \tilde{t} \,
		\langle \, cV_A (t_1) \,
		\mathrm{e}^{\tilde{t} \, b_{-1} +t \, L_{-1}} \cdot cV_B (0) \, cV_C (t_2) \, cV_D (t_3) \, \rangle \,,
	\end{split}
	\label{eq:bosamp}
\end{equation}
where we used
\begin{equation}
[ \, L_{-1}, b_{-1} \, ] = 0
\end{equation}
to combine the two exponential factors into one exponential factor in the last step.
In~\cite{Witten:2012bh}, the BRST operator is extended to act also on $t$ and $\tilde{t}$.
Let us denote the extended BRST operator by $Q'$.
It acts on states of the boundary CFT in the same way as the ordinary BRST operator $Q$,
and its actions on $t$ and $\tilde{t}$ are given by
\begin{equation}
	[ \, Q', t \, ] = \tilde{t} \,, \qquad \{ \, Q', \tilde{t} \, \} = 0 \,.
\end{equation}
It is easy to see that $Q'$ is nilpotent:
\begin{equation}
Q'^{\, 2} = 0 \,.
\end{equation}
We can express $Q'$ as\footnote{
In \cite{Witten:2012bh} the extended BRST operator is introduced as the sum of the BRST operator $Q$
and the exterior derivative on the moduli space or the supermoduli space.}
\begin{equation}
Q' = Q +\tilde{t} \, \partial_t \,.
\label{eq:Qprimedef}
\end{equation}
Since
\begin{equation}
\{ Q', b_{-1} \, \} = L_{-1} \,, \qquad [ \, Q', L_{-1} \, ] = 0 \,,
\end{equation}
the operator $\tilde{t} \, b_{-1} +t \, L_{-1}$
is invariant under the extended BRST transformation,
\begin{equation}
[ \, Q', \tilde{t} \, b_{-1} +t \, L_{-1} \, ] = 0 \,,
\end{equation}
and the operator $\mathrm{e}^{\tilde{t} \, b_{-1} +t \, L_{-1}}$ is also invariant:
\begin{equation}
[ \, Q', \mathrm{e}^{\tilde{t} \, b_{-1} +t \, L_{-1}} \, ] = 0 \,.
\end{equation}
In fact, the operator $\tilde{t} \, b_{-1} +t \, L_{-1}$
can be written in the form of an anticommutator with $Q'$ as
\begin{equation}
\tilde{t} \, b_{-1} +t \, L_{-1}
= \{ \, Q', t \, b_{-1} \, \} \,,
\end{equation}
and the integral over~$t$ takes the following form:
\begin{equation}
-\int_{t_1}^{t_2} \mathrm{d}t \int \mathrm{d} \tilde{t} \,
\langle \, cV_A (t_1) \,
\mathrm{e}^{\{ \, Q',\, t \, b_{-1} \, \}} \cdot cV_B (0) \, cV_C (t_2) \, cV_D (t_3) \, \rangle \,.
\end{equation}
This is the form we will use for generalization to the NS three-point function
in the superstring.

Let us next see the decoupling of BRST-exact states based on this form.
Unintegrated vertex operators are invariant under the extended BRST transformation:
\begin{equation}
Q' \cdot cV_A (t_1) = 0 \,, \quad
Q' \cdot cV_B (0) = 0 \,, \quad
Q' \cdot cV_C (t_2) = 0 \,, \quad
Q' \cdot cV_D (t_3) = 0 \,.
\end{equation}
Again consider the case where $cV_D$ is BRST exact. It can be written as
\begin{equation}
cV_D (t_3) = Q \cdot \Lambda (t_3) = Q' \cdot \Lambda (t_3) \,.
\end{equation}
We then find
\begin{equation}
\begin{split}
& -\int_{t_1}^{t_2} \mathrm{d}t \int \mathrm{d} \tilde{t} \,
\langle \, cV_A (t_1) \,
\mathrm{e}^{\{ \, Q',\, t \, b_{-1} \, \}} \cdot cV_B (0) \, cV_C (t_2) \, Q' \cdot \Lambda (t_3) \, \rangle \\
& = \int_{t_1}^{t_2} \mathrm{d}t \int \mathrm{d} \tilde{t} \,
\langle \, Q' \cdot ( \, cV_A (t_1) \,
\mathrm{e}^{\{ \, Q',\, t \, b_{-1} \, \}} \cdot cV_B (0) \, cV_C (t_2) \, \Lambda (t_3) \, ) \, \rangle \,.
\end{split}
\label{Q'-exact}
\end{equation}
Since
\begin{equation}
\langle \, Q \cdot ( \, cV_A (t_1) \,
\mathrm{e}^{\{ \, Q',\, t \, b_{-1} \, \}} \cdot cV_B (0) \, cV_C (t_2) \, \Lambda (t_3) \, ) \, \rangle = 0 \,,
\end{equation}
the extended BRST operator $Q'$ on the right-hand side of~\eqref{Q'-exact}
can be replaced by $\tilde{t} \, \partial_t$ to give
\begin{equation}
\begin{split}
& \int_{t_1}^{t_2} \mathrm{d}t \int \mathrm{d} \tilde{t} \,
\tilde{t} \, \partial_t \, \langle \, cV_A (t_1) \,
\mathrm{e}^{\{ \, Q',\, t \, b_{-1} \, \}} \cdot cV_B (0) \, cV_C (t_2) \, \Lambda (t_3) \, \rangle \\
& = \int_{t_1}^{t_2} \mathrm{d}t \, \partial_t \,
\langle \, cV_A (t_1) \, cV_B (t) \, cV_C (t_2) \, \Lambda (t_3) \, \rangle \,.
\end{split}
\end{equation}
We have thus reproduced the expression~\eqref{decoupling}.
As can be seen from this example, we in general have
\begin{equation}
\int \mathrm{d} \tilde{t} \, \langle \, Q' \cdot f(t, \tilde{t} \, ) \, \rangle
= \int \mathrm{d} \tilde{t} \, \tilde{t} \, \partial_t \, \langle \, f(t, \tilde{t} \, ) \, \rangle
= \partial_t \, \langle \, f(t, 0 ) \, \rangle
\end{equation}
for any function $f(t, \tilde{t} \, )$ of $t$ and $\tilde{t}$.
Any correlation function with an overall action of $Q'$
is thus a total derivative in the modulus~$t$
after integrating over $\tilde{t}$.
We will use this property in what follows.

\subsection{Three-point functions in the open superstring}
\label{sec:super3pt}

In the preceding subsection,
we considered
the integration over the bosonic modulus~$t$ in the form
\begin{equation}
\int_{t_1}^{t_2} \mathrm{d}t \, \mathrm{e}^{t \, L_{-1}} \,.
\end{equation}
We have learned that this can be combined
with the associated ghost insertion as
\begin{equation}
\int_{t_1}^{t_2} \mathrm{d}t \int \mathrm{d} \tilde{t} \, \mathrm{e}^{\tilde{t} \, b_{-1} +t \, L_{-1}} \,,
\end{equation}
where $\tilde{t}$ is an odd variable, and this can be written
in terms of the extended BRST operator $Q'$ as
\begin{equation}
\int_{t_1}^{t_2} \mathrm{d}t \int \mathrm{d} \tilde{t} \, \mathrm{e}^{\{ \, Q', \, t \, b_{-1} \, \}} \,.
\end{equation}

Let us generalize this to on-shell three-point amplitudes in the NS sector of the open superstring
at the tree level. These amplitudes can be obtained by integrating CFT three-point
functions over the supermoduli space.
Disks with three NS punctures have one odd modulus,
and we denote it by $\zeta$.
Consider the integration over $\zeta$ in the form
\begin{equation}
\int \mathrm{d} \zeta \, \mathrm{e}^{\zeta \, G_{-1/2}} \quad \text{with} \quad
G_{-1/2} = \oint \frac{\mathrm{d}z}{2 \pi \mathrm{i}} \, T_F (z) \,,
\label{zeta-integration}
\end{equation}
where $T_F (z)$ is the supercurrent.
This is motivated by the fact that the operator $G_{-1/2}$ generates the translation along the odd direction of the superspace.
Following~\cite{Witten:2012bh}, we introduce a Grassmann-even variable $\tilde{\zeta}$
and define the actions of the extended BRST operator $Q'$
on $\zeta$ and $\tilde{\zeta}$ by
\begin{equation}
\{ \, Q', \zeta \, \} = \tilde{\zeta} \,, \qquad
[ \, Q', \tilde{\zeta} \, ] = 0 \,.
\end{equation}
As in~\eqref{eq:Qprimedef}, the operator $Q'$ in this case can also be expressed as
\begin{equation}
	Q'=Q+\tilde{\zeta}\partial_\zeta.
\end{equation}
The action of~$Q'$ on states of the boundary CFT
is the same as that of the BRST operator~$Q$.
In particular, we have
\begin{equation}
[ \, Q', \beta_{-1/2} \, ] = G_{-1/2} \,, \qquad
\{ \, Q', G_{-1/2} \, \} = 0 \,,
\end{equation}
where $\beta_{-1/2}$ is a mode of the $\beta$ ghost $\beta (z)$ given by
\begin{equation}
\beta_{-1/2} = \oint \frac{\mathrm{d}z}{2 \pi \mathrm{i}} \, \beta (z) \,.
\end{equation}
The integration over $\zeta$ in~\eqref{zeta-integration}
can be combined with the associated ghost insertion
to yield the following operator:
\begin{equation}
X = \int \mathrm{d} \zeta \mathrm{d} \tilde{\zeta} \, \mathcal{X} ( \zeta, \tilde{\zeta} ) \,,
\label{X-definition}
\end{equation}
where
\begin{equation}
\mathcal{X} ( \zeta, \tilde{\zeta} )= \mathrm{e}^{-\{ \, Q', \, \zeta \, \beta_{-1/2} \, \}}
= \mathrm{e}^{-\{ \, Q', \, \zeta \, \} \, \beta_{-1/2} +\zeta \, [ \, Q', \, \beta_{-1/2} \, ]}
= \mathrm{e}^{-\tilde{\zeta} \, \beta_{-1/2} +\zeta \, G_{-1/2}} \,.
\end{equation}
Note that the operator $X$ commutes with the BRST operator $Q$ since
the integrand $\mathcal{X}(\zeta,\tilde{\zeta})$
is annihilated by $Q'$ and thus
\begin{equation}
	\begin{split}
		[ \, Q, X \, ] &= \int \mathrm{d} \zeta \mathrm{d} \tilde{\zeta} \, [ \, Q,\mathcal{X} ( \zeta, \tilde{\zeta} ) \, ] \\
		      &= -\int \mathrm{d} \zeta \mathrm{d} \tilde{\zeta} \, \tilde{\zeta}\partial_\zeta\mathcal{X} ( \zeta, \tilde{\zeta} ) \\
		      &= 0 \,.
	\end{split}
\end{equation}

Let us carry out the integration over $\zeta$ in the definition of the operator $X$ explicitly.
We use the commutation relations
\begin{equation}
[ \, \beta_{-1/2}, G_{-1/2} \, ] = 2 \, b_{-1} \,, \qquad
[ \, \beta_{-1/2}, b_{-1} \, ] = 0 \,, \qquad
\{ \, G_{-1/2}, b_{-1} \, \} = 0 \,,
\end{equation}
and the Baker-Campbell-Hausdorff formula
to expand $\mathcal{X} ( \zeta, \tilde{\zeta} )$ in $\zeta$ as follows:
\begin{equation}
\begin{split}
\mathcal{X} ( \zeta, \tilde{\zeta} )
& = \mathrm{e}^{-\tilde{\zeta} \, \beta_{-1/2} +\zeta \, G_{-1/2}}
= \mathrm{e}^{-\frac{1}{2} \, [ \, \tilde{\zeta} \, \beta_{-1/2}, \, \zeta \, G_{-1/2} \, ]} \,
\mathrm{e}^{\zeta \, G_{-1/2}} \, \mathrm{e}^{-\tilde{\zeta} \, \beta_{-1/2}} \\
& = \mathrm{e}^{-\tilde{\zeta} \zeta \, b_{-1}} \,
\mathrm{e}^{\zeta \, G_{-1/2}} \, \mathrm{e}^{-\tilde{\zeta} \, \beta_{-1/2}}
= ( \, 1 -\tilde{\zeta} \zeta \, b_{-1} +\zeta \, G_{-1/2} \, ) \, \mathrm{e}^{-\tilde{\zeta} \, \beta_{-1/2}} \,.
\end{split}
\end{equation}
Therefore, we obtain the following expression for $X$
after integrating over $\zeta$:
\begin{equation}
X = G_{-1/2} \, \delta ( \beta_{-1/2} ) +b_{-1} \, \delta' ( \beta_{-1/2} ) \,,
\end{equation}
where the delta function operators are defined by
\begin{equation}
\delta ( \beta_{-1/2} ) = \int \mathrm{d} \tilde{\zeta} \, \mathrm{e}^{-\tilde{\zeta} \, \beta_{-1/2}} \,, \qquad
\delta' ( \beta_{-1/2} ) = -\int \mathrm{d} \tilde{\zeta} \, \tilde{\zeta} \, \mathrm{e}^{-\tilde{\zeta} \, \beta_{-1/2}} \,.
\end{equation}
As emphasized in subsection~3.2.2 of~\cite{Witten:2012bh},
the integration over the Grassmann-even variable $\tilde{\zeta}$
should not be understood as an ordinary integral,
and it should be understood as an algebraic operation
analogous to the integration over Grassmann-odd variables.
Similarly, the delta function operator $\delta(\gamma(t))$ of the $\gamma$ ghost
used in unintegrated vertex operators in the $-1$ picture
is also defined in terms of an integral over a Grassmann-even variable
as an algebraic operation:
\begin{equation}
\delta ( \gamma (t) ) = \int \mathrm{d} \sigma \, \mathrm{e}^{\, \sigma \gamma (t)} \,,
\end{equation}
where $\sigma$ is the Grassmann-even variable.
Based on the prescription for the integration over Grassmann-even variables
in~\cite{Witten:2012bh}, we show in appendix~\ref{sec:beta-gamma} that
\begin{equation}
	\delta(\beta_{-1/2})\cdot \delta(\gamma(0)) =1
	\label{eq:delbetagamma}
\end{equation}
and
\begin{equation}
	\delta'(\beta_{-1/2})\cdot \delta(\gamma(0)) =\gamma(0) \,.
	\label{eq:delprimebetagamma}
\end{equation}
These properties are in accord with the notation in terms of the delta function.
The proofs in appendix~\ref{sec:beta-gamma} can be intuitively understood as
\begin{equation}
	\delta(\beta_{-1/2})\cdot \delta(\gamma(0))
	= \int \mathrm{d} \tau \, \mathrm{e}^{-\beta_{-1/2} \, \tau}\cdot \delta(\gamma(0))
	= \int \mathrm{d} \tau \delta(\gamma(0)+\tau) =1
\end{equation}
and
\begin{equation}
	\delta'(\beta_{-1/2})\cdot \delta(\gamma(0))
	= -\int \mathrm{d} \tau \, \tau\mathrm{e}^{-\beta_{-1/2} \, \tau}\cdot \delta(\gamma(0))
	= -  \int \mathrm{d} \tau \, \tau \delta(\gamma(0)+\tau) =\gamma(0) \,,
\end{equation}
where we have used the fact that the action of $\beta_{-1/2}$ on an arbitrary function $f(\gamma(0))$ of $\gamma(0)$ is
\begin{align}
	\beta_{-1/2}\cdot f(\gamma(0)) = - f'(\gamma(0))
\end{align}
and thus
\begin{align}
	\mathrm{e}^{-\beta_{-1/2} \, \tau}\cdot f(\gamma(0)) =  f(\gamma(0)+\tau) \,.
\end{align}

\begin{figure}
\begin{align*}
	\mathcal{A}= 
	\int \mathrm{d}\zeta\mathrm{d}\tilde{\zeta}\,\,
	\raisebox{-.5\height}{\includegraphics{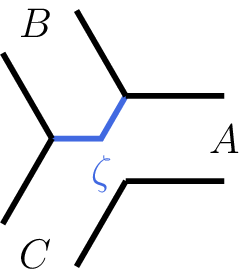}}
\end{align*}
\caption{
The form of on-shell three-point amplitudes in the open superstring.
The blue line represents the insertion of $\mathcal{X} ( \zeta, \tilde{\zeta} )$.
}
\label{fig:3pt1}
\end{figure}

Since the integration over the odd modulus~$\zeta$ can be implemented
by an insertion of the operator~$X$, we expect that on-shell three-point amplitudes
in the open superstring can be written as
\begin{equation}
\begin{split}
\mathcal{A} & = \langle \, \mathcal{V}^{\, (-1)}_A (t_1) \,
X \cdot \mathcal{V}^{\, (-1)}_B (t_2) \,
\mathcal{V}^{\, (-1)}_C (t_3) \, \rangle \\
& = \int \mathrm{d} \zeta \mathrm{d} \tilde{\zeta} \, \langle \,
\mathcal{V}^{\, (-1)}_A (t_1) \,
\mathcal{X} ( \zeta, \tilde{\zeta} ) \cdot \mathcal{V}^{\, (-1)}_B (t_2) \,
\mathcal{V}^{\, (-1)}_C (t_3) \, \rangle \,,
\end{split}
\end{equation}
where $\mathcal{V}^{\, (-1)}_A$, $\mathcal{V}^{\, (-1)}_B$,
and $\mathcal{V}^{\, (-1)}_C$ are on-shell unintegrated vertex operators
in the $-1$ picture given by
\begin{equation}
\mathcal{V}^{\, (-1)}_A = c \delta ( \gamma ) V_A \,, \qquad
\mathcal{V}^{\, (-1)}_B = c \delta ( \gamma ) V_B \,, \qquad
\mathcal{V}^{\, (-1)}_C = c \delta ( \gamma ) V_C
\end{equation}
with $V_A$, $V_B$ and $V_C$ being superconformal primary fields
of weight $1/2$ in the matter sector.
The form of the amplitude $\mathcal{A}$ is illustrated in figure~\ref{fig:3pt1}.
Let us demonstrate that this form of the amplitudes is equivalent
to the standard form given by
\begin{equation}
\mathcal{A} = \langle \,
\mathcal{V}^{\, (-1)}_A (t_1) \,
\mathcal{V}^{\, (0)}_B (t_2) \,
\mathcal{V}^{\, (-1)}_C (t_3) \, \rangle \,,
\end{equation}
where $\mathcal{V}^{\, (0)}_B$ is the on-shell unintegrated vertex operator
in the $0$ picture given by
\begin{equation}
\mathcal{V}^{\, (0)}_B = c G_{-1/2} \cdot V_B +\gamma V_B \,.
\end{equation}
Using the relations~\eqref{eq:delbetagamma} and~\eqref{eq:delprimebetagamma},
the action of $X$ on the unintegrated vertex operator~$\mathcal{V}^{\, (-1)}$
in the $-1$ picture with $V$ being a superconformal primary field of weight $1/2$
in the matter sector can be calculated as
\begin{equation}
	\begin{split}
	X \cdot \mathcal{V}^{\, (-1)} (0) 
	& = \left( G_{-1/2} \, \delta(\beta_{-1/2}) +b_{-1} \, \delta'(\beta_{-1/2}) \right)
	\cdot c(0) \delta(\gamma(0)) V(0) \\
	& = {}-G_{-1/2}\cdot cV(0) - b_{-1}\cdot c\gamma V(0) \\
	& = cG_{-1/2}\cdot V(0) + \gamma V(0) \,,
	\end{split}
\end{equation}
where we have used $G_{-1/2}\cdot c(0) = {}-2 \, \gamma(0)$.
Therefore we have
\begin{equation}
\mathcal{A}
= \langle \,
\mathcal{V}^{\, (-1)}_A (t_1) \,
X \cdot \mathcal{V}^{\, (-1)}_B (t_2) \,
\mathcal{V}^{\, (-1)}_C (t_3) \, \rangle \\
= \langle \,
\mathcal{V}^{\, (-1)}_A (t_1) \,
\mathcal{V}^{\, (0)}_B (t_2) \,
\mathcal{V}^{\, (-1)}_C (t_3) \, \rangle \,.
\end{equation}

We have seen that the operator $X$ changes an on-shell vertex operator $\mathcal{V}^{\, (-1)}$
in the $-1$ picture to an on-shell vertex operator $\mathcal{V}^{\, (0)}$ in the $0$ picture:
\begin{equation}
	X \cdot \mathcal{V}^{\, (-1)} (0) = \mathcal{V}^{\, (0)} (0) \,.
\label{eq:XdotV}
\end{equation}
Its role is similar to that of the picture-changing operator $X_\mathrm{PCO} (z)$,
\begin{equation}
\lim_{z \to 0} X_\mathrm{PCO} (z) \, \mathcal{V}^{\, (-1)} (0) = \mathcal{V}^{\, (0)} (0) \,,
\end{equation}
or that of the zero mode of the picture-changing operator:
\begin{equation}
\oint \frac{\mathrm{d}z}{2 \pi \mathrm{i}} \, \frac{1}{z} \, X_\mathrm{PCO} (z) \, \mathcal{V}^{\, (-1)} (0)
= \mathcal{V}^{\, (0)} (0) \,.
\end{equation}
In fact,  the picture-changing operator $X_\mathrm{PCO} (z)$ is essentially given by
replacing $\beta_{-1/2}$ with $\beta (z)$ in the definition of $X$
with the singularities from coincident operators regularized appropriately
to make the operator invariant under the BRST transformation.
This corresponds to parameterizing the moduli space as
$\mathrm{e}^{\, \zeta G (z)} $ in terms of the local operator $G(z)$
instead of the line integral $G_{-1/2}$\,,
and the zero mode of the picture-changing operator corresponds
to taking a weighted average over the location $z$ of $G(z)$.

The key property of the operator $\mathcal{X} ( \zeta, \tilde{\zeta} )$
is that it is invariant under the extended BRST transformation.
We have seen in the preceding subsection
that the invariance of the operator $\mathrm{e}^{\{Q', \, t \, b_{-1} \}}$
under the extended BRST transformation
ensures the decoupling of BRST-exact states,
and we can similarly show the decoupling of BRST-exact states
in the current case.
Suppose that the operator $\mathcal{V}_C^{\, (-1)}$ is BRST exact
and can be written as
\begin{equation}
\mathcal{V}_C^{\, (-1)} (t_3) = Q \cdot \Lambda (t_3) = Q' \cdot \Lambda (t_3) \,.
\end{equation}
The amplitude~$\mathcal{A}$ is then
\begin{equation}
	\begin{split}
		\mathcal{A} &= \int \mathrm{d} \zeta \mathrm{d} \tilde{\zeta} \, \langle \,
		\mathcal{V}^{\, (-1)}_A (t_1) \,
		\mathcal{X} ( \zeta, \tilde{\zeta} ) \cdot \mathcal{V}^{\, (-1)}_B (t_2) \,
		Q' \cdot \Lambda (t_3) \, \rangle \, \\
		&= \int \mathrm{d} \zeta \mathrm{d} \tilde{\zeta} \, \langle \,
		Q'\cdot\left(\mathcal{V}^{\, (-1)}_A (t_1) \,
			\mathcal{X} ( \zeta, \tilde{\zeta} ) \cdot \mathcal{V}^{\, (-1)}_B (t_2) \,
		\Lambda (t_3) \right) \, \rangle \,\\
		&= \int \mathrm{d} \zeta \mathrm{d} \tilde{\zeta} \, \tilde{\zeta} \partial_{\zeta} \, \langle \,
		\mathcal{V}^{\, (-1)}_A (t_1) \,
		\mathcal{X} ( \zeta, \tilde{\zeta} ) \cdot \mathcal{V}^{\, (-1)}_B (t_2) \,
		\Lambda (t_3) \, \rangle \,\\
		&=0 \,,
	\end{split}
\end{equation}
which vanishes because the integrand is a total derivative in $\zeta$.

\subsection{The definition of the cubic vertex}
\label{sec:cubic-vertex}

Motivated by the discussion in the preceding subsection,
we define the two-string product by
\begin{equation}
[ \, A_1, A_2 \, ]
= \frac{1}{3} \, \Bigl[ \,
X^\star ( A_1 \ast A_2 ) +X A_1 \ast A_2 +A_1 \ast X A_2 \, \Bigr]
\label{eq:cubvertex2}
\end{equation}
with $X$ introduced in~\eqref{X-definition}:
\begin{align}
X = \int \mathrm{d} \zeta \mathrm{d} \tilde{\zeta} \, \mathcal{X} ( \zeta, \tilde{\zeta} )
= \int \mathrm{d} \zeta \mathrm{d} \tilde{\zeta} \, \mathrm{e}^{-\{ \, Q', \, \zeta \, \beta_{-1/2} \, \}}
= \int \mathrm{d} \zeta \mathrm{d} \tilde{\zeta} \, \mathrm{e}^{-\tilde{\zeta} \, \beta_{-1/2} +\zeta \, G_{-1/2}} \,.
\label{eq:Xdef}
\end{align}
Since $X$ commutes with the BRST operator,
this two-string product satisfies the relation~\eqref{derivation}
as we discussed in section~\ref{sec:A_infinity}.
The cyclic property~\eqref{two-string-cyclicity} is also satisfied,
as can be seen from the general discussion in section~\ref{sec:A_infinity}.
Therefore, the action with the cubic term constructed from this two-string product
is gauge invariant up to~$O(g)$.
It is also obvious from the discussion in the preceding subsection
that on-shell three-point amplitudes in the world-sheet theory
are correctly reproduced.

\begin{figure}
\begin{equation*}
	\dfrac{1}{3} \Biggl(
	\raisebox{-.5\height}{\includegraphics{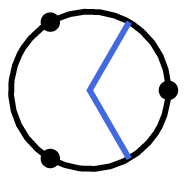}}
	+
	\raisebox{-.5\height}{\includegraphics{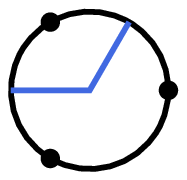}}
	+
	\raisebox{-.5\height}{\includegraphics{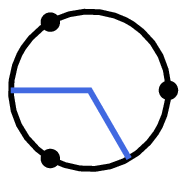}}
	\Biggr)
\end{equation*}
\caption{
A pictorial representation of $\langle \, A_1, [ \, A_2, A_3 \, ] \, \rangle$.
The insertions of the operators
$f_1 \circ \mathcal{V}_1 (0)$, $f_2 \circ \mathcal{V}_2 (0)$, and $f_3 \circ \mathcal{V}_3 (0)$
are represented by dots on the boundary of the disk,
and the insertions of the line integrals $X [ \, f_1 \, ]$, $X [ \, f_2 \, ]$, and $X [ \, f_3 \, ]$
are represented by blue lines.
}
\label{fig:3ptcir}
\end{figure}

It will be useful to present the definition of~$\langle \, A_1, [ \, A_2, A_3 \, ] \, \rangle$
for arbitrary states $A_1$, $A_2$, and $A_3$ in terms of a CFT correlation function
on the disk.
First, the BPZ inner product with the star product $\langle \, A_1, A_2 \ast A_3 \, \rangle$
is defined by
\begin{equation}
	\langle \, A_1, A_2 \ast A_3 \, \rangle =
	\langle \, f_1\circ \mathcal{V}_1 (0) \, f_2 \circ \mathcal{V}_2 (0) \, f_3\circ \mathcal{V}_3 (0) \, \rangle_D \,,
\end{equation}
where $\mathcal{V}_1 (0)$, $\mathcal{V}_2 (0)$, and $\mathcal{V}_3 (0)$
are the operators corresponding to $A_1$, $A_2$, and $A_3$, respectively,
in the state-operator correspondence,
and $f_i \circ \mathcal{V}_i (0)$ with $i=1,2,3$ denotes
the operator mapped from $\mathcal{V}_i (0)$ by the conformal transformation $f_i (\xi)$
defined by
\begin{equation}
	f_1 (\xi) = \left(\frac{1+\mathrm{i}\xi}{1-\mathrm{i}\xi}\right)^{\frac{2}{3}} \,, \quad
	f_2 (\xi) = \mathrm{e}^{\frac{2}{3} \pi \mathrm{i}}\left(\frac{1+\mathrm{i}\xi}{1-\mathrm{i}\xi}\right)^{\frac{2}{3}} \,, \quad
	f_3 (\xi) = \mathrm{e}^{\frac{4}{3} \pi \mathrm{i}}\left(\frac{1+\mathrm{i}\xi}{1-\mathrm{i}\xi}\right)^{\frac{2}{3}} \,.
\end{equation}
The subscript $D$ stands for a correlation function evaluated
on the unit disk.
Then the BPZ inner product~$\langle \, A_1, [ \, A_2, A_3 \, ] \, \rangle$ can be expressed as
\begin{equation}
	\begin{split}
	\langle \, A_1, \, [ \, A_2, A_3 \, ] \, \rangle
	& = \frac{1}{3} \, \bigl( \, \langle \, f_1 \circ ( X \cdot \mathcal{V}_1 (0) ) \,
	f_2 \circ \mathcal{V}_2 (0) \, f_3 \circ \mathcal{V}_3 (0) \, \rangle_D \\
	& \qquad \quad +\langle \, f_1 \circ \mathcal{V}_1 (0) \,
	f_2 \circ ( X \cdot \mathcal{V}_2 (0) ) \, f_3 \circ \mathcal{V}_3 (0) \, \rangle_D \\
	& \qquad \quad +\langle \, f_1 \circ \mathcal{V}_1 (0) \,
	f_2 \circ \mathcal{V}_2 (0) \, f_3 \circ ( X \cdot \mathcal{V}_3 (0) ) \, \rangle_D \, \bigr) \,.
	\end{split}
\end{equation}
The operators $f_i \circ (X \cdot \mathcal{V}_i (0) )$ with $i=1,2,3$ can be written
in terms of line integrals obtained from $X$ by conformal transformations.
We define the line integral $X [ \, f \, ]$ for a conformal transformation $f(\xi)$ by
\begin{equation}
X [ \, f \, ]
= \int \mathrm{d}\zeta \mathrm{d}\tilde{\zeta} \, \mathrm{e}^{-\{Q', \, \zeta\, \beta [ \, f \, ] \, \}} \,,
	\label{eq:defX}
\end{equation}
where $\beta [ \, f \, ]$ is the line integral mapped from $\beta_{-1/2}$ by the conformal transformation $f(\xi)$:
\begin{equation}
\beta [ \, f \, ]
= \oint \frac{\mathrm{d}z}{2 \pi \mathrm{i}} \,
\Bigl( \, \frac{\mathrm{d} f^{-1} (z)}{\mathrm{d}z} \, \Bigr)^{-1/2} \, \beta (z) \quad
\text{with} \quad z = f(\xi) \,.
\end{equation}
Then, we have
\begin{equation}
f_i \circ (X \cdot \mathcal{V}_i (0) ) = X [ \, f_i \, ] \, f_i \circ \mathcal{V}_i (0)
\end{equation}
for $i=1,2,3$,
and $\langle \, A_1, [ \, A_2, A_3 \, ] \, \rangle$ can be written as follows:
\begin{equation}
	\langle \, A_1, \, [ \, A_2, A_3 \, ] \, \rangle =
	\frac{1}{3} \, \langle \, ( \, X [ \, f_1\, ] +X [ \, f_2 \, ] +X [ \, f_3 \, ] \, ) \,
	f_1\circ \mathcal{V}_1 (0) \, f_2\circ \mathcal{V}_2 (0) \, f_3\circ \mathcal{V}_3 (0) \, \rangle_D \,.
\label{cubic-vertex}
\end{equation}
This form of $\langle \, A_1, \, [ \, A_2, A_3 \, ] \, \rangle$ is pictorially represented in figure~\ref{fig:3ptcir}.

\section{Quartic interaction}
\label{sec:quartic}

In section~\ref{sec:cubic} we constructed
a two-string product $[ \, A_1, A_2 \, ]$ for a pair of states $A_1$ and $A_2$
in such a way that the associated cubic vertex $\langle \, A_1, [ \, A_2, A_3 \, ] \, \rangle$
defined for arbitrary states $A_1$, $A_2$, and $A_3$ 
takes the form of an integral over an odd modulus $\zeta$ of disks with three NS punctures.
Then Feynman diagrams with two cubic vertices and one propagator
are expressed as integrals over two odd moduli from the vertices
and one even modulus from the propagator,
and contributions from these Feynman diagrams partially cover
the supermoduli space of disks with four NS punctures.
The goal of this section is to construct a three-string product $[ \, A_1, A_2, A_3 \, ]$
defined for three states $A_1$, $A_2$, and $A_3$
satisfying the relations~\eqref{A_infinity} and~\eqref{three-string-cyclicity}
in such a way that the associated quartic vertex
$\langle \, A_1, [ \, A_2, A_3, A_4 \, ] \, \rangle$
defined for arbitrary states $A_1$, $A_2$, $A_3$, and $A_4$ 
takes the form of an integral over the remaining region
of the supermoduli space of disks with four NS punctures.

\subsection{Nonassociativity of the two-string product}
\label{sec:non-assoc}

Let us first see that the two-string product~\eqref{eq:cubvertex2} is not associative
and therefore we need a three-string product to satisfy the relation~\eqref{A_infinity}.
There are two terms in~\eqref{A_infinity} where the two-string product is used twice.
One of them is $[ \, [ \, A_1, A_2 \, ], A_3 \, ]$ and is given by
\begin{equation}
	\begin{split}
		& [ \, [ \, A_1, A_2 \, ], A_3 \, ] \\
		& = \frac{1}{9} \, \Bigl( \,
		X^\star ( X^\star ( A_1 \ast A_2 ) \ast A_3 )
		+X X^\star ( A_1 \ast A_2 ) \ast A_3
	       +X^\star ( A_1 \ast A_2 ) \ast X A_3 \\
	       & \qquad \quad
	       +X^\star ( X A_1 \ast A_2 \ast A_3 )
		+X ( X A_1 \ast A_2 ) \ast A_3
	       +X A_1 \ast A_2 \ast X A_3 \\
	       & \qquad \quad
	       +X^\star ( A_1 \ast X A_2 \ast A_3 ) 
		+X ( A_1 \ast X A_2 ) \ast A_3
	       +A_1 \ast X A_2 \ast X A_3 \, \Bigr) \,.
		\label{eq:iterat1}
	\end{split}
\end{equation}
The other term is $[ \, A_1, [ \, A_2, A_3 \, ] \, ]$ and is given by
\begin{equation}
	\begin{split}
		& [ \, A_1, [ \, A_2, A_3 \, ] \, ] \\
		& = \frac{1}{9} \, \Bigl( \,
		X^\star ( A_1 \ast X^\star ( A_2 \ast A_3 ) )
		+X A_1 \ast X^\star ( A_2 \ast A_3 )
		+A_1 \ast X X^\star ( A_2\ast A_3 ) \\
		& \qquad \quad
		+X^\star ( A_1 \ast X A_2 \ast A_3 )
		+X A_1 \ast X A_2 \ast A_3
		+A_1 \ast X ( X A_2 \ast A_3 ) \\
		& \qquad \quad
		+X^\star ( A_1 \ast A_2 \ast X A_3 )
		+X A_1 \ast A_2 \ast X A_3
		+A_1 \ast X ( A_2 \ast X A_3 ) \, \Bigr) \,.
	\end{split}
		\label{eq:iterat2}
\end{equation}
Since the operators $X$ and $X^\star$ are inserted differently, 
$[ \, [ \, A_1, A_2 \, ], A_3 \, ]$ and $[ \, A_1, [ \, A_2, A_3 \, ] \, ]$ do not coincide
and the two-string product is not associative:
\begin{equation}
[ \, [ \, A_1, A_2 \, ], A_3 \, ] \ne [ \, A_1, [ \, A_2, A_3 \, ] \, ] \,.
\end{equation}

\begin{figure}
\begin{center}
    \includegraphics[scale=1]{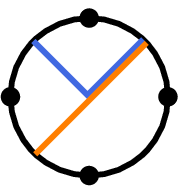}
\end{center}
\caption{
A pictorial representation of $X [ \, g_2 \, ]$ and $X [ \, h_2 \, ]$ in~\eqref{eq:4ptXX}.
The blue line represents $X [ \, g_2 \, ]$ and the orange line is for $X [ \, h_2 \, ]$.
}
\label{fig:4ptcir1}
\end{figure}

It will be convenient to express associated quartic vertices
$\langle \, A_1, [ \, [ \, A_2, A_3 \, ], A_4 \, ] \, \rangle$
and $\langle \, A_1, [ \, A_2, [ \, A_3, A_4 \, ] \, ] \, \rangle$
defined for four states $A_1$, $A_2$, $A_3$, and $A_4$
in terms of CFT correlation functions
on the disk.
First, the BPZ inner product $\langle \, A_1, A_2 \ast A_3 \ast A_4 \, \rangle$
can be expressed as
\begin{equation}
	\langle \, A_1, A_2 \ast A_3 \ast A_4 \, \rangle
	= \langle \, g_1\circ \mathcal{V}_1 (0) \, g_2 \circ \mathcal{V}_2 (0)\,
	g_3 \circ \mathcal{V}_3 (0) \, g_4 \circ \mathcal{V}_4 (0) \, \rangle_D \,,
\end{equation}
where $\mathcal{V}_1 (0)$, $\mathcal{V}_2 (0)$, $\mathcal{V}_3 (0)$, and $\mathcal{V}_4 (0)$
are the operators corresponding to $A_1$, $A_2$, $A_3$, and $A_4$, respectively,
in the state-operator correspondence,
and $g_i \circ \mathcal{V}_i (0)$ for $i=1,2,3,4$ denotes
the operator mapped from $\mathcal{V}_i (0)$ by the conformal transformation $g_i (\xi)$
defined by
\begin{equation}
\begin{split}
	& g_1 (\xi) = \left(\frac{1+\mathrm{i}\xi}{1-\mathrm{i}\xi}\right)^{\frac{1}{2}} \,, \quad
	g_2 (\xi) =
	\mathrm{e}^{\frac{1}{2} \pi \mathrm{i}}\left(\frac{1+\mathrm{i}\xi}{1-\mathrm{i}\xi}\right)^{\frac{1}{2}} \,, \\
	& g_3 (\xi) =
	\mathrm{e}^{\pi \mathrm{i}}\left(\frac{1+\mathrm{i}\xi}{1-\mathrm{i}\xi}\right)^{\frac{1}{2}} \,, \quad
	g_4 (\xi) =
	\mathrm{e}^{\frac{3}{2} \pi \mathrm{i}}\left(\frac{1+\mathrm{i}\xi}{1-\mathrm{i}\xi}\right)^{\frac{1}{2}} \,.
\end{split}
\end{equation}
Then the term $\langle \, A_1,  X A_2 \ast X A_3 \ast A_4 \, \rangle$
in $\langle \, A_1, [ \, A_2, [ \, A_3, A_4 \, ] \, ] \, \rangle$, for example, can be expressed
using line integrals $X [ \, g_2 \, ]$ and $X [ \, g_3 \, ]$ introduced in the preceding section:\footnote{
To be precise, we should specify a way
to avoid the collision of contours of the line integrals $X [ \, g_2 \, ]$ and  $X [ \, g_3 \, ]$.
One natural way would be to use star products with stubs as discussed, for example, in~\cite{Erler:2014eba}.
We can locate the contours within the stubs and take the limit of vanishing stubs.
This procedure induces a way to deform the contours in general.
In this paper we instead specify a way to deform the contours explicitly.
For $X [ \, g_i \, ]$,
we deform $g_i (\xi)$ to $g_i^\epsilon (\xi) = g_i ( \mathrm{e}^{-\epsilon} \xi)$
with $\epsilon$ being a positive constant which is sufficiently small.
We will implicitly use the same prescription in the rest of the paper when needed.
}
\begin{equation}
	\langle \, A_1,  X A_2 \ast X A_3 \ast A_4 \, \rangle
	= \langle \, X [ \, g_2 \, ] \, X [ \, g_3 \, ] \,\, g_1\circ \mathcal{V}_1 (0) \,
	g_2 \circ \mathcal{V}_2 (0) \, g_3 \circ \mathcal{V}_3 (0) \, g_4 \circ \mathcal{V}_4 (0) \, \rangle_D.
\end{equation}
We also need $X [ \, h_i \, ]$ for $i = 1, 2, 3, 4$ with $h_i (\xi)$ defined by
\begin{equation}
\begin{split}
	h_1 (\xi) & = \mathrm{e}^{\frac{1}{4} \pi \mathrm{i}} \left(\frac{1+\mathrm{i}\xi}{1-\mathrm{i}\xi}\right), \quad
	h_2 (\xi ) = \mathrm{e}^{\frac{3}{4} \pi \mathrm{i}} \left(\frac{1+\mathrm{i}\xi}{1-\mathrm{i}\xi}\right), \\
	h_3 (\xi ) & = \mathrm{e}^{\frac{5}{4} \pi \mathrm{i}} \left(\frac{1+\mathrm{i}\xi}{1-\mathrm{i}\xi}\right), \quad
	h_4 (\xi ) = \mathrm{e}^{\frac{7}{4} \pi \mathrm{i}} \left(\frac{1+\mathrm{i}\xi}{1-\mathrm{i}\xi}\right).
\end{split}
\end{equation}
We then have
\begin{equation}
\begin{split}
	\langle \, A_1,  A_2 \ast X^\star ( A_3 \ast A_4 ) \, \rangle
	& = \langle \, X [ \, h_1 \, ] \,\, g_1\circ \mathcal{V}_1 (0) \,
	g_2 \circ \mathcal{V}_2 (0) \, g_3 \circ \mathcal{V}_3 (0) \, g_4 \circ \mathcal{V}_4 (0) \, \rangle_D \,, \\
	\langle \, A_1,  X ( A_2 \ast A_3 ) \ast A_4 \, \rangle
	& = \langle \, X [ \, h_2 \, ] \,\, g_1\circ \mathcal{V}_1 (0) \,
	g_2 \circ \mathcal{V}_2 (0) \, g_3 \circ \mathcal{V}_3 (0) \, g_4 \circ \mathcal{V}_4 (0) \, \rangle_D \,, \\
	\langle \, A_1,  A_2 \ast X ( A_3 \ast A_4 ) \, \rangle
	& = \langle \, X [ \, h_3 \, ] \,\, g_1\circ \mathcal{V}_1 (0) \,
	g_2 \circ \mathcal{V}_2 (0) \, g_3 \circ \mathcal{V}_3 (0) \, g_4 \circ \mathcal{V}_4 (0) \, \rangle_D \,, \\
	\langle \, A_1,  X^\star ( A_2 \ast A_3 ) \ast A_4 \, \rangle
	& = \langle \, X [ \, h_4 \, ] \,\, g_1\circ \mathcal{V}_1 (0) \,
	g_2 \circ \mathcal{V}_2 (0) \, g_3 \circ \mathcal{V}_3 (0) \, g_4 \circ \mathcal{V}_4 (0) \, \rangle_D \,.
\end{split}
\end{equation}
The term $\langle \, A_1,  X ( X A_2 \ast A_3 ) \ast A_4 \, \rangle$
in $\langle \, A_1, [ \, [ \, A_2, A_3 \, ], A_4 \, ] \, \rangle$, for example, can be expressed as\footnote{
As we mentioned before, we can use star products with stubs
and take the limit of vanishing stubs
to induce a way to avoid the collision of the contours
for $X [ \, h_2 \, ]$ and $X [ \, g_2 \, ]$.
In this paper we instead specify a way to deform the contours explicitly.
For $X [ \, h_i \, ]$,
we deform $h_i (\xi)$ to $h_i^\epsilon (\xi) =h_i ( e^\epsilon \xi)$
with $\epsilon$ being a positive constant which is sufficiently small.
(Note that we use $h_i ( e^\epsilon \xi)$ instead of $h_i ( \mathrm{e}^{-\epsilon} \xi)$.)
We will implicitly use the same prescription in the rest of the paper when needed.
}
\begin{equation}
	\langle \, A_1,  X ( X A_2 \ast A_3 ) \ast A_4 \, \rangle
	= \langle \, X [ \, h_2 \, ] \, X [ \, g_2 \, ] \,\,
	g_1\circ \mathcal{V}_1 (0) \, g_2 \circ \mathcal{V}_2 (0) \,
	g_3 \circ \mathcal{V}_3 (0) \, g_4 \circ \mathcal{V}_4 (0) \, \rangle_D \,.
	\label{eq:4ptXX}
\end{equation}
The insertions of $X [ \, g_2 \, ]$ and $X [ \, h_2 \, ]$ are pictorially represented in figure~\ref{fig:4ptcir1}.
This way we can express all terms of $\langle \, A_1, [ \, [ \, A_2, A_3 \, ], A_4 \, ] \, \rangle$
and $\langle \, A_1, [ \, A_2, [ \, A_3, A_4 \, ]  \, ] \, \rangle$ using the line integrals.
For $\langle \, A_1, [ \, [ \, A_2, A_3 \, ], A_4 \, ] \, \rangle$,  we have
\begin{equation}
	\langle \, A_1, [ \, [ \, A_2, A_3 \, ], A_4 \, ]\, \rangle
	= \langle \, X_t \,\,
	g_1\circ \mathcal{V}_1 (0) \, g_2 \circ \mathcal{V}_2 (0) \,
	g_3 \circ \mathcal{V}_3 (0) \, g_4 \circ \mathcal{V}_4 (0) \, \rangle_D
\end{equation}
with
\begin{equation}
X_t = \frac{1}{9} \, \Bigl( \, X [ \, h_4 \, ] +X [ \, g_2 \, ] +X [ \, g_3 \, ] \, \Bigr) \,
\Bigl( \, X [ \, g_1 \, ] +X [ \, h_2 \, ] +X [ \, g_4 \, ] \, \Bigr) \,.
\end{equation}
For $\langle \, A_1, [ \, A_2, [ \, A_3, A_4 \, ]  \, ] \, \rangle$,  we have
\begin{equation}
	\langle \, A_1, [ \, A_2, [ \, A_3, A_4 \, ]  \, ] \, \rangle
	= \langle \, X_s \,\,
	g_1\circ \mathcal{V}_1 (0) \, g_2 \circ \mathcal{V}_2 (0) \,
	g_3 \circ \mathcal{V}_3 (0) \, g_4 \circ \mathcal{V}_4 (0) \, \rangle_D
\end{equation}
with
\begin{equation}
X_s = \frac{1}{9} \, \Bigl( \, X [ \, h_1 \, ] +X [ \, g_3 \, ] +X [ \, g_4 \, ] \, \Bigr) \,
\Bigl( \, X [ \, g_1 \, ] +X [ \, g_2 \, ] +X [ \, h_3 \, ] \, \Bigr) \,.
\end{equation}
The difference between $[ \, [ \, A_2, A_3 \, ], A_4 \, ]$ and $[ \, A_2, [ \, A_3, A_4 \, ] \, ]$ is
therefore captured by the operator $X_t-X_s$.

\subsection{Interpolation in four-point functions of the bosonic string}
\label{sec:interpolation-bosonic}

We have calculated $[ \, [ \, A_1, A_2 \, ], A_3 \, ] -[ \, A_1, [ \, A_2, A_3 \, ] \, ]$
and the next step is to construct a three-string product $[ \, A_1, A_2, A_3 \, ]$
which satisfies the relation~\eqref{A_infinity}.
When $X$ is a line integral of the picture-changing operator,
such a three-string product was constructed in~\cite{Erler:2013xta}
using the fact that $X$ can be written
as an anticommutator with the BRST operator, $X = \{ \, Q, \xi \, \}$,
if we allow $\xi$ to be an operator
in the large Hilbert space of the superconformal ghost sector.
The main point of this paper is that we construct such a three-string product
in the perspective of the supermoduli space of super-Riemann surfaces.
The construction is based on the $\beta \gamma$ ghosts
and we will not use the large Hilbert space at any intermediate stage.
Our construction can be thought of as a generalization
of the construction of a three-string product
in open bosonic string field theory when we use a nonassociative two-string product
instead of the associative star product.
In this subsection we explain relevant aspects of four-point functions
of the open bosonic string
in the language which can be generalized to the superstring.

Let us consider the following integral over the moduli space
for on-shell four-point functions in the open bosonic string:\footnote{
Strictly speaking, the point at infinity is outside the coordinate patch,
and the operator $\mathcal{V}_D (\infty)$ should be understood as being defined
using another coordinate patch.
We will not encounter any subtlety coming from this simplified treatment.
}
\begin{equation}
	\int_0^1 \mathrm{d}t \,
	\langle \, \mathcal{V}_A (0) \, b_{-1} \, \mathrm{e}^{t \, L_{-1}} \cdot \mathcal{V}_B (0) \,
\mathcal{V}_C (1) \, \mathcal{V}_D (\infty) \, \rangle \,,
\end{equation}
where $\mathcal{V}_A$, $\mathcal{V}_B$, $\mathcal{V}_C$, and $\mathcal{V}_D$
are unintegrated vertex operators annihilated by the BRST operator.
In the context of string field theory, this integral is obtained from
Feynman diagrams with two cubic vertices and one propagator.
Suppose that the contribution $\mathcal{A}_s$ in the $s$ channel covers the region
\begin{equation}
	\mathcal{A}_s = \int_0^{1/2} \mathrm{d}t \,
	\langle \, \mathcal{V}_A (0) \, b_{-1} \, \mathrm{e}^{t \, L_{-1}} \cdot \mathcal{V}_B (0) \,
\mathcal{V}_C (1) \, \mathcal{V}_D (\infty) \, \rangle
\end{equation}
and the contribution $\mathcal{A}_t$ in the $t$ channel covers the region
\begin{equation}
	\mathcal{A}_t = \int_{1/2}^1 \mathrm{d}t \,
	\langle \, \mathcal{V}_A (0) \, b_{-1} \, \mathrm{e}^{t \, L_{-1}} \cdot \mathcal{V}_B (0) \,
\mathcal{V}_C (1) \, \mathcal{V}_D (\infty) \, \rangle \,.
\end{equation}
In this case, the sum of the two regions covers the whole moduli space
and correct on-shell amplitudes are reproduced
without any contributions from Feynman diagrams with quartic vertices
which we will denote by $\mathcal{A}_4$.
This is the situation in open bosonic string field theory
with the associative star product.
If we use a nonassociative two-string product, the sum of $\mathcal{A}_s$ and $\mathcal{A}_t$
does not cover the whole moduli space.
Let us model the situation in the following way.
Suppose that $\mathcal{A}_s$ and $\mathcal{A}_t$ are given by
\begin{align}
	\mathcal{A}_s & = \int_0^{t_-} \mathrm{d}t \,
	\langle \, \mathcal{V}_A (0) \, b_{-1} \, \mathrm{e}^{t \, L_{-1}} \cdot \mathcal{V}_B (0) \,
\mathcal{V}_C (1) \, \mathcal{V}_D (\infty) \, \rangle \,,
\label{A_-} \\
\mathcal{A}_t & = \int_{t_+}^1 \mathrm{d}t \,
\langle \, \mathcal{V}_A (0) \, b_{-1} \, \mathrm{e}^{t \, L_{-1}} \cdot \mathcal{V}_B (0) \,
\mathcal{V}_C (1) \, \mathcal{V}_D (\infty) \, \rangle \,,
\label{A_+-1}
\end{align}
where $0 < t_- < t_+ < 1$.
In this case, the region $t_- < t < t_+$ of the moduli space is missing,
and the decoupling of unphysical states discussed in subsection~\ref{sec:bos4pt} does not work.
As in subsection~\ref{sec:bos4pt},
consider the case where the vertex operator $\mathcal{V}_D$ is BRST exact and is written as
\begin{equation}
\mathcal{V}_D (\infty) = Q \cdot \Lambda (\infty) \,.
\end{equation}
Then the integrals $\mathcal{A}_s$ and $\mathcal{A}_t$ localize at the boundaries of the integral regions as
\begin{equation}
\begin{split}
\mathcal{A}_s & = \langle \, \mathcal{V}_A (0) \, \mathcal{V}_B (t_-) \,
\mathcal{V}_C (1) \, \Lambda (\infty) \, \rangle \,, \\
\mathcal{A}_t & = {}-\langle \, \mathcal{V}_A (0) \, \mathcal{V}_B (t_+) \,
\mathcal{V}_C (1) \, \Lambda (\infty) \, \rangle \,,
\end{split}
\end{equation}
and these surface terms do not cancel because $t_- \ne t_+$.

In this case it is easy to remedy the problem.
We can add $\mathcal{A}_4$ given by
\begin{equation}
	\mathcal{A}_4 = \int_{t_-}^{t_+} \mathrm{d}t \,
	\langle \, \mathcal{V}_A (0) \, b_{-1} \, \mathrm{e}^{t \, L_{-1}} \cdot \mathcal{V}_B (0) \,
\mathcal{V}_C (1) \, \mathcal{V}_D (\infty) \, \rangle
\label{A_mid-1}
\end{equation}
to cover the missing region.
When $\mathcal{V}_D (\infty) = Q \cdot \Lambda (\infty)$,
the integral $\mathcal{A}_4$ localizes as
\begin{equation}
\mathcal{A}_4
= {}-\langle \, \mathcal{V}_A (0) \, \mathcal{V}_B (t_-) \, \mathcal{V}_C (1) \, \Lambda (\infty) \, \rangle
+\langle \, \mathcal{V}_A (0) \, \mathcal{V}_B (t_+) \, \mathcal{V}_C (1) \, \Lambda (\infty) \, \rangle \,,
\end{equation}
and the sum of all the surface terms vanishes.
In fact, this is the essence of the connection between the covering of the moduli space
and the relation~\eqref{A_infinity}.
In this context, the operators $\mathcal{V}_A$, $\mathcal{V}_B$, $\mathcal{V}_C$,
and $\mathcal{V}_D$ are off shell and their ghost numbers can be arbitrary.
The key relation is
\begin{equation}
\begin{split}
	& \int_{t_-}^{t_+} \mathrm{d}t \,
	\langle \, \mathcal{V}_A (0) \, \{ Q, b_{-1} \, \mathrm{e}^{t \, L_{-1}} \, \} \cdot \mathcal{V}_B (0) \,
\mathcal{V}_C (1) \, \mathcal{V}_D (\infty) \, \rangle \\
& = \langle \, \mathcal{V}_A (0) \, \mathcal{V}_B (t_+) \,
\mathcal{V}_C (1) \, \mathcal{V}_D (\infty) \, \rangle
-\langle \, \mathcal{V}_A (0) \, \mathcal{V}_B (t_-) \,
\mathcal{V}_C (1) \, \mathcal{V}_D (\infty) \, \rangle \,.
\end{split}
\label{bosonic-key-relation-1}
\end{equation}
The left-hand side corresponds to the sum of the four terms in~\eqref{A_infinity}
which contain the BRST operator,
and the right-hand side corresponds
to the nonassociativity of the two-string product in~\eqref{A_infinity}.
Therefore, when the nonassociativity of the two-string product in~\eqref{A_infinity}
is expressed in terms of four-point functions as
\begin{equation}
\langle \, \mathcal{V}_A (0) \, \mathcal{V}_B (t_+) \,
\mathcal{V}_C (1) \, \mathcal{V}_D (\infty) \, \rangle
-\langle \, \mathcal{V}_A (0) \, \mathcal{V}_B (t_-) \,
\mathcal{V}_C (1) \, \mathcal{V}_D (\infty) \, \rangle \,,
\end{equation}
we can construct a three-string product based on $\mathcal{A}_4$
which takes the form of an integral over the missing region
of the moduli space.

In open bosonic string field theory with a nonassociative two-string product,
the construction of a three-string product satisfying~\eqref{A_infinity}
is more complicated.
Let us therefore consider a more nontrivial case.
Suppose that the contribution $\mathcal{A}_s$ is the same as before,
but $\mathcal{A}_t$ is different and is given by
\begin{equation}
	\mathcal{A}_t = {}-\int_1^{1/t_+} \mathrm{d}t \,
\langle \, \mathcal{V}_A (0) \, \mathcal{V}_B (1) \,
b_{-1} \, \mathrm{e}^{t \, L_{-1}} \cdot \mathcal{V}_C (0) \, \mathcal{V}_D (\infty) \, \rangle \,.
\label{A_+-2}
\end{equation}
As can be seen from the calculation of the cross ratio,
the contribution $\mathcal{A}_t$ covers the region $t_+ \le t \le 1$ of the moduli space,
and the missing region is again $t_- < t < t_+$.
When $\mathcal{V}_D (\infty) = Q \cdot \Lambda (\infty)$,
the integral $\mathcal{A}_t$ localizes as
\begin{equation}
\mathcal{A}_t = {}-\langle \, \mathcal{V}_A (0) \, \mathcal{V}_B (1) \,
\mathcal{V}_C (1/t_+) \, \Lambda (\infty) \, \rangle \,.
\end{equation}
In the context of constructing a three-string product satisfying~\eqref{A_infinity},
the nonassociativity of the two-string product now corresponds to
\begin{equation}
\langle \, \mathcal{V}_A (0) \, \mathcal{V}_B (1) \,
\mathcal{V}_C (1/t_+) \, \mathcal{V}_D (\infty) \, \rangle
-\langle \, \mathcal{V}_A (0) \, \mathcal{V}_B (t_-) \,
\mathcal{V}_C (1) \, \mathcal{V}_D (\infty) \, \rangle \,.
\label{bosonic-example-2}
\end{equation}
How can we interpolate between these terms?

In the previous case, the terms we need to interpolate can be written as
\begin{equation}
\begin{split}
& \langle \, \mathcal{V}_A (0) \, \mathcal{V}_B (t_+) \,
\mathcal{V}_C (1) \, \mathcal{V}_D (\infty) \, \rangle
-\langle \, \mathcal{V}_A (0) \, \mathcal{V}_B (t_-) \,
\mathcal{V}_C (1) \, \mathcal{V}_D (\infty) \, \rangle \\
& = \langle \, \mathcal{V}_A (0) \, \mathrm{e}^{(t_+ -t_0) \, L_{-1}} \cdot \mathcal{V}_B (t_0) \,
\mathcal{V}_C (1) \, \mathcal{V}_D (\infty) \, \rangle \\
& \quad~ {}-\langle \, \mathcal{V}_A (0) \, \mathrm{e}^{(t_{-}-t_0) \, L_{-1}} \cdot \mathcal{V}_B (t_0) \,
\mathcal{V}_C (1) \, \mathcal{V}_D (\infty) \, \rangle \,,
\end{split}
\end{equation}
where $t_0$ in the range $0 < t_0 < t_-$ is an arbitrary reference point.
We further express these terms in terms of the line integral $L [ \, f \, ]$
defined for a conformal transformation $f(\xi)$ by
\begin{equation}
	L [ \, f \, ] = \oint \frac{\mathrm{d}z}{2 \pi \mathrm{i}} \,
	\Bigl( \, \frac{\mathrm{d} f^{-1} (z)}{\mathrm{d}z} \, \Bigr)^{-1} \, T(z) \quad
\text{with} \quad z = f(\xi) \,.
\end{equation}
We write
\begin{equation}
\begin{split}
& \langle \, \mathcal{V}_A (0) \, \mathcal{V}_B (t_+) \,
\mathcal{V}_C (1) \, \mathcal{V}_D (\infty) \, \rangle
-\langle \, \mathcal{V}_A (0) \, \mathcal{V}_B (t_-) \,
\mathcal{V}_C (1) \, \mathcal{V}_D (\infty) \, \rangle \\
& = \langle \, \mathcal{V}_A (0) \, ( \mathcal{L}_t -\mathcal{L}_s ) \, \mathcal{V}_B (t_0) \,
\mathcal{V}_C (1) \, \mathcal{V}_D (\infty) \, \rangle \,,
\end{split}
\end{equation}
where $\mathcal{L}_t$ and $\mathcal{L}_s$ are
\begin{equation}
	\mathcal{L}_t = \mathrm{e}^{(t_+ -t_0) \, L [ \, f_0 \, ]} \,, \qquad
	\mathcal{L}_s = \mathrm{e}^{(t_- -t_0) \, L [ \, f_0 \, ]}
\end{equation}
with $f_0 (\xi) = \xi+t_0$.
Then the key relation~\eqref{bosonic-key-relation-1} can be stated as
\begin{equation}
Q \cdot \mathcal{B} = \mathcal{L}_t -\mathcal{L}_s \,,
\end{equation}
where
\begin{equation}
	\mathcal{B} = \int_{t_-}^{t_+} \mathrm{d}t \, b [ \, f_0 \, ] \, \mathrm{e}^{(t -t_0) \, L [ \, f_0 \, ]}
\end{equation}
with the line integral $b [ \, f \, ]$
being defined for a conformal transformation $f(\xi)$ by
\begin{equation}
	b [ \, f \, ] = \oint \frac{\mathrm{d}z}{2 \pi \mathrm{i}} \,
	\Bigl( \, \frac{\mathrm{d} f^{-1} (z)}{\mathrm{d}z} \, \Bigr)^{-1} \, b(z) \quad
\text{with} \quad z = f(\xi) \,.
\end{equation}
In the case with~\eqref{bosonic-example-2},
we need to solve a more general problem.
In general, the problem we need to solve is to find an appropriate operator $B$
which satisfies
\begin{equation}
\begin{split}
& \langle \, \mathcal{V}_A (t_1) \, \{ \, Q, B \} \cdot \mathcal{V}_B (t_2) \,
\mathcal{V}_C (t_3) \, \mathcal{V}_D (t_4) \, \rangle \\
& = \langle \, \mathcal{V}_A (t_1) \, \mathrm{e}^{\, p_+ L_{-1}} \cdot \mathcal{V}_B (t_2) \,
\mathrm{e}^{\, q_+ L_{-1}} \cdot \mathcal{V}_C (t_3) \, \mathcal{V}_D (t_4) \, \rangle \\
& \quad~ {}-\langle \, \mathcal{V}_A (t_1) \, \mathrm{e}^{\, p_- L_{-1}} \cdot \mathcal{V}_B (t_2) \,
\mathrm{e}^{\, q_- L_{-1}} \cdot \mathcal{V}_C (t_3) \, \mathcal{V}_D (t_4) \, \rangle \,,
\end{split}
\label{interpolation-model}
\end{equation}
where $p_+$, $q_+$, $q_-$, and $p_-$ are some constants.\footnote{
We consider a simplified situation where the conformal transformation
is generated by $L_{-1}$,
but we need to consider a more general linear combination of Virasoro generators
in open bosonic string field theory with a nonassociative two-string product.
This generalization is straightforward.
Furthermore, all vertex operators are in general transformed,
while we consider a simplified situation where only two vertex operators
are transformed.
Once we understand how we interpolate between the two terms
on the right-hand side of~\eqref{interpolation-model},
the extension to more general cases is straightforward.
We will also need an analogous interpolation
for the open superstring in the next subsection.
Since this is our motivation
to consider the interpolation for~\eqref{interpolation-model},
we do not divide the interpolation into two steps
where we interpolate between the terms for $\mathcal{V}_B$
in the first step using the previous procedure
and we interpolate between the terms for $\mathcal{V}_C$ in the second step.
}
In the language of line integrals, this can be translated as finding $\mathcal{B}$
which satisfies
\begin{equation}
Q \cdot \mathcal{B}
= \mathrm{e}^{\, p_+ \, L [ \, f_2 \, ]} \, \mathrm{e}^{\, q_+ \, L [ \, f_3 \, ]}
-\mathrm{e}^{\, p_- \, L [ \, f_2 \, ]} \, \mathrm{e}^{\, q_- \, L [ \, f_3 \, ]}
\label{bosonic-key-relation-2}
\end{equation}
with $f_2 (\xi) = \xi +t_2$ and $f_3 (\xi) = \xi +t_3$.
To construct $\mathcal{B}$ satisfying this relation,
it is convenient to use the extended BRST transformation.
It follows from the general consideration on appropriate measures
for integrals over the moduli space in~\cite{Witten:2012bh} that
$\mathcal{B}$ satisfying
\begin{equation}
Q \cdot \mathcal{B}
= \mathrm{e}^{\{ \, Q,\, \mathcal{F} (t_+) \, \}} -\mathrm{e}^{\{ \, Q,\, \mathcal{F} (t_-) \, \}}
\label{interpolation-identity}
\end{equation}
for a Grassmann-odd operator $\mathcal{F} (t)$ can be constructed as
\begin{equation}
	\mathcal{B} = \int_{t_-}^{t_+} \mathrm{d}t \int \mathrm{d} \tilde{t} \, \mathrm{e}^{\{ \, Q',\, \mathcal{F} (t) \, \}} \,.
\end{equation}
The structure $\{ \, Q',\, \mathcal{F} (t) \, \}$ in the exponent
and the integrals over $t$ and $\tilde{t}$
are the ingredients we expect from the world-sheet path integral in\cite{Witten:2012bh}.
As we discussed in subsection~\ref{sec:bos4pt}, this can be shown in the following way:
\begin{equation}
\begin{split}
& Q \cdot \mathcal{B}
	= -\int_{t_-}^{t_+} \mathrm{d}t \int \mathrm{d} \tilde{t} \, \Bigl( \, Q \cdot \mathrm{e}^{\{ \, Q',\, \mathcal{F} (t) \, \}} \, \Bigr)
	= \int_{t_-}^{t_+} \mathrm{d}t \int \mathrm{d} \tilde{t} \, \tilde{t} \, \partial_t \, \mathrm{e}^{\{ \, Q',\, \mathcal{F} (t) \, \}} \\
	& = \int_{t_-}^{t_+} \mathrm{d}t \, \partial_t \, \mathrm{e}^{\{ \, Q',\, \mathcal{F} (t) \, \}} \biggr|_{\tilde{t}=0}
	= \int_{t_-}^{t_+} \mathrm{d}t \, \partial_t \, \mathrm{e}^{\{ \, Q,\, \mathcal{F} (t) \, \}}
	= \mathrm{e}^{\{ \, Q,\, \mathcal{F} (t_+) \, \}} -\mathrm{e}^{\{ \, Q,\, \mathcal{F} (t_-) \, \}} \,.
\end{split}
\end{equation}
We used the fact that $\mathrm{e}^{\{ \, Q',\, \mathcal{F} (t) \, \}}$ is annihilated
by $Q' = Q +\tilde{t} \, \partial_t$ so we replaced the action of $Q$
on $\mathrm{e}^{\{ \, Q',\, \mathcal{F} (t) \, \}}$ by that of ${}-\tilde{t} \, \partial_t$.
Note that the relation~\eqref{interpolation-identity} holds
for any choice of the interpolation $\mathcal{F} (t)$
between $\mathcal{F} (t_+)$ and $\mathcal{F} (t_-)$.
For~\eqref{bosonic-key-relation-2}, we can rewrite the right-hand side in the form of \eqref{interpolation-identity} as
\begin{equation}
\begin{split}
	& \mathrm{e}^{\, p_+ \, L [ \, f_2 \, ]} \, \mathrm{e}^{\, q_+ \, L [ \, f_3 \, ]}
	-\mathrm{e}^{\, p_- \, L [ \, f_2 \, ]} \, \mathrm{e}^{\, q_- \, L [ \, f_3 \, ]} \\
	& = \mathrm{e}^{\{ Q, \, ( \, t \, p_+ +(1-t) \, p_- \, ) \, b [ \, f_2 \, ] \, \}} \, 
	\mathrm{e}^{\{ Q, \, ( \, t \, q_+ +(1-t) \, q_- \, ) \, b [ \, f_3 \, ] \, \}} \Bigr|_{t=1} \\
	& \quad~ {}-\mathrm{e}^{\{ Q, \, ( \, t \, p_+ +(1-t) \, p_- \, ) \, b [ \, f_2 \, ] \, \}} \, 
	\mathrm{e}^{\{ Q, \, ( \, t \, q_+ +(1-t) \, q_- \, ) \, b [ \, f_3 \, ] \, \}} \Bigr|_{t=0} \,,
\end{split}
\end{equation}
where we set $t_+=1$ and $t_-=0$.
Therefore, the operator $\mathcal{B}$ which satisfies the relation~\eqref{bosonic-key-relation-2}
can be constructed as
\begin{equation}
	\mathcal{B} = \int_0^1 \mathrm{d}t \int \mathrm{d} \tilde{t} \,
	\mathrm{e}^{\{ Q', \, ( \, t \, p_+ +(1-t) \, p_- \, ) \, b [ \, f_2 \, ] \, \}} \, 
	\mathrm{e}^{\{ Q', \, ( \, t \, q_+ +(1-t) \, q_- \, ) \, b [ \, f_3 \, ] \, \}} \,.
\label{B-example-2}
\end{equation}
Let us confirm that the relation~\eqref{bosonic-key-relation-2} is satisfied
by explicitly carrying out the integral over $\tilde{t}$.
Since
\begin{equation}
\begin{split}
\{ Q', \, ( \, t \, p_+ +(1-t) \, p_- \, ) \, b [ \, f_2 \, ] \, \}
& = \tilde{t} \, (p_+ -p_-) \, b [ \, f_2 \, ]
+( \, t \, p_+ +(1-t) \, p_- \, ) \, L [ \, f_2 \, ] \,, \\
\{ Q', \, ( \, t \, q_+ +(1-t) \, q_- \, ) \, b [ \, f_3 \, ] \, \}
& = \tilde{t} \, (q_+ -q_-) \, b [ \, f_3 \, ]
+( \, t \, q_+ +(1-t) \, q_- \, ) \, L [ \, f_3 \, ] \,,
\end{split}
\end{equation}
we obtain
\begin{equation}
\begin{split}
\mathcal{B}
& = \int_0^1 \mathrm{d}t \,
(p_+ -p_-) \, b [ \, f_2 \, ] \, \mathrm{e}^{( \, t \, p_+ +(1-t) \, p_- \, ) \, L [ \, f_2 \, ]} \, 
\mathrm{e}^{( \, t \, q_+ +(1-t) \, q_- \, ) \, L [ \, f_3 \, ]} \\
& \quad ~ +\int_0^1 \mathrm{d}t \,
(q_+ -q_-) \, b [ \, f_3 \, ] \, \mathrm{e}^{( \, t \, p_+ +(1-t) \, p_- \, ) \, L [ \, f_2 \, ]} \, 
\mathrm{e}^{( \, t \, q_+ +(1-t) \, q_- \, ) \, L [ \, f_3 \, ]} \,.
\end{split}
\label{t-tilde-integrated}
\end{equation}
The BRST transformation of $\mathcal{B}$ is then
\begin{equation}
\begin{split}
Q \cdot \mathcal{B}
& = \int_0^1 \mathrm{d}t \,
(p_+ -p_-) \, L [ \, f_2 \, ] \, \mathrm{e}^{( \, t \, p_+ +(1-t) \, p_- \, ) \, L [ \, f_2 \, ]} \, 
\mathrm{e}^{( \, t \, q_+ +(1-t) \, q_- \, ) \, L [ \, f_3 \, ]} \\
& \quad ~ +\int_0^1 \mathrm{d}t \,
(q_+ -q_-) \, L [ \, f_3 \, ] \, \mathrm{e}^{( \, t \, p_+ +(1-t) \, p_- \, ) \, L [ \, f_2 \, ]} \, 
\mathrm{e}^{( \, t \, q_+ +(1-t) \, q_- \, ) \, L [ \, f_3 \, ]} \\
& = \int_0^1 \mathrm{d}t \, \partial_t \,
\mathrm{e}^{( \, t \, p_+ +(1-t) \, p_- \, ) \, L [ \, f_2 \, ]} \, 
\mathrm{e}^{( \, t \, q_+ +(1-t) \, q_- \, ) \, L [ \, f_3 \, ]} \\
& = \mathrm{e}^{\, p_+ \, L [ \, f_2 \, ]} \, \mathrm{e}^{\, q_+ \, L [ \, f_3 \, ]}
-\mathrm{e}^{\, p_- \, L [ \, f_2 \, ]} \, \mathrm{e}^{\, q_- \, L [ \, f_3 \, ]} \,.
\end{split}
\end{equation}
We have thus confirmed the relation~\eqref{bosonic-key-relation-2}.

In this example, the form of the $b$-ghost insertion in~\eqref{t-tilde-integrated}
is not very complicated and we may be able to construct it by inspection.
However, the form of $\mathcal{B}$ in~\eqref{B-example-2}
in terms of the extended BRST operator $Q'$
and the integrals over $t$ an $\tilde{t}$
exhibits an important structure which can be generalized to the superstring.
The moduli space of disks with four punctures has one even modulus,
and we have considered a one-dimensional integral over a region of this moduli space.
When we consider an integral over a higher-dimensional region $\Sigma$,
the right-hand side of~\eqref{interpolation-identity} will be replaced
by an integral over the boundary $\partial \Sigma$ of the region $\Sigma$
and the BRST operator $Q$ on the right-hand side of~\eqref{interpolation-identity}
should be replaced by $Q'$ for the integral over $\partial \Sigma$.
Then the relation~\eqref{interpolation-identity} can be schematically generalized as
\begin{equation}
Q \cdot \mathcal{B}
= \int_{\partial \Sigma} \mathrm{d}t_1 \mathrm{d} \tilde{t}_1 \ldots \mathrm{d}t_{n-1} \mathrm{d} \tilde{t}_{n-1} \,
\mathrm{e}^{\{ \, Q',\, \mathcal{F} (t) \, \}} \quad
\text{for} \quad
\mathcal{B}
= \int_{\Sigma} \mathrm{d}t_1 \mathrm{d} \tilde{t}_1 \ldots \mathrm{d}t_n \mathrm{d} \tilde{t}_n \,
\mathrm{e}^{\{ \, Q',\, \mathcal{F} (t) \, \}} \,.
\end{equation}
We will use this insight for the generalization to the superstring.

\subsection{Interpolation in four-point functions of the superstring}
\label{sec:interpolation-super}

Disks with four NS punctures have one even modulus and two odd moduli,
and on-shell scattering amplitudes of four NS states at the tree level in the open superstring
take the following form:
\begin{equation}
	\int_0^1 \mathrm{d}t \int \mathrm{d} \zeta_1 \mathrm{d} \tilde{\zeta}_1 \mathrm{d} \zeta_2 \mathrm{d} \tilde{\zeta}_2 \,
\langle \,  \mathcal{X} (\zeta_1, \tilde{\zeta}_1) \cdot \mathcal{V}_A (0) \,
b_{-1} \mathrm{e}^{t \, L_{-1}} \mathcal{X} (\zeta_2, \tilde{\zeta}_2) \cdot \mathcal{V}_B (0) \,
\mathcal{V}_C (1) \, \mathcal{V}_D (\infty) \, \rangle \,,
\end{equation}
where $t$ is the even modulus, $\zeta_1$ and $\zeta_2$ are the odd moduli,
and $\mathcal{V}_A$, $\mathcal{V}_B$, $\mathcal{V}_C$, and $\mathcal{V}_D$
are unintegrated vertex operators annihilated by the BRST operator.
The cubic vertex we constructed takes the form of an integral over an odd modulus,
and so contributions to the scattering amplitudes from Feynman diagrams
with two cubic vertices and one propagator in open superstring field theory
also take the form of an integral over one even modulus from the propagator
and two odd moduli from the two cubic vertices.
Let us consider the following toy model
for the contribution $\mathcal{A}_s$ in the $s$ channel
and the contribution $\mathcal{A}_t$ in the $t$ channel:
\begin{equation}
\begin{split}
	\mathcal{A}_s & = \int_0^{1/2} \mathrm{d}t \int \mathrm{d} \zeta_1 \mathrm{d} \tilde{\zeta}_1 \mathrm{d} \zeta_2 \mathrm{d} \tilde{\zeta}_2 \,
\langle \,  \mathcal{X} (\zeta_1, \tilde{\zeta}_1) \cdot \mathcal{V}_A (0) \,
b_{-1} \mathrm{e}^{t \, L_{-1}} \mathcal{X} (\zeta_2, \tilde{\zeta}_2) \cdot \mathcal{V}_B (0) \,
\mathcal{V}_C (1) \, \mathcal{V}_D (\infty) \, \rangle \,, \\
\mathcal{A}_t & = \int_{1/2}^1 \mathrm{d}t \int \mathrm{d} \zeta_1 \mathrm{d} \tilde{\zeta}_1 \mathrm{d} \zeta_2 \mathrm{d} \tilde{\zeta}_2 \,
\langle \,  \mathcal{V}_A (0) \,
b_{-1} \mathrm{e}^{t \, L_{-1}} \mathcal{X} (\zeta_2, \tilde{\zeta}_2) \cdot \mathcal{V}_B (0) \,
\mathcal{X} (\zeta_1, \tilde{\zeta}_1) \cdot \mathcal{V}_C (1) \, \mathcal{V}_D (\infty) \, \rangle \,.
\end{split}
\label{A_s-A_t-model}
\end{equation}
If we honestly calculate $\mathcal{A}_s$ and $\mathcal{A}_t$ using the cubic vertex
we constructed, there will be various terms for each channel
where the locations
of $\mathcal{X} (\zeta_1, \tilde{\zeta}_1)$ and $\mathcal{X} (\zeta_2, \tilde{\zeta}_2)$
are different.
Since the cubic vertex is constructed from the star product,
the region $0 \le t \le 1$ is covered when we add $\mathcal{A}_s$ and $\mathcal{A}_t$,
but the actions of $\mathcal{X} (\zeta_1, \tilde{\zeta}_1)$ and $\mathcal{X} (\zeta_2, \tilde{\zeta}_2)$
are generically different in $\mathcal{A}_s$ and $\mathcal{A}_t$.
We presented a simplified situation in~\eqref{A_s-A_t-model},
where the action of $\mathcal{X} (\zeta_2, \tilde{\zeta}_2)$
is the same in both channels
but the action of $\mathcal{X} (\zeta_1, \tilde{\zeta}_1)$ is different.

Let us discuss the decoupling of unphysical states
based on this toy model.
When $\mathcal{V}_D (\infty) = Q \cdot \Lambda (\infty)$,
the integrals $\mathcal{A}_s$ and $\mathcal{A}_t$ localize as
\begin{equation}
\begin{split}
	\mathcal{A}_s & = \int \mathrm{d} \zeta_1 \mathrm{d} \tilde{\zeta}_1 \mathrm{d} \zeta_2 \mathrm{d} \tilde{\zeta}_2 \,
\langle \,  \mathcal{X} (\zeta_1, \tilde{\zeta}_1) \cdot \mathcal{V}_A (0) \,
\mathcal{X} (\zeta_2, \tilde{\zeta}_2) \cdot \mathcal{V}_B (1/2) \,
\mathcal{V}_C (1) \, \Lambda (\infty) \, \rangle \,, \\
\mathcal{A}_t & = -\int \mathrm{d} \zeta_1 \mathrm{d} \tilde{\zeta}_1 \mathrm{d} \zeta_2 \mathrm{d} \tilde{\zeta}_2 \,
\langle \,  \mathcal{V}_A (0) \,
\mathcal{X} (\zeta_2, \tilde{\zeta}_2) \cdot \mathcal{V}_B (1/2) \,
\mathcal{X} (\zeta_1, \tilde{\zeta}_1) \cdot \mathcal{V}_C (1) \, \Lambda (\infty) \, \rangle \,.
\end{split}
\end{equation}
The location of $\mathcal{X} (\zeta_1, \tilde{\zeta}_1)$ is different,
and these terms do not cancel.
While the region $0 \le t \le 1$ for the even modulus $t$ is covered,
BRST-exact states do not decouple because of this difference
about the location of $\mathcal{X} (\zeta_1, \tilde{\zeta}_1)$,
and this is a concrete example of the general discussion explained in section~7 of~\cite{Sen:2015hia}.
We will elaborate on the relation in subsection~\ref{sec:supermoduli}.

In the context of construction of string products satisfying the $A_\infty$ relations,
this is the reason we need a three-string product to satisfy the relation~\eqref{A_infinity}.
As in the case of the bosonic string,
we consider off-shell vertex operators in this context
and their ghost numbers are arbitrary.
We then have
\begin{equation}
\begin{split}
	& \int \mathrm{d} \zeta_1 \mathrm{d} \tilde{\zeta}_1 \mathrm{d} \zeta_2 \mathrm{d} \tilde{\zeta}_2 \,
\langle \,  \mathcal{X} (\zeta_1, \tilde{\zeta}_1) \cdot \mathcal{V}_A (0) \,
\mathcal{X} (\zeta_2, \tilde{\zeta}_2) \cdot \mathcal{V}_B (1/2) \,
\mathcal{V}_C (1) \, \mathcal{V}_D (\infty) \, \rangle \\
& -\int \mathrm{d} \zeta_1 \mathrm{d} \tilde{\zeta}_1 \mathrm{d} \zeta_2 \mathrm{d} \tilde{\zeta}_2 \,
\langle \,  \mathcal{V}_A (0) \,
\mathcal{X} (\zeta_2, \tilde{\zeta}_2) \cdot \mathcal{V}_B (1/2) \,
\mathcal{X} (\zeta_1, \tilde{\zeta}_1) \cdot \mathcal{V}_C (1) \, \mathcal{V}_D (\infty) \, \rangle \\
& = \langle \,  X \cdot \mathcal{V}_A (0) \, X \cdot \mathcal{V}_B (1/2) \,
\mathcal{V}_C (1) \, \mathcal{V}_D (\infty) \, \rangle
-\langle \,  \mathcal{V}_A (0) \, X \cdot \mathcal{V}_B (1/2) \,
X \cdot \mathcal{V}_C (1) \, \mathcal{V}_D (\infty) \, \rangle \,,
\end{split}
\end{equation}
and this is the structure we found
in subsection~\ref{sec:non-assoc}
for the nonassociativity of the two-string product.
These terms are expressed as integrals over the odd moduli,
and a three-string product can be constructed based on the perspective
of the integration over the missing region of the supermoduli space.
We will discuss this geometric interpretation in subsection~\ref{sec:supermoduli}.

It will be useful to express these terms in terms of line integrals.
We have
\begin{equation}
\begin{split}
	& \quad~ \int \mathrm{d} \zeta_1 \mathrm{d} \tilde{\zeta}_1 \mathrm{d} \zeta_2 \mathrm{d} \tilde{\zeta}_2 \,
\langle \,  \mathcal{X} (\zeta_1, \tilde{\zeta}_1) \cdot \mathcal{V}_A (0) \,
\mathcal{X} (\zeta_2, \tilde{\zeta}_2) \cdot \mathcal{V}_B (1/2) \,
\mathcal{V}_C (1) \, \mathcal{V}_D (\infty) \, \rangle \\
& \quad~ -\int \mathrm{d} \zeta_1 \mathrm{d} \tilde{\zeta}_1 \mathrm{d} \zeta_2 \mathrm{d} \tilde{\zeta}_2 \,
\langle \,  \mathcal{V}_A (0) \,
\mathcal{X} (\zeta_2, \tilde{\zeta}_2) \cdot \mathcal{V}_B (1/2) \,
\mathcal{X} (\zeta_1, \tilde{\zeta}_1) \cdot \mathcal{V}_C (1) \, \mathcal{V}_D (\infty) \, \rangle \\
& = \int \mathrm{d} \zeta_1 \mathrm{d} \tilde{\zeta}_1 \mathrm{d} \zeta_2 \mathrm{d} \tilde{\zeta}_2 \,
\langle \,  \mathrm{e}^{-\{ \, Q',\, \zeta_1 \beta [ \, f_1 \, ] \, \}} \, \mathcal{V}_A (0) \,
\mathrm{e}^{-\{ \, Q',\, \zeta_2 \beta [ \, f_2 \, ] \, \}} \, \mathcal{V}_B (1/2) \,
\mathcal{V}_C (1) \, \mathcal{V}_D (\infty) \, \rangle \\
& \quad~ -\int \mathrm{d} \zeta_1 \mathrm{d} \tilde{\zeta}_1 \mathrm{d} \zeta_2 \mathrm{d} \tilde{\zeta}_2 \,
\langle \,  \mathcal{V}_A (0) \,
\mathrm{e}^{-\{ \, Q',\, \zeta_2 \beta [ \, f_2 \, ] \, \}} \, \mathcal{V}_B (1/2) \,
\mathrm{e}^{-\{ \, Q',\, \zeta_1 \beta [ \, f_3 \, ] \, \}} \, \mathcal{V}_C (1) \, \mathcal{V}_D (\infty) \, \rangle \,,
\end{split}
\end{equation}
where
\begin{equation}
f_1 (\xi) = \xi \,, \qquad
f_2 (\xi) = \xi +\frac{1}{2} \,, \qquad
f_3 (\xi) = \xi +1 \,.
\end{equation}
Therefore, the problem of constructing a three-string product
reduces to the construction of $\Xi$ satisfying
\begin{equation}
Q \cdot \Xi
= \int \mathrm{d} \zeta_1 \mathrm{d} \tilde{\zeta}_1 \,
\mathrm{e}^{-\{ \, Q',\, \zeta_1 \beta [ \, f_3 \, ] \, \}}
-\int \mathrm{d} \zeta_1 \mathrm{d} \tilde{\zeta}_1 \,
\mathrm{e}^{-\{ \, Q',\, \zeta_1 \beta [ \, f_1 \, ] \, \}} \,.
\end{equation}
It is straightforward to apply the strategy
developed in the preceding subsection
for the bosonic string to the current case, and we obtain
\begin{equation}
	\Xi = \int_0^1 \mathrm{d}t' \int \mathrm{d} \tilde{t'} \int \mathrm{d} \zeta_1 \mathrm{d} \tilde{\zeta}_1 \,
	\mathrm{e}^{-\{ \, Q',\, t' \zeta_1 \beta [ \, f_3 \, ] \, \}} \,
	\mathrm{e}^{-\{ \, Q',\, (1-t') \, \zeta_1 \beta [ \, f_1 \, ] \, \}} \,,
\end{equation}
where $t'$ should be regarded as an even modulus
and the definition of $Q'$ has been generalized as
\begin{equation}
Q' = Q +\tilde{\zeta}_1 \partial_{\zeta_1}
 +\tilde{\zeta}_2 \partial_{\zeta_2} +\tilde{t'} \partial_{t'} \,.
\end{equation}

Let us also present an expression of $\Xi$ after integrating
over the odd variables $\tilde{t'}$ and $\zeta_1$.
Since
\begin{equation}
\begin{split}
{}-\{ \, Q',\, t' \zeta_1 \beta [ \, f_3 \, ] \, \}
& = {}-\tilde{t'} \zeta_1 \beta [ \, f_3 \, ] -t' \tilde{\zeta_1} \beta [ \, f_3 \, ] +t' \zeta_1 G [ \, f_3 \, ]\,, \\
{}-\{ \, Q',\, (1-t') \, \zeta_1 \beta [ \, f_1 \, ] \, \}
& = \tilde{t'} \zeta_1 \beta [ \, f_1 \, ] -(1-t') \, \tilde{\zeta_1} \beta [ \, f_1 \, ] +(1-t') \, \zeta_1 G [ \, f_1 \, ] \,,
\end{split}
\end{equation}
where $G [ \, f \, ]$ for $f(\xi)$ is defined by
\begin{equation}
	G [ \, f \, ] = \oint \frac{\mathrm{d}z}{2 \pi \mathrm{i}} \,
	\Bigl( \, \frac{\mathrm{d} f^{-1} (z)}{\mathrm{d}z} \, \Bigr)^{-1/2} \, G (z) \quad
\text{with} \quad z = f(\xi) \,,
\end{equation}
we find
\begin{equation}
\begin{split}
	\Xi & = \int_0^1 \mathrm{d}t' \int \mathrm{d} \tilde{\zeta}_1 \,
( \, \beta [ \, f_3 \, ] -\beta [ \, f_1 \, ] \, ) \,
\mathrm{e}^{-t' \tilde{\zeta_1} \beta [ \, f_3 \, ]} \,
\mathrm{e}^{-(1-t') \, \tilde{\zeta_1} \beta [ \, f_1 \, ]} \\
& = \int_0^1 \mathrm{d}t' \,
( \, \beta [ \, f_3 \, ] -\beta [ \, f_1 \, ] \, ) \,
\delta ( \, t' \beta [ \, f_3 \, ] +(1-t') \, \beta [ \, f_1 \, ] \, ) \,.
\end{split}
\end{equation}
As correlation functions involving the operator $\delta ( \, t' \beta [ \, f_3 \, ] +(1-t') \, \beta [ \, f_1 \, ] \, )$
look unfamiliar, we will present an example of explicit calculations of such correlation functions
in appendix~\ref{sec:beta-gamma}.
We will find in the example that the correlation function develops a double pole
when $t' = 1/2$.
In general, when the gauge fixing of the world-sheet supergravity fails,
the superconformal ghost sector develops a pole
and such a pole is called a spurious pole.\footnote{
When local picture-changing operators are used for the gauge fixing,
spurious poles do not arise for correlation functions
on the disk.
See subsection~4.5 of~\cite{Witten:2012bh}.}
In our case, the double pole at $t' = 1/2$ can be avoided
by deforming the contour for the integral over $t'$ to be complex.
The resulting integral does not depend on how we avoid the singularity
because the singularity consists of a double pole and does not contain a single pole.
Note also that the decoupling of BRST-exact states depends only on the endpoints
of the integral and is unaffected by the deformation of the contour.

We have learned from this toy model that, in general, $\Xi [ \, f_1, f_2 \, ]$ defined
for a pair of conformal transformations $f_1 (\xi)$ and $f_2 (\xi)$ by
\begin{equation}
\Xi [ \, f_1, f_2 \, ]
= \int_0^1 \mathrm{d}t' \int \mathrm{d} \tilde{t'} \int \mathrm{d} \zeta \mathrm{d} \tilde{\zeta} \,
\mathrm{e}^{-\{ \, Q',\, t' \zeta \beta [ \, f_1 \, ] \, \}} \,
\mathrm{e}^{-\{ \, Q',\, (1-t') \, \zeta \beta [ \, f_2 \, ] \, \}}
\end{equation}
plays an important role in the construction of a three-string product based on our approach.
The expression of $\Xi [ \, f_1, f_2 \, ]$ after integrating over $\tilde{t'}$ and $\zeta$ is
\begin{equation}
\Xi [ \, f_1, f_2 \, ]
= \int_0^1 \mathrm{d}t' \,
( \, \beta [ \, f_1 \, ] -\beta [ \, f_2 \, ] \, ) \,
\delta ( \, t' \beta [ \, f_1 \, ] +(1-t') \, \beta [ \, f_2 \, ] \, ) \,.
\label{Xi-after-integration}
\end{equation}
Two basic properties of $\Xi [ \, f_1, f_2 \, ]$ are
\begin{equation}
Q \cdot \Xi [ \, f_1, f_2 \, ]
= X [ \, f_1 \, ] -X [ \, f_2 \, ] \,, \qquad
\Xi [ \, f_1, f_2 \, ] = {}-\Xi [ \, f_2, f_1 \, ] \,.
\end{equation}
In the next subsection we will use $\Xi [ \, f_1, f_2 \, ]$
and construct a three-string product satisfying the relation~\eqref{A_infinity}. 

\subsection{The definition of the quartic vertex}
\label{sec:quartic-vertex}

In subsection~\ref{sec:non-assoc} we calculated
$\langle \, A_1, [ \, [ \, A_2, A_3 \, ], A_4 \, ] \, \rangle$
and $\langle \, A_1, [ \, A_2, [ \, A_3, A_4 \, ]  \, ] \, \rangle$
for four states $A_1$, $A_2$, $A_3$, and $A_4$, and we found
\begin{equation}
\begin{split}
	\langle \, A_1, [ \, [ \, A_2, A_3 \, ], A_4 \, ]\, \rangle
	& = \langle \, X_t \,\,
	g_1\circ \mathcal{V}_1 (0) \, g_2 \circ \mathcal{V}_2 (0) \,
	g_3 \circ \mathcal{V}_3 (0) \, g_4 \circ \mathcal{V}_4 (0) \, \rangle_D \,, \\
	\langle \, A_1, [ \, A_2, [ \, A_3, A_4 \, ]  \, ] \, \rangle
	& = \langle \, X_s \,\,
	g_1\circ \mathcal{V}_1 (0) \, g_2 \circ \mathcal{V}_2 (0) \,
	g_3 \circ \mathcal{V}_3 (0) \, g_4 \circ \mathcal{V}_4 (0) \, \rangle_D \,,
\end{split}
\end{equation}
where
\begin{equation}
\begin{split}
		X_t & = \frac{1}{9} \, \Bigl( \, X [ \, h_2 \, ] +X [ \, g_4 \, ] +X [ \, g_1 \, ] \, \Bigr) \,
		\Bigl( \, X [ \, h_4 \, ] +X [ \, g_2 \, ] +X [ \, g_3 \, ] \, \Bigr) \,, \\
		X_s & = \frac{1}{9} \, \Bigl( \, X [ \, h_1 \, ] +X [ \, g_3 \, ] +X [ \, g_4 \, ] \, \Bigr) \,
		\Bigl( \, X [ \, h_3 \, ] +X [ \, g_1 \, ] +X [ \, g_2 \, ] \, \Bigr) \,.
\end{split}
\end{equation}
If we can find a Grassmann-odd operator $\Xi$ which satisfies
\begin{equation}
Q \cdot \Xi = X_t -X_s \,,
\end{equation}
a three-string product satisfying the relation~\eqref{A_infinity} can be defined by
\begin{equation}
\begin{split}
	\langle \, A_1, [ \, A_2, A_3, A_4 \, ]\, \rangle
	& = \langle \,
	g_1\circ \mathcal{V}_1 (0) \,\, \Xi \,\, g_2 \circ \mathcal{V}_2 (0) \,
	g_3 \circ \mathcal{V}_3 (0) \, g_4 \circ \mathcal{V}_4 (0) \, \rangle_D \\
	& = (-1)^{A_1} \langle \, \Xi \,\,
	g_1\circ \mathcal{V}_1 (0) \, g_2 \circ \mathcal{V}_2 (0) \,
	g_3 \circ \mathcal{V}_3 (0) \, g_4 \circ \mathcal{V}_4 (0) \, \rangle_D \,.
\end{split}
\label{quartic-vertex-Xi}	
\end{equation}
For the three-string product to have the cyclic property in~\eqref{eq:cyclicity},
we require the operator $\Xi$ to satisfy
\begin{equation}
\omega \circ \Xi = {}-\Xi \,,
\end{equation}
where $\omega (z)$ is the conformal transformation for the rotation of degree $\pi / 2$
on the disk $D$,
\begin{equation}
	\omega (z) = \mathrm{e}^{\frac{2 \pi \mathrm{i}}{4}} z = \mathrm{i} \, z \,,
\end{equation}
and here and in what follows
we denote the operator transformed from $\mathcal{O}$ under the rotation $\omega (z)$
by $\omega \circ \mathcal{O}$.

We can use $\Xi [ \, f_1, f_2 \, ]$ introduced in the preceding subsection
as building blocks for the construction of $\Xi$.
Since $X_s$ and $X_t$ consist of several terms,
let us first consider how we should interpolate these terms.
It will be useful to consider the transformations of $X [ \, g_i \, ]$ and $X [ \, h_i \, ]$
under the rotation $\omega (z)$. We have
\begin{equation}
\omega \circ X [ \, g_i \, ] = X [ \, g_{i+1} \, ] \,, \qquad
\omega \circ X [ \, h_i \, ] = X [ \, h_{i+1} \, ]
\end{equation}
for $i= 1, 2, 3, 4$ with the understanding that $g_5 = g_1$ and $h_5 = h_1$.
As the $s$ channel and the $t$ channel are related by the rotation $\omega (z)$,
we find
\begin{equation}
X_t = \omega \circ X_s \,, \qquad X_s = \omega \circ X_t \,.
\label{Xtsrotate}
\end{equation}
The connection between $X_s$ and $X_t$ via the rotation $\omega (z)$
suggests one natural way to interpolate between $X_s$ and $X_t$.
For example, the term $X [ \, g_1 \, ] \, X [ \, h_1 \, ]$ in $X_s$
is mapped to $X [ \, g_2 \, ] \, X [ \, h_2 \, ]$ in $X_t$ under the rotation $\omega (z)$
so that we can choose to interpolate between $X [ \, g_1 \, ] \, X [ \, h_1 \, ]$
and $X [ \, g_2 \, ] \, X [ \, h_2 \, ]$.
Even if we decided to interpolate between $X [ \, g_1 \, ] \, X [ \, h_1 \, ]$
and $X [ \, g_2 \, ] \, X [ \, h_2 \, ]$ given by
\begin{equation}
\begin{split}
X [ \, g_1 \, ] \, X [ \, h_1 \, ]
& =  \int \mathrm{d} \zeta_1 \mathrm{d} \tilde{\zeta}_1 \mathrm{d} \zeta_2 \mathrm{d} \tilde{\zeta}_2 \,
\mathrm{e}^{-\{ \, Q',\, \zeta_1 \beta [ \, g_1 \, ] \, \}} \,
\mathrm{e}^{-\{ \, Q',\, \zeta_2 \beta [ \, h_1 \, ] \, \}} \,, \\
X [ \, g_2 \, ] \, X [ \, h_2 \, ]
& =  \int \mathrm{d} \zeta_1 \mathrm{d} \tilde{\zeta}_1 \mathrm{d} \zeta_2 \mathrm{d} \tilde{\zeta}_2 \,
\mathrm{e}^{-\{ \, Q',\, \zeta_1 \beta [ \, g_2 \, ] \, \}} \,
\mathrm{e}^{-\{ \, Q',\, \zeta_2 \beta [ \, h_2 \, ] \, \}} \,,
\end{split}
\end{equation}
there are still various ways for the interpolation.
One canonical way would be the following simultaneous interpolation:
\begin{equation}
\begin{split}
	\int_0^1 \mathrm{d}t' \int \mathrm{d} \tilde{t'}
	\int \mathrm{d} \zeta_1 \mathrm{d} \tilde{\zeta}_1 \mathrm{d} \zeta_2 \mathrm{d} \tilde{\zeta}_2 \,
	& \mathrm{e}^{-\{ \, Q',\, t' \zeta_1 \beta [ \, g_1 \, ] \, \}} \,
	\mathrm{e}^{-\{ \, Q',\, t' \zeta_2 \beta [ \, h_1 \, ] \, \}} \\
	& \times \mathrm{e}^{-\{ \, Q',\, (1-t') \, \zeta_1 \beta [ \, g_2 \, ] \, \}} \,
	\mathrm{e}^{-\{ \, Q',\, (1-t') \, \zeta_2 \beta [ \, h_2 \, ] \, \}} \,.
\end{split}
\end{equation}
The extended BRST transformation in this interpolation
generates potentially singular products of line integrals with intersecting contours
such as $G [ \, h_1 \, ] \, \delta ( \, \beta [ \, h_2 \, ] \, )$.
While it may be possible to define properly normal-ordered products of these line integrals,
in this paper we adopt a safer method for the interpolation.
Instead of the simultaneous interpolation,
we interpolate between $X [ \, g_1 \, ] \, X [ \, h_1 \, ]$
and $X [ \, g_2 \, ] \, X [ \, h_2 \, ]$ as
\begin{equation}
\Xi [ \, g_1, g_2 \, ] \, X [ \, h_1 \, ] +X [ \, g_2 \, ] \, \Xi [ \, h_1, h_2 \, ]
\label{g_1-to-g_2-then-h_1-to-h_2}
\end{equation}
or as
\begin{equation}
X [ \, g_1 \, ] \, \Xi [ \, h_1, h_2 \, ] +\Xi [ \, g_1, g_2 \, ] \, X [ \, h_2 \, ] \,. 
\label{h_1-to-h_2-then-g_1-to-g_2}
\end{equation}
As we have seen in the preceding subsection,
the operator $\Xi [ \, f_1, f_2 \, ]$ can be written
only in terms of $\beta [ \, f_1 \, ]$ and $\beta [ \, f_2 \, ]$
after the integrations over the odd variables
so that it is free from singularities
even for $\Xi [ \, h_1, h_2 \, ]$
where the contours of $\beta [ \, h_1 \, ]$ and $\beta [ \, h_2 \, ]$ intersect.
We average the two ways
and define
$\Xi [ \, a_1, a_2; b_1, b_2 \, ]$
in general
for conformal transformations $a_1 (\xi)$, $a_2 (\xi)$, $b_1 (\xi)$, and $b_2 (\xi)$ by
\begin{equation}
\begin{split}
\Xi [ \, a_1, a_2; b_1, b_2 \, ]
& = \frac{1}{2} \, \Bigl( \,
\Xi [ \, a_1, b_1 \, ] \, X [ \, a_2 \, ]
+X [ \, b_1 \, ] \, \Xi [ \, a_2, b_2 \, ] \\
& \qquad \quad +X [ \, a_1 \, ] \, \Xi [ \, a_2, b_2 \, ]
+\Xi [ \, a_1, b_1 \, ] \, X [ \, b_2 \, ] \, \Bigr) \,.
\end{split}
\label{Xi-a_1-a_2-b_1-b_2}
\end{equation}
The basic relation of $\Xi [ \, a_1, a_2; b_1, b_2 \, ]$ is
\begin{equation}
	Q \cdot \Xi [ \, a_1, a_2; b_1, b_2 \, ]
	= X [ \, a_1 \, ] \, X [ \, a_2 \, ] -X [ \, b_1 \, ] \, X [ \, b_2 \, ] \,,
	\label{eq:QXifour}
\end{equation}
and it has the following property:
\begin{equation}
\Xi [ \, a_1, a_2; b_1, b_2 \, ] = \Xi [ \, a_2, a_1; b_2, b_1 \, ] \,.
\end{equation}
Following this strategy for the interpolation,
let us expand $X_s$ and $X_t$ and consider
the interpolation $\Xi [ \, t;  s \, ]$
given by
\begin{equation}
\begin{split}
\Xi [ \, t;  s \, ] = \frac{1}{9} \, \Bigl( \,
& \Xi [ \, h_2, h_4; h_3, h_1 \, ]
+\Xi [ \, h_2, g_2; h_3, g_3 \, ]
+\Xi [ \, h_2, g_3; h_3, g_4 \, ] \\
& +\Xi [ \, g_4, h_4; g_1, h_1 \, ]
+\Xi [ \, g_4, g_2; g_1, g_3 \, ]
+\Xi [ \, g_4, g_3; g_1, g_4 \, ] \\
& +\Xi [ \, g_1, h_4; g_2, h_1 \, ]
+\Xi [ \, g_1, g_2; g_2, g_3 \, ]
+\Xi [ \, g_1, g_3; g_2, g_4 \, ] \, \Bigr) \,.
\end{split}
\end{equation}
By construction this operator $\Xi [ \, t;  s \, ]$ satisfies the relation
\begin{equation}
Q \cdot \Xi [ \, t;  s \, ] = X_t -X_s \,,
\end{equation}
but it does not have the cyclic property
required for the identification of $\Xi [ \, t;  s \, ]$ as $\Xi$.
In fact, under the rotation $\omega (z)$ it is transformed as
\begin{equation}
\begin{split}
\omega \circ \Xi [ \, t;  s \, ] = \frac{1}{9} \, \Bigl( \,
& \Xi [ \, h_3, h_1; h_4, h_2 \, ]
+\Xi [ \, h_3, g_3; h_4, g_4 \, ]
+\Xi [ \, h_3, g_4; h_4, g_1 \, ] \\
& +\Xi [ \, g_1, h_1; g_2, h_2 \, ]
+\Xi [ \, g_1, g_3; g_2, g_4 \, ]
+\Xi [ \, g_1, g_4; g_2, g_1 \, ] \\
& +\Xi [ \, g_2, h_1; g_3, h_2 \, ]
+\Xi [ \, g_2, g_3; g_3, g_4 \, ]
+\Xi [ \, g_2, g_4; g_3, g_1 \, ] \, \Bigr) \,,
\end{split}
\end{equation}
and we find
\begin{equation}
\omega \circ \Xi [ \, t;  s \, ] \ne {}-\Xi [ \, t;  s \, ] \,.
\end{equation}
We can improve $\Xi [ \, t;  s \, ]$ to satisfy the required cyclic property
based on the following general discussion.
As required for the compatibility of the relation
$Q \cdot \Xi = X_t -X_s$
and the cyclic property $\omega \circ \Xi = {}-\Xi$,
we should have
\begin{equation}
\omega \circ ( \, X_t -X_s \, ) = {}-( \, X_t -X_s \, ) \, , 
\end{equation}
which indeed follows from~\eqref{Xtsrotate}.
We therefore have
\begin{equation}
Q \cdot ( \, \omega \circ \Xi [ \, t;  s \, ] \, ) = {}-( \, X_t -X_s \, ) \,.
\end{equation}
This means that when the BRST transformation of $\Xi [ \, t;  s \, ]$ is $X_t -X_s$,
the BRST transformation of ${}-\omega \circ \Xi [ \, t;  s \, ]$
is also given by $X_t -X_s$.
More generally, we have
\begin{equation}
Q \cdot ( \, (-1)^n \omega^n \circ \Xi [ \, t;  s \, ] \, ) = X_t -X_s
\end{equation}
for any integer $n$, where $\omega^n (z) = \mathrm{e}^{\frac{2 \pi n \mathrm{i}}{4}} z$
and we denoted the operator transformed from $\Xi [ \, t;  s \, ]$
under the rotation $\omega^n (z)$ by $\omega^n \circ \Xi [ \, t;  s \, ]$.
We then define $\Xi$ by
\begin{equation}
\Xi = \frac{1}{4} \, \Bigl( \,
\Xi [ \, t;  s \, ] -\omega \circ \Xi [ \, t;  s \, ]
+\omega^2 \circ \Xi [ \, t;  s \, ] -\omega^3 \circ \Xi [ \, t;  s \, ] \, \Bigr) \,.
\end{equation}
This operator $\Xi$ satisfies
\begin{equation}
Q \cdot \Xi = X_t -X_s \,.
\end{equation}
Furthermore, since
\begin{equation}
\omega \circ \Xi = \frac{1}{4} \, \Bigl( \,
\omega \circ \Xi [ \, t;  s \, ] -\omega^2 \circ \Xi [ \, t;  s \, ]
+\omega^3 \circ \Xi [ \, t;  s \, ] -\Xi [ \, t;  s \, ] \, \Bigr) \,,
\end{equation}
this operator $\Xi$ has the required cyclic property:
\begin{equation}
\omega \circ \Xi = {}-\Xi \,.
\end{equation}
Actually, we note that
\begin{equation}
\omega^2 \circ \Xi [ \, t;  s \, ] =  \Xi [ \, t;  s \, ]
\end{equation}
so that the operator $\Xi$ simplifies to
\begin{equation}
\Xi = \frac{1}{2} \, \Bigl( \,
\Xi [ \, t;  s \, ] -\omega \circ \Xi [ \, t;  s \, ] \, \Bigr) \,.
\end{equation}
Its explicit expression is
\begin{equation}
\begin{split}
\Xi = \frac{1}{18} \, \Bigl( \,
& \Xi [ \, h_2, h_4; h_3, h_1 \, ]
+\Xi [ \, h_2, g_2; h_3, g_3 \, ]
+\Xi [ \, h_2, g_3; h_3, g_4 \, ] \\
& +\Xi [ \, g_4, h_4; g_1, h_1 \, ]
+\Xi [ \, g_4, g_2; g_1, g_3 \, ]
+\Xi [ \, g_4, g_3; g_1, g_4 \, ] \\
& +\Xi [ \, g_1, h_4; g_2, h_1 \, ]
+\Xi [ \, g_1, g_2; g_2, g_3 \, ]
+\Xi [ \, g_1, g_3; g_2, g_4 \, ] \\
& -\Xi [ \, h_3, h_1; h_4, h_2 \, ]
-\Xi [ \, h_3, g_3; h_4, g_4 \, ]
-\Xi [ \, h_3, g_4; h_4, g_1 \, ] \\
& -\Xi [ \, g_1, h_1; g_2, h_2 \, ]
-\Xi [ \, g_1, g_3; g_2, g_4 \, ]
-\Xi [ \, g_1, g_4; g_2, g_1 \, ] \\
& -\Xi [ \, g_2, h_1; g_3, h_2 \, ]
-\Xi [ \, g_2, g_3; g_3, g_4 \, ]
-\Xi [ \, g_2, g_4; g_3, g_1 \, ] \, \Bigr) \,.
\end{split}
\label{explicit-Xi}
\end{equation}

To summarize, we defined a three-string product
in terms of $\langle \, A_1, [ \, A_2, A_3, A_4 \, ]\, \rangle$
for arbitrary string fields $A_1$, $A_2$, $A_3$, and $A_4$
in~\eqref{quartic-vertex-Xi} 
with $\Xi$ given by~\eqref{explicit-Xi},
which is written in terms of
$\Xi [ \, a_1, a_2; b_1, b_2 \, ]$ in~\eqref{Xi-a_1-a_2-b_1-b_2}
with $\Xi [ \, f_1, f_2 \, ]$ defined in the preceding subsection.
This three-string product satisfies the relation~\eqref{A_infinity}
and has the cyclic property~\eqref{eq:cyclicity}.
This is one of the main results of this paper.

\subsection{The relation to the covering of the supermoduli space}
\label{sec:supermoduli}

In subsection~\ref{sec:interpolation-bosonic} we considered four-point functions of the bosonic string
and discussed a toy model where the sum of $\mathcal{A}_s$ and $\mathcal{A}_t$
does not cover the whole region of the moduli space of disks with four punctures.
Our main focus was the construction of the integral of off-shell amplitudes denoted by $\mathcal{A}_4$
which interpolates the boundary of the moduli space covered by $\mathcal{A}_s$
and the boundary of the moduli space covered by $\mathcal{A}_t$.
In the context of the covering of the moduli space of disks with four punctures,
the contribution $\mathcal{A}_4$ covers the region
which is not covered by $\mathcal{A}_s$ and $\mathcal{A}_t$.

We then generalized the discussion to four-point functions of the superstring.
In the toy model we considered in subsection~\ref{sec:interpolation-super},
the sum of $\mathcal{A}_s$ and $\mathcal{A}_t$
covers the bosonic moduli space of disks with four punctures,
but we argued that an additional contribution is necessary.
In subsection~\ref{sec:quartic-vertex} we used this insight
and constructed the three-string product which satisfies the $A_\infty$ relation~\eqref{A_infinity}.
In this subsection we discuss the additional contribution
for four-point functions of the superstring
in the context of the covering of the supermoduli space
of disks with four NS punctures.

Let us start with the bosonic string.
While our main focus in subsection~\ref{sec:interpolation-bosonic}
was the construction of the off-shell amplitudes $\mathcal{A}_4$,
it is sufficient to consider on-shell amplitudes in the context of the covering of the moduli space.
Then the relevant information is the set of coordinates of the four punctures which we denote
by $z_1$, $z_2$, $z_3$, and $z_4$.
The moduli space of disks with four punctures can be parameterized
by the following cross ratio:
\begin{equation}
\Phi_s = \frac{(z_1-z_2)(z_3-z_4)}{(z_1-z_3)(z_2-z_4)} \,.
\end{equation}
In subsection~\ref{sec:interpolation-bosonic}
we chose the conditions $z_1 = 0$, $z_3 = 1$, and $z_4 \to \infty$,
and the cross ratio in this case is given by $\Phi_s = z_2$.
The moduli space for the cyclic ordering we are considering
corresponds to the region $0 \le \Phi_s \le 1$.
Suppose that the contribution in the $s$ channel covers the region $0 \le \Phi_s \le 1/2$.
Then the contribution in the $t$ channel would cover the region $0 \le \Phi_t \le 1/2$,
where $\Phi_t$ is the cross ratio obtained from $\Phi_s$
by the cyclic permutation $z_i \to z_{i+1}$ with the understanding that $z_5= z_1$
and is given by
\begin{equation}
\Phi_t = \frac{(z_2-z_3)(z_4-z_1)}{(z_2-z_4)(z_3-z_1)} \,.
\end{equation}
The two cross ratios $\Phi_s$ and $\Phi_t$ are not independent,
and they are related as follows:
\begin{equation}
\Phi_s +\Phi_t =1 \,.
\end{equation}
Therefore, the region covered by the contribution in the $t$ channel
can be described in terms of $\Phi_s$ as $1/2 \le \Phi_s \le 1$.
In this case the sum of the contributions in the $s$ channel and in the $t$ channel
covers the whole moduli space.

Let us next consider the superstring.
In the context of the covering of the supermoduli space of disks with four NS punctures,
it is sufficient to consider on-shell amplitudes,
and the relevant information is the set of supercoordinates of four NS punctures
which we denote by $(z_1 | \theta_1)$, $(z_2 | \theta_2)$, $(z_3 | \theta_3)$, and $(z_4 | \theta_4)$.
Choosing a vertex operator to be in the $-1$ picture corresponds
to setting the odd coordinate to zero,
and choosing a vertex operator to be in the $0$ picture corresponds
to integrating over the odd coordinate.
The bosonic direction of the supermoduli space can be parameterized
by the following super cross ratio:
\begin{equation}
\widehat{\Phi}_s = \frac{(z_1-z_2-\theta_1 \theta_2)(z_3-z_4-\theta_3 \theta_4)}
{(z_1-z_3-\theta_1 \theta_3)(z_2-z_4-\theta_2 \theta_4)} \,.
\end{equation}
If we impose the conditions
$(z_1 | \theta_1) = (0|0)$, $(z_3 | \theta_3) = (1| \theta_3)$, and $(z_4 | \theta_4) \to (\infty |0)$,
the super cross ratio is given by $\widehat{\Phi}_s = z_2$.
As in the case of the bosonic string,
consider the super cross ratio $\widehat{\Phi}_t$ obtained from $\widehat{\Phi}_s$
by the cyclic permutation $(z_i | \theta_i) \to (z_{i+1} | \theta_{i+1})$
with the understanding that $(z_5 | \theta_5) = (z_1 | \theta_1)$.
It is given by
\begin{equation}
\widehat{\Phi}_t = \frac{(z_2-z_3-\theta_2 \theta_3)(z_4-z_1-\theta_4 \theta_1)}
{(z_2-z_4-\theta_2 \theta_4)(z_3-z_1-\theta_3 \theta_1)} \,.
\end{equation}
Compared to the case of the bosonic string,
the important difference which was emphasized in~subsection 5.1.3 of~\cite{Witten:2012ga} is that
\begin{equation}
\widehat{\Phi}_s +\widehat{\Phi}_t \neq 1 \,.
\end{equation}
For example, suppose that the coordinates of the boundary in the $s$ channel are given by
\begin{equation}
(z_1 | \theta_1) = (0|0) \,, \quad (z_2 | \theta_2) = (1/2 \, | \zeta_2) \,, \quad
(z_3 | \theta_3) = (1| \zeta_3) \,, \quad (z_4 | \theta_4) = (\infty |0) \,,
\label{coordinates-before-permutation}
\end{equation}
where $\zeta_2$ and $\zeta_3$ are Grassmann-odd constants.
The value of the super cross ratio $\widehat{\Phi}_s$
for the set of coordinates~\eqref{coordinates-before-permutation} is given by
\begin{equation}
\widehat{\Phi}_s = \frac{1}{2} \,.
\end{equation}
The coordinates of the boundary in the $t$ channel would be obtained
by the permutation $(z_i | \theta_i) \to (z_{i+1} | \theta_{i+1})$
with the understanding that $(z_5 | \theta_5) = (z_1 | \theta_1)$
and are given by
\begin{equation}
(z_1 | \theta_1) = (\infty |0) \,, \quad (z_2 | \theta_2) = (0|0) \,, \quad
(z_3 | \theta_3) = (1/2 \, | \zeta_2) \,, \quad (z_4 | \theta_4) = (1| \zeta_3) \,.
\label{coordinates-after-permutation}
\end{equation}
The value of $\widehat{\Phi}_t$ for the set of coordinates~\eqref{coordinates-after-permutation}
 is by construction the same as the value of $\widehat{\Phi}_s$
for the set of coordinates~\eqref{coordinates-before-permutation}
and is given by
\begin{equation}
\widehat{\Phi}_t = \frac{1}{2} \,,
\end{equation}
but the value of $\widehat{\Phi}_s$
for the set of coordinates~\eqref{coordinates-after-permutation} is
different from $1/2$
and is given by
\begin{equation}
\widehat{\Phi}_s = \frac{1}{2} +\zeta_2 \zeta_3 \,.
\end{equation}
In this example, we indeed found that
\begin{equation}
\widehat{\Phi}_s +\widehat{\Phi}_t = 1 +\zeta_2 \zeta_3 \neq 1 \,.
\end{equation}
While the value of $\Phi_s$ for~\eqref{coordinates-before-permutation}
and the value of $\Phi_s$ for~\eqref{coordinates-after-permutation} coincide,
the value of $\widehat\Phi_s$ for~\eqref{coordinates-before-permutation}
and the value of $\widehat\Phi_s$ for~\eqref{coordinates-after-permutation} are different.
This is the reason why an additional contribution is necessary
to cover the supermoduli space of disks with four NS punctures.

This conclusion based on the super cross ratio can also be understood in the following way.
Using superconformal transformations on the disk, we can always bring
four supercoordinates to the form
\begin{equation}
(z_1 | \theta_1) = (0|0) \,, \quad (z_2 | \theta_2) = (t \, | \zeta_2) \,, \quad
(z_3 | \theta_3) = (1| \zeta_3) \,, \quad (z_4 | \theta_4) = (\infty |0) \,.
\end{equation}
This corresponds to choosing the first and fourth vertex operators
to be in the $-1$ picture
and the second and third vertex operators
to be in the $0$ picture.
Then the on-shell scattering amplitude takes the following form:
\begin{equation}
	\int_0^1 \mathrm{d}t \int \mathrm{d} \zeta_2 \mathrm{d} \zeta_3 \, \mathcal{F}(t; 0, \zeta_2, \zeta_3, 0) \,,
\end{equation}
where an explicit form of $\mathcal{F}(t; \theta_1, \theta_2, \theta_3, \theta_4)$
can be expressed in terms of a boundary CFT correlation function. 
Suppose that the pictures of the vertex operators are chosen this way
in the $s$ channel and the contribution in the $s$ channel is given by
\begin{equation}
	\int_0^{1/2} \mathrm{d}t \int \mathrm{d} \zeta_2 \mathrm{d} \zeta_3 \, \mathcal{F}(t; 0, \zeta_2, \zeta_3,  0) \,.
\end{equation}
In the $t$ channel, on the other hand, suppose that
the third and fourth vertex operators are chosen
to be in the $-1$ picture
and the first and second vertex operators are chosen
to be in the $0$ picture.
Then we have
\begin{equation}
(z_1 | \theta_1) = (0|\zeta_2) \,, \quad (z_2 | \theta_2) = (t \, |\zeta_3) \,, \quad
(z_3 | \theta_3) = (1| 0) \,, \quad (z_4 | \theta_4) = (\infty | 0) \,. 
\end{equation}
While we can set $\theta_1 = 0$ keeping the condition $\theta_4 = 0$
by a superconformal transformation,
the value of $z_2$ is necessarily changed by the superconformal transformation.
We in fact know from the discussion based on the super cross ratio that
the value of $z_2$ is changed from $t$ to $t +\zeta_2 \zeta_3$.
The contribution in the $t$ channel could be schematically written as\footnote{
As we mentioned in the introduction,
the covering of the supermoduli space of super-Riemann surfaces
should not be understood as the covering of integration regions of ordinary integrals,
and it is manifested in this schematic expression.
One approach to making sense of this schematic expression
is to use a partition of unity as explained in section~7 of~\cite{Sen:2015hia}.
See below for our approach.
}
\begin{equation}
	\int \mathrm{d} \zeta_2 \mathrm{d} \zeta_3
	\int_{1/2+\zeta_2 \zeta_3}^1 \mathrm{d}t \, \mathcal{F}(t; 0, \zeta_2, \zeta_3,  0) \,.
\label{A_t-schematical}
\end{equation}

When there are at least two odd coordinates,
mixing of even coordinates and odd coordinates such as $t +\zeta_2 \zeta_3$
is unavoidable in the theory of supermanifolds and integration on them.
While it would be interesting to describe the missing region schematically as
\begin{equation}
 	\int \mathrm{d} \zeta_2 \mathrm{d} \zeta_3
	\int_{1/2}^{1/2+\zeta_2 \zeta_3} \mathrm{d}t \, \mathcal{F}(t; 0, \zeta_2, \zeta_3,  0) \,,
\end{equation}
the form of the contribution in the $t$ channel which we naturally obtain
in open superstring field theory is different from the form in~\eqref{A_t-schematical}.
In subsection~\ref{sec:interpolation-super} we instead considered
the contribution in the $t$ channel of the form
\begin{equation}
	\int_{1/2}^1 \mathrm{d}t \int \mathrm{d} \zeta_2 \mathrm{d} \zeta_3 \, \mathcal{F}(t; \zeta_3, \zeta_2, 0, 0) \,,
\end{equation}
and we interpolated between the boundaries in the $s$ channel and in the $t$ channel as follows:
\begin{equation}
	\int_0^1 \mathrm{d}t'
	\int \mathrm{d} \zeta_2 \mathrm{d} \zeta_3 \, \mathcal{F}(1/2; t' \zeta_3, \zeta_2, (1-t') \, \zeta_3, 0) \,.
\end{equation}
We did not encounter the mixing of even and odd coordinates of the form $t+\zeta_2 \zeta_3$,
but the odd coordinate $\zeta_3$ is mixed with the even coordinate $t'$
as $t' \zeta_3$ or as $(1-t') \, \zeta_3$.
Furthermore, we can evaluate the super cross ratio at the boundaries
of the integration regions
and confirm that $\widehat{\Phi}_s = 1/2$ at the boundary in the $s$ channel
and $\widehat{\Phi}_s = 1/2 +\zeta_3 \zeta_2$ at the boundary in the $t$ channel.
Therefore there is a missing region which is not covered
and the additional contribution precisely covers the missing region.

Our approach is based on the parameterization of the odd direction
of the supermoduli space in terms of a line integral $G [ \, f \, ]$,
and we interpolate
\begin{equation}
	X [ \, f_1 \, ] = \int \mathrm{d} \zeta \mathrm{d} \tilde{\zeta} \,
	\mathrm{e}^{-\{ \, Q',\, \zeta \beta [ \, f_1 \, ] \, \}}
\end{equation}
and
\begin{equation}
	X [ \, f_2 \, ] = \int \mathrm{d} \zeta \mathrm{d} \tilde{\zeta} \,
	\mathrm{e}^{-\{ \, Q',\, \zeta \beta [ \, f_2 \, ] \, \}}
\end{equation}
as follows:
\begin{equation}
	\int_0^1 \mathrm{d}t' \int \mathrm{d}\tilde{t'}
	\int \mathrm{d} \zeta \mathrm{d} \tilde{\zeta} \,
	\mathrm{e}^{-\{ \, Q',\, t' \zeta \beta [ \, f_1 \, ] \, \}} \,
	\mathrm{e}^{-\{ \, Q',\, (1-t') \, \zeta \beta [ \, f_2 \, ] \, \}} \,.
\end{equation}
If we instead use a local insertion of the supercurrent $G(z)$ to parameterize
the odd direction of the supermoduli space,
we will need to interpolate between two local insertions of the picture-changing operators.
In the description of the superconformal ghost sector in terms of $\xi (z)$, $\eta (z)$, and $\phi (z)$,
the picture-changing operator $X_\mathrm{PCO} (z)$ can be written as
\begin{equation}
X_\mathrm{PCO} (z) = Q \cdot \xi (z) \,.
\end{equation}
Then we can interpolate between $X_\mathrm{PCO} (z_1)$ and $X_\mathrm{PCO} (z_2)$ as
\begin{equation}
X_\mathrm{PCO} (z_1) -X_\mathrm{PCO} (z_2)
= Q \cdot ( \, \xi (z_1) -\xi (z_2) \, )
= Q \cdot \int_{z_2}^{z_1} \mathrm{d}z \, \partial \xi (z) \,.
\end{equation}
Note that although the operator $\xi (z)$ is not in the small Hilbert space,
the integrand of the interpolation is written in terms of its derivative $\partial \xi (z)$
and so the interpolation is within the small Hilbert space.
This interpolation is called {\normalfont \itshape vertical integration} in the approach
developed in \cite{Sen:2014pia,Sen:2015hia}.

In the language of the $\beta \gamma$ system,
the operator $\xi (z)$ is usually denoted as $\Theta ( \beta (z) )$, which is defined by
\begin{equation}
\xi (z) = \Theta ( \beta (z) ) = -\int \mathrm{d} \tau \, \frac{1}{\tau} \, \exp \, ( -\beta (z) \, \tau ) \,.
\end{equation}
The existence of a singular factor $1 / \tau$ corresponds
to the fact that the operator $\xi (z)$ is not in the small Hilbert space.
Similarly, the operator $X [ \, f \, ]$ can be written as
\begin{equation}
X [ \, f \, ] = Q \cdot \Theta ( \beta [ \, f \, ] ) \,,
\end{equation}
where $\Theta ( \beta [ \, f \, ] )$ is defined by
\begin{equation}
\Theta ( \beta [ \, f \, ] ) = -\int \mathrm{d} \tau \,  \frac{1}{\tau} \, \exp \, ( -\beta [ \, f \, ] \, \tau ) \,.
\end{equation}
We can then express $X [ \, f_1 \, ] -X [ \, f_2 \, ]$ as
\begin{equation}
X [ \, f_1 \, ] -X [ \, f_2 \, ]
= Q \cdot (\, \Theta ( \beta [ \, f_1 \, ] ) -\Theta ( \beta [ \, f_2 \, ] ) \, ) \,.
\end{equation}
As in the case of $\xi (z_1) -\xi (z_2)$,
the difference $\Theta ( \beta [ \, f_1 \, ] ) -\Theta ( \beta [ \, f_2 \, ] )$ is in the small Hilbert space.
We can see the cancellation of the singular factor $1 / \tau$ in the following way:
\begin{equation}
\begin{split}
& \Theta ( \beta [ \, f_1 \, ] ) -\Theta ( \beta [ \, f_2 \, ] )
= -\int \mathrm{d} \tau \,  \frac{1}{\tau} \, \exp \, ( -\beta [ \, f_1 \, ] \, \tau )
+\int \mathrm{d} \tau \,  \frac{1}{\tau} \, \exp \, ( -\beta [ \, f_2 \, ] \, \tau ) \\
& = -\int_0^1 \mathrm{d} t' \, \frac{\partial}{\partial t'}
\int \mathrm{d} \tau \,  \frac{1}{\tau} \,
\exp \, ( \, ( -t' \, \beta [ \, f_1 \, ] -(1-t') \, \beta [ \, f_2 \, ] \, ) \, \tau \, ) \\
& = \int_0^1 \mathrm{d} t' \, \int \mathrm{d} \tau \, 
( \, \beta [ \, f_1 \, ] -\beta [ \, f_2 \, ] \, ) \,
\exp \, ( \, ( -t' \, \beta [ \, f_1 \, ] -(1-t') \, \beta [ \, f_2 \, ] \, ) \, \tau \, ) \\
& = \int_0^1 \mathrm{d} t' \,
( \, \beta [ \, f_1 \, ] -\beta [ \, f_2 \, ] \, ) \,
\delta ( \, t' \, \beta [ \, f_1 \, ] +(1-t') \, \beta [ \, f_2 \, ] \, ) \,.
\end{split}
\end{equation}
This is exactly the expression~\eqref{Xi-after-integration} of $\Xi [ \, f_1, f_2 \, ]$.
We have therefore found\footnote{
We thank Ted Erler for pointing out this relation to us.
}
\begin{equation}
\Xi [ \, f_1, f_2 \, ] = \Theta ( \beta [ \, f_1 \, ] ) -\Theta ( \beta [ \, f_2 \, ] ) \,.
\end{equation}
If we allow ourselves to use $\Theta ( \beta [ \, f \, ] )$ at intermediate steps,
we can apply the powerful method of generating multi-string products
satisfying the $A_\infty$ relations by field redefinition outside the small Hilbert space
developed in~\cite{Erler:2013xta}.
On the other hand, use of $\Theta ( \beta [ \, f \, ] )$ obscures
the relation to the supermoduli space of disks with NS punctures.

\section{Conclusions and discussion}
\label{sec:conclusions-discussion}

We presented a new approach to formulating open superstring field theory
based on the covering of the supermoduli space of super-Riemann surfaces.
We considered the NS sector
and constructed the cubic vertex $\langle \, A_1, \, [ \, A_2, A_3 \, ] \, \rangle$
for three states $A_1$, $A_2$, and $A_3$ given in~\eqref{cubic-vertex}
and the quartic vertex $\langle \, A_1, \, [ \, A_2, A_3, A_4 \, ] \, \rangle$
for four states $A_1$, $A_2$, $A_3$, and $A_4$ given in~\eqref{quartic-vertex-Xi}.
The associated two-string and three-string products
satisfy the $A_\infty$ relation~\eqref{A_infinity}
and have the cyclic properties in~\eqref{eq:cyclicity}.
The cubic vertex~\eqref{cubic-vertex}
is constructed by an insertion of a line integral $X [ \, f \, ]$ defined by
\begin{equation}
	X [ \, f \, ]
	= \int \mathrm{d}\zeta \mathrm{d}\tilde{\zeta} \,
	\mathrm{e}^{-\{Q', \, \zeta\, \beta [ \, f \, ] \, \}} \,.
\end{equation}
It takes the form of an integral over $\zeta$,
and we can identify $\zeta$ with an odd modulus of disks with three NS punctures.
The quartic vertex~\eqref{quartic-vertex-Xi}
is constructed by an insertion of the operator~$\Xi$.
The operator $\Xi$ is a linear combination of products
of the form $\Xi [ \, f_1, f_2 \, ] \, X [ \, f_3 \, ]$,
where $\Xi [ \, f_1, f_2 \, ]$ is defined by
\begin{equation}
\Xi [ \, f_1, f_2 \, ]
= \int_0^1 \mathrm{d}t' \int \mathrm{d} \tilde{t'} \int \mathrm{d} \zeta \mathrm{d} \tilde{\zeta} \,
\mathrm{e}^{-\{ \, Q',\, t' \zeta \beta [ \, f_1 \, ] \, \}} \,
\mathrm{e}^{-\{ \, Q',\, (1-t') \, \zeta \beta [ \, f_2 \, ] \, \}}
\,.
\end{equation}
This takes
the form of an integral over one even variable $t'$ and two odd variables,
where one of them is from $\Xi [ \, f_1, f_2 \, ]$ and the other is from $X [ \, f_3 \, ]$.
We can interpret $\Xi$ as the integral
over the region of the supermoduli space of disks with four NS punctures
which is not covered by Feynman diagrams with two cubic vertices and one propagator.
The relation~\eqref{A_infinity} follows from this construction.

It is conceptually straightforward to generalize our construction to higher-order vertices.
For the construction of the quartic vertex $\langle \, A_1, \, [ \, A_2, A_3, A_4 \, ] \, \rangle$,
we first expressed $\langle \, A_1, [ \, [ \, A_2, A_3 \, ], A_4 \, ] \, \rangle$
and $\langle \, A_1, [ \, A_2, [ \, A_3, A_4 \, ]  \, ] \, \rangle$
in terms of products of line integrals of the form $X [ \, f_1 \, ] \, X [ \, f_2 \, ]$.
For the construction of the quintic vertex $\langle \, A_1, \, [ \, A_2, A_3, A_4, A_5 \, ] \, \rangle$,
we will need to express $\langle \, A_1, [ \, [ \, A_2, A_3 \, ], A_4, A_5 \, ] \, \rangle$
and other similar terms
in terms of products of operators of the form $\Xi [ \, f_1, f_2 \, ] \, X [ \, f_3 \, ] \, X [ \, f_4 \, ]$.
We then need to construct an operator
such that its BRST transformation
gives the resulting linear combination
of operators of the form $\Xi [ \, f_1, f_2 \, ] \, X [ \, f_3 \, ] \, X [ \, f_4 \, ]$.
Such an operator takes the form of an integral over two even variables and three odd variables,
and it should be related to the region of the supermoduli space of disks
with five NS punctures
which is not covered by Feynman diagrams with only cubic and quartic vertices.
While the idea itself is simple,
the construction of the quintic vertex with an appropriate cyclic property
by explicitly implementing the idea
would be combinatorially complicated,
and there would be a lot of ambiguities.
It is an important open problem
to establish a systematic way to implement the idea
and find a closed-form expression for higher-order vertices.\footnote{
In~\cite{Sen:2015hia} local picture-changing operators instead of line integrals are used,
but the analysis in~\cite{Sen:2015hia} would be very useful
in generalizing our construction to higher orders.
In particular, one picture-changing operator is moved at a time
to avoid spurious singularities in~\cite{Sen:2015hia},
which is analogous to~\eqref{g_1-to-g_2-then-h_1-to-h_2} or~\eqref{h_1-to-h_2-then-g_1-to-g_2}
in our case.
We thank Ashoke Sen for useful discussions on this.
}

Another important direction for the generalization
is to incorporate the Ramond sector.
In the recent constructions of a gauge-invariant action
of open superstring field theory
including both the NS sector and the Ramond sector~\cite{Kunitomo:2015usa, Erler:2016ybs, Konopka:2016grr},
the interpretation of the covariant kinetic term for the Ramond sector in~\cite{Kunitomo:2015usa}
in the context of the supermoduli space of super-Riemann surfaces
has played an important role.
In fact, the operator $X_\mathrm{Ramond}$ used in~\cite{Kunitomo:2015usa}
to characterize the constraint on the Ramond sector
is given by
\begin{equation}
	X_\mathrm{Ramond}
	= \int \mathrm{d} \zeta \mathrm{d} \tilde{\zeta} \, \mathrm{e}^{-\{ Q',\, \zeta \, \beta_0 \}}
	= \int \mathrm{d} \zeta \mathrm{d} \tilde{\zeta} \, \mathrm{e}^{-\tilde{\zeta} \beta_0 +\zeta G_0}
	= G_0 \, \delta ( \beta_0 ) +b_0 \, \delta' ( \beta_0 ) \,,
\end{equation}
where $\beta_0$ is the zero mode of the $\beta$ ghost
and $G_0$ is the zero mode of the supercurrent,
and it has the same structure as $X$ we used for the NS sector in this paper.
To describe the constraint imposed on the Ramond sector,
it is easier to use the small Hilbert space in terms of
the $\beta \gamma$ ghosts
and it is awkward to use the large Hilbert space in terms of $\xi$, $\eta$, and $\phi$.
On the other hand, the large Hilbert space was used
for the constructions of interaction terms in~\cite{Kunitomo:2015usa, Erler:2016ybs, Konopka:2016grr},
and thus the whole constructions are somewhat hybrid
between the small Hilbert space and the large Hilbert space.
The approach to the construction of open superstring field theory
presented in this paper is based on the perspective
of the supermoduli space of super-Riemann surfaces,
and we will be able to construct open superstring field theory
including both the NS sector and the Ramond sector
within the framework of the $\beta \gamma$ ghosts.
The construction would be conceptually straightforward
even with the Ramond sector,
but again an important open problem would be
to find a closed-form expression for higher-order vertices.
\\

\bigskip
\noindent
{\normalfont \bfseries \large Acknowledgments}

\medskip
We would like to thank Yuji Tachikawa for useful discussions
and Ted Erler and Ashoke Sen for valuable comments
on an earlier version of this paper.
The work of K.O.\ was supported in part
by the Programs for Leading Graduate Schools, MEXT,
Japan, via the Advanced Leading Graduate Course for Photon Science
and by JSPS Research Fellowship for Young Scientists.
K.O.\ also gratefully acknowledges support from the Institute for Advanced Study.
The work of Y.O.\ was supported in part
by Grant-in-Aid for Scientific Research~(B) No.~25287049,
Grant-in-Aid for Scientific Research~(C) No.~24540254,
and Grant-in-Aid for Scientific Research~(C) 17K05408
from the Japan Society for the Promotion of Science (JSPS).

\appendix

\section{Correlation functions of the $\beta \gamma$ system}
\label{sec:beta-gamma}

A primary role of the $\beta \gamma$ ghosts is to provide
appropriate measures for integration on the supermoduli space
of super-Riemann surfaces.
As was emphasized in~\cite{Witten:2012bh},
the $\beta \gamma$ ghosts are Grassmann even
but they do not obey any reality condition,
and the path integral of the $\beta \gamma$ system
should be understood as a formal algebraic operation
analogous to the integral over Grassmann-odd variables.
Furthermore, we use the operator usually denoted by $\delta ( \gamma (z) )$
for vertex operators in the $-1$ picture
and operators such as $\delta ( \beta_{-1/2} )$ play a crucial role
in the approach developed in this paper,
so we need to define these operators
in the treatment of the path integral of the $\beta \gamma$ system
as an algebraic operation.
In this appendix we briefly review the algebraic treatment
of the path integral of the $\beta \gamma$ system
following section 10 of~\cite{Witten:2012bh},
and we explain how the relations~\eqref{eq:delbetagamma}
and~\eqref{eq:delprimebetagamma} are derived in this treatment
and how we calculate correlation functions including an insertion of $\Xi [ \, f_1, f_2 \, ]$.\footnote{
Correlation functions of the superconformal ghost sector
based on the description
in terms of $\xi$, $\eta$, and $\phi$
exhibit an apparently different structure,
and the transformation to the form
which is natural in the $\beta \gamma$ system
was discussed in~\cite{Lechtenfeld:1989wu, Morozov:1989ma}.
}

\subsection{The definition of the Gaussian integral of Grassmann-even variables}

Let us begin with the definition of the integration
of Grassmann-even variables as an algebraic operation
when the number of variables is finite.
We only need to define the following Gaussian integral
for Grassmann-even variables $x_1, x_2, \ldots, x_{2n}$:
\begin{equation}
	\int \mathrm{d}^{2n} x \, \exp \biggl( \, -\frac{1}{2} \sum_{a,b} x_a N_{ab} \, x_b \, \biggr) \,.
\label{Gaussian-even}
\end{equation}
As we mentioned before, its definition is analogous to the integration
of Grassmann-odd variables.
The integration of a Grassmann-odd variable $x^*$ has the following properties:
\begin{align}
	\int \mathrm{d}x^* \, \partial_{x^*} f (x^*) & = 0 \,, \\
	\int \mathrm{d}x^* \, f (x^*+a^*) & = \int \mathrm{d}x^* \, f (x^*) \,,
\label{x^*+a^*}
\end{align}
where $a^*$ is an arbitrary Grassmann-odd constant.
We require these properties for the integration of a Grassmann-even variable $x$ as well:
\begin{align}
	\int \mathrm{d}x \, \partial_x f (x) & = 0 \,,
\label{integration-by-parts} \\
\int \mathrm{d}x \, f (x+a) & = \int \mathrm{d}x \, f (x) \,,
\label{x+a}
\end{align}
where $a$ is an arbitrary Grassmann-even constant.
Consider the following Gaussian integral for Grassmann-odd variables
$x^*_1, x^*_2, \ldots , x^*_{2n}$:
\begin{equation}
	\int \mathrm{d}^{2n} x^* \, \exp \biggl( \, -\frac{1}{2} \sum_{a,b} x^*_a \widetilde{N}_{ab} \, x^*_b \, \biggr) \,.
\end{equation}
We consider the case where the variables $x^*_a$ split into variables $\beta^*_i$ of ghost number $-1$
and variables $\gamma^*_i$ of ghost number $1$,
and we assume that the antisymmetric matrix $\widetilde{N}$ takes the following form:
\begin{equation}
\widetilde{N} = \biggl(
\begin{array}{cc}
0 & M \\
-M^t & 0
\end{array} \biggr) \,,
\end{equation}
where $M^t$ is the transpose of $M$.
The Gaussian integral is then given by
\begin{equation}
	\int \mathrm{d}^{2n} x^* \, \exp \biggl( \, -\frac{1}{2} \sum_{a,b} x^*_a \widetilde{N}_{ab} \, x^*_b \, \biggr)
	= \int \mathrm{d}^n \beta^* \mathrm{d}^n \gamma^* \, \exp \biggl( \, -\sum_{i,j} \beta^*_i M_{ij} \gamma^*_j \, \biggr)
\end{equation}
up to a possible sign by reordering from $\mathrm{d}^{2n} x^*$
to $\mathrm{d}^n \beta^* \mathrm{d}^n \gamma^*$,
where $\mathrm{d}^n \beta^* \mathrm{d}^n \gamma^*$ is defined by
$\mathrm{d}^n \beta^* \mathrm{d}^n \gamma^*
= \mathrm{d} \beta^*_1 \mathrm{d} \gamma^*_1 \mathrm{d} \beta^*_2 \mathrm{d} \gamma^*_2
\ldots \mathrm{d} \beta^*_n \mathrm{d} \gamma^*_n$.
We normalize the Gaussian integral as follows:
\begin{equation}
	\int \mathrm{d}^n \beta^* \mathrm{d}^n \gamma^* \, \exp \biggl( \, -\sum_{i,j} \beta^*_i M_{ij} \gamma^*_j \, \biggr)
= \det M \,.
\label{determinant-odd}
\end{equation}
For the Gaussian integral~\eqref{Gaussian-even} for Grassmann-even variables,
we also consider the case where the variables $x_a$ split into variables $\beta_i$ of ghost number $-1$
and variables $\gamma_i$ of ghost number $1$,
and we assume that the symmetric matrix $N$ takes the following form:
\begin{equation}
N = \biggl(
\begin{array}{cc}
0 & M \\
M^t & 0
\end{array} \biggr) \,.
\end{equation}
The Gaussian integral is then given by
\begin{equation}
	\int \mathrm{d}^{2n} x \, \exp \biggl( \, -\frac{1}{2} \sum_{a,b} x_a N_{ab} \, x_b \, \biggr)
	= \int \mathrm{d}^n \beta \mathrm{d}^n \gamma \, \exp \biggl( \, -\sum_{i,j} \beta_i M_{ij} \gamma_j \, \biggr)
\end{equation}
up to a possible sign by reordering from $\mathrm{d}^{2n} x$
to $\mathrm{d}^n \beta \mathrm{d}^n \gamma$,
where $\mathrm{d}^n \beta \mathrm{d}^n \gamma$ is defined by
$\mathrm{d}^n \beta \mathrm{d}^n \gamma
= \mathrm{d} \beta_1 \mathrm{d} \gamma_1 \mathrm{d} \beta_2 \mathrm{d} \gamma_2
\ldots \mathrm{d} \beta_n \mathrm{d} \gamma_n$.\footnote{
We treat $\mathrm{d}x_a$ as a Grassmann-odd object. We will elaborate on this shortly.
}
We define the Gaussian integral as follows:
\begin{equation}
	\int \mathrm{d}^n \beta \mathrm{d}^n \gamma \, \exp \biggl( \, -\sum_{i,j} \beta_i M_{ij} \gamma_j \, \biggr)
= \frac{1}{\det M} \,.
\label{determinant-even}
\end{equation}
Note that the overall sign of the determinant depends on how we order the variables.
If we order the variables, for example, as
\begin{equation}
-( \begin{array}{cc}
\beta_1 & \beta_2
\end{array} )
\biggl( \begin{array}{cc}
M_{11} & M_{12} \\
M_{21} & M_{22}
\end{array} \biggr)
\biggl( \begin{array}{c}
\gamma_1 \\
\gamma_2
\end{array} \biggr) \,,
\end{equation}
the determinant is given by
\begin{equation}
\biggl| \begin{array}{cc}
M_{11} & M_{12} \\
M_{21} & M_{22}
\end{array} \biggr|
= M_{11} M_{22} -M_{12} M_{21} \,,
\end{equation}
but if we order the variables as
\begin{equation}
-( \begin{array}{cc}
\beta_2 & \beta_1
\end{array} )
\biggl( \begin{array}{cc}
M_{21} & M_{22} \\
M_{11} & M_{12}
\end{array} \biggr)
\biggl( \begin{array}{c}
\gamma_1 \\
\gamma_2
\end{array} \biggr) \,,
\end{equation}
the determinant is given by
\begin{equation}
\biggl| \begin{array}{cc}
M_{21} & M_{22} \\
M_{11} & M_{12}
\end{array} \biggr|
= M_{21} M_{12} -M_{22} M_{11}
= {}-( M_{11} M_{22} -M_{12} M_{21} ) \,.
\end{equation}
Our prescription is to correlate this ordering of the quadratic form
with the ordering of the measure
$\mathrm{d} \beta_1 \mathrm{d} \gamma_1 \mathrm{d} \beta_2 \mathrm{d} \gamma_2 
\ldots \mathrm{d} \beta_n \mathrm{d} \gamma_n$
and to treat $\mathrm{d} \beta_i$ and $\mathrm{d} \gamma_j$ as Grassmann-odd objects.

The correspondence between~\eqref{determinant-even} and~\eqref{determinant-odd}
also provides an efficient way to calculate integrals of Grassmann-even variables
in terms of integrals of Grassmann-odd variables.
First, we have
\begin{equation}
\int \mathrm{d}^n \beta \mathrm{d}^n \gamma \, \exp \biggl( \, -\sum_{i,j} \beta_i M_{ij} \gamma_j \, \biggr)
= \biggl[ \, \int \mathrm{d}^n \beta^* \mathrm{d}^n \gamma^* \, \exp \biggl( \, -\sum_{i,j} \beta^*_i M_{ij} \gamma^*_j \, \biggr) \,
\biggr]^{-1} \,.
\label{determinant-formula}
\end{equation}

Second, consider the following integral:
\begin{equation}
	\int \mathrm{d}^n \beta \mathrm{d}^n \gamma \,
\exp \biggl( \, -\sum_{i,j} \beta_i M_{ij} \gamma_j \, \biggr) \, \gamma_k \, \beta_l \,.
\end{equation}
Because of the property~\eqref{x+a}, we have
\begin{equation}
	\frac{\displaystyle{\int \mathrm{d}^n \beta \mathrm{d}^n \gamma \,
\exp \biggl( \, -\sum_{i,j} \beta_i M_{ij} \gamma_j \, \biggr) \, \gamma_k \, \beta_l}}
{\displaystyle{\int \mathrm{d}^n \beta \mathrm{d}^n \gamma \,
\exp \biggl( \, -\sum_{i,j} \beta_i M_{ij} \gamma_j \, \biggr)}}
= ( M^{-1} )_{kl} \,.
\end{equation}
Because of the property~\eqref{x^*+a^*}, we also have
\begin{equation}
	\frac{\displaystyle{\int \mathrm{d}^n \beta^* \mathrm{d}^n \gamma^* \,
\exp \biggl( \, -\sum_{i,j} \beta^*_i M_{ij} \gamma^*_j \, \biggr) \, \gamma^*_k \, \beta^*_l}}
{\displaystyle{\int \mathrm{d}^n \beta^* \mathrm{d}^n \gamma^* \,
\exp \biggl( \, -\sum_{i,j} \beta^*_i M_{ij} \gamma^*_j \, \biggr)}}
= ( M^{-1} )_{kl} \,.
\label{eq:Minverse}
\end{equation}
We therefore have
\begin{equation}
	\frac{\displaystyle{\int \mathrm{d}^n \beta \mathrm{d}^n \gamma \,
\exp \biggl( \, -\sum_{i,j} \beta_i M_{ij} \gamma_j \, \biggr) \, \gamma_k \, \beta_l}}
{\displaystyle{\int \mathrm{d}^n \beta \mathrm{d}^n \gamma \,
\exp \biggl( \, -\sum_{i,j} \beta_i M_{ij} \gamma_j \, \biggr)}}
= \frac{\displaystyle{\int \mathrm{d}^n \beta^* \mathrm{d}^n \gamma^* \,
\exp \biggl( \, -\sum_{i,j} \beta^*_i M_{ij} \gamma^*_j \, \biggr) \, \gamma^*_k \, \beta^*_l}}
{\displaystyle{\int \mathrm{d}^n \beta^* \mathrm{d}^n \gamma^* \,
\exp \biggl( \, -\sum_{i,j} \beta^*_i M_{ij} \gamma^*_j \, \biggr)}} \,.
\end{equation}
Using the formula~\eqref{determinant-formula}, we find
\begin{equation}
	\int \mathrm{d}^n \beta \mathrm{d}^n \gamma \,
\exp \biggl( \, -\sum_{i,j} \beta_i M_{ij} \gamma_j \, \biggr) \, \gamma_k \, \beta_l
= \frac{\displaystyle{\int \mathrm{d}^n \beta^* \mathrm{d}^n \gamma^* \,
\exp \biggl( \, -\sum_{i,j} \beta^*_i M_{ij} \gamma^*_j \, \biggr) \, \gamma^*_k \, \beta^*_l}}
{\displaystyle{\biggl[ \,
\int \mathrm{d}^n \beta^* \mathrm{d}^n \gamma^* \,
\exp \biggl( \, -\sum_{i,j} \beta^*_i M_{ij} \gamma^*_j \, \biggr) \, \biggr]^2}} \,.
\label{propagator-formula}
\end{equation}

Third, consider the following integral:
\begin{equation}
	\int \mathrm{d}^n \beta \mathrm{d}^n \gamma \,
\exp \biggl( \, -\sum_{i,j} \beta_i M_{ij} \gamma_j \, \biggr) \,
\gamma_k \, \beta_l \, \gamma_m \, \beta_n \,.
\end{equation}
Because of the property~\eqref{x+a}, we have
\begin{equation}
	\frac{\displaystyle{\int \mathrm{d}^n \beta \mathrm{d}^n \gamma \,
\exp \biggl( \, -\sum_{i,j} \beta_i M_{ij} \gamma_j \, \biggr) \,
\gamma_k \, \beta_l \, \gamma_m \, \beta_n}}
{\displaystyle{\int \mathrm{d}^n \beta \mathrm{d}^n \gamma \,
\exp \biggl( \, -\sum_{i,j} \beta_i M_{ij} \gamma_j \, \biggr)}}
= ( M^{-1} )_{kl} \, ( M^{-1} )_{mn} +( M^{-1} )_{kn} \, ( M^{-1} )_{ml} \,.
\end{equation}
Using the formulas~\eqref{determinant-formula} and~\eqref{eq:Minverse}, we find
\begin{equation}
\int \mathrm{d}^n \beta \mathrm{d}^n \gamma \,
\exp \biggl( \, -\sum_{i,j} \beta_i M_{ij} \gamma_j \, \biggr) \,
\gamma_k \, \beta_l \, \gamma_m \, \beta_n
= \frac{\langle \, \gamma^*_k \, \beta^*_l \, \rangle \,
\langle \, \gamma^*_m \, \beta^*_n \, \rangle
+\langle \, \gamma^*_k \, \beta^*_n \, \rangle \,
\langle \, \gamma^*_m \, \beta^*_l \, \rangle}
{\displaystyle{\biggl[ \,
\int \mathrm{d}^n \beta^* \mathrm{d}^n \gamma^* \,
\exp \biggl( \, -\sum_{i,j} \beta^*_i M_{ij} \gamma^*_j \, \biggr) \, \biggr]^3}} \,,
	\label{four-point-formula}	
\end{equation}
where
\begin{equation}
\langle \, \gamma^*_k \, \beta^*_l \, \rangle
= \int \mathrm{d}^n \beta^* \mathrm{d}^n \gamma^* \,
\exp \biggl( \, -\sum_{i,j} \beta^*_i M_{ij} \gamma^*_j \, \biggr) \, \gamma^*_k \, \beta^*_l \,.
\end{equation}
Generalization to
\begin{equation}
	\int \mathrm{d}^n \beta \mathrm{d}^n \gamma \,
\exp \biggl( \, -\sum_{i,j} \beta_i M_{ij} \gamma_j \, \biggr) \,
\prod_{i=1}^N
\gamma_{k_i} \, \beta_{l_i}
\end{equation}
is straightforward.

\subsection{The path integral of the $\beta \gamma$ system}

Let us now consider the path integral of the $\beta \gamma$ system.
The action is given by
\begin{equation}
	I_{\beta \gamma} = \frac{1}{2 \pi} \int \mathrm{d}^2 z \, \beta \partial_{\tilde{z}} \gamma \,.
\end{equation}
Based on the discussion so far, let us also consider Grassmann-odd fields
$\beta^*$ and $\gamma^*$ with the following action:
\begin{equation}
	I_{\beta^* \gamma^*} = \frac{1}{2 \pi} \int \mathrm{d}^2 z \, \beta^* \partial_{\tilde{z}} \gamma^* \,.
\end{equation}
The $\beta^* \gamma^*$ system is the same as the familiar $bc$ system
except that the weight of $\beta^*$ is $3/2$ and the weight of $\gamma^*$ is $-1/2$.
As our primary interest in the $\beta \gamma$ system and the $\beta^* \gamma^*$ system
is related to the construction of a classical action
for open superstring field theory,
let us consider correlation functions
on the disk.
We use the doubling trick and in this appendix
all the correlation functions are defined on the whole complex plane.
There are two zero modes on the disk for the equation of motion
for $\gamma^*$, and we need two insertions of $\gamma^*$
for nonvanishing correlation functions.
We normalize the path integral as follows:
\begin{equation}
\langle \, \gamma^* (u_1) \, \gamma^* (u_2) \, \rangle
= \int \mathcal{D} \beta^* \, \mathcal{D} \gamma^* \, \exp \, ( -I_{\beta^* \gamma^*} ) \,
\gamma^* (u_1) \, \gamma^* (u_2)
= u_1 -u_2 \,.
\end{equation}
We can express an insertion of $\gamma^* (u)$
using a Grassmann-odd variable $\sigma^*$ as
\begin{equation}
	\gamma^* (u) = \int \mathrm{d} \sigma^* \, \exp \, ( \sigma^* \gamma^* (u) ) \,,
\end{equation}
and then the two-point function is written as
\begin{equation}
\begin{split}
\langle \, \gamma^* (u_1) \, \gamma^* (u_2) \, \rangle
& = \int \mathrm{d} \sigma^*_1 \mathrm{d} \sigma^*_2 \,
\langle \, \exp \, ( \sigma^*_1 \gamma^* (u_1) ) \, \exp \, ( \sigma^*_2 \gamma^* (u_2) ) \, \rangle \\
& = \int \mathcal{D} \beta^* \, \mathcal{D} \gamma^* \, \mathrm{d} \sigma^*_1 \mathrm{d} \sigma^*_2 \,
\exp \, ( -I_{\beta^* \gamma^*}
+\sigma^*_1 \gamma^* (u_1) +\sigma^*_2 \gamma^* (u_2) ) \,.
\end{split}
\label{gamma^*-gamma^*}
\end{equation}
This takes the form of~\eqref{determinant-odd}
when we regard $( \beta^* (z), \sigma^*_1, \sigma^*_2 )$
as variables of ghost number $-1$
and $\gamma^* (z)$ as variables of ghost number $1$.
Let us now consider the corresponding integral
for Grassmann-even variables:
\begin{equation}
	\int \mathcal{D} \beta \, \mathcal{D} \gamma \, \mathrm{d} \sigma_1 \mathrm{d} \sigma_2 \,
\exp \, ( -I_{\beta \gamma}
+\sigma_1 \gamma (u_1) +\sigma_2 \gamma (u_2) ) \,,
\end{equation}
where we regard $( \beta (z), \sigma_1, \sigma_2 )$
as variables of ghost number $-1$
and $\gamma (z)$ as variables of ghost number $1$.
This integral takes the form of~\eqref{determinant-even}.
It follows from the formula~\eqref{determinant-formula}
that this integral is related to~\eqref{gamma^*-gamma^*} as
\begin{equation}
\begin{split}
	& \int \mathcal{D} \beta \, \mathcal{D} \gamma \, \mathrm{d} \sigma_1 \mathrm{d} \sigma_2 \,
\exp \, ( -I_{\beta \gamma}
+\sigma_1 \gamma (u_1) +\sigma_2 \gamma (u_2) ) \\
& = \biggl[ \, \int \mathcal{D} \beta^* \, \mathcal{D} \gamma^* \, \mathrm{d} \sigma^*_1 \mathrm{d} \sigma^*_2 \,
\exp \, ( -I_{\beta^* \gamma^*}
+\sigma^*_1 \gamma^* (u_1) +\sigma^*_2 \gamma^* (u_2) ) \, \biggr]^{-1} \,.
\end{split}
\end{equation}
If we define $\delta ( \gamma (u) )$ by
\begin{equation}
	\delta ( \gamma (u) ) = \int \mathrm{d} \sigma \, \exp \, ( \sigma \gamma (u) ) \,,
\label{delta(gamma)}
\end{equation}
we therefore have
\begin{equation}
\begin{split}
\langle \, \delta ( \gamma (u_1) ) \, \delta ( \gamma (u_2) ) \, \rangle
& = \int \mathcal{D} \beta \, \mathcal{D} \gamma \, \mathrm{d} \sigma_1 \mathrm{d} \sigma_2 \,
\exp \, ( -I_{\beta \gamma} +\sigma_1 \gamma (u_1) +\sigma_2 \gamma (u_2) ) \\
& = \frac{1}{u_1-u_2} \,.
\end{split}
\label{delta(gamma)^2}
\end{equation}
The integration of the Grassmann-even variable $\sigma$ in~\eqref{delta(gamma)}
should be understood as an algebraic operation.
As we explained before, we treat $\mathrm{d} \sigma$ in~\eqref{delta(gamma)}
as a Grassmann-odd object.
Then the operator $\delta ( \gamma (u) )$ behaves as a Grassmann-odd operator,
which is consistent with the antisymmetry of $1/(u_1-u_2)$ in~\eqref{delta(gamma)^2}
under the exchange of $u_1$ and $u_2$.
It is also consistent with the description in terms of $\xi$, $\eta$, and $\phi$,
where $\delta (\gamma (u))$ is mapped to $\mathrm{e}^{-\phi (u)}$,
which is Grassmann odd.

It is straightforward to generalize this
to correlation functions with more operators.
Consider, for example, the following four-point function:
\begin{equation}
\langle \, \beta^* (w) \, \gamma^* (u_1) \, \gamma^* (u_2) \, \gamma^* (u_3) \, \rangle
= \int \mathcal{D} \beta^* \, \mathcal{D} \gamma^* \, \exp \, ( -I_{\beta^* \gamma^*} ) \,
\beta^* (w) \, \prod_{i=1}^3 \gamma^* (u_i) \,.
\end{equation}
We can express an insertion of $\beta^* (w)$
using a Grassmann-odd variable $\tau^*$ as
\begin{equation}
	\beta^* (w) = \int \mathrm{d} \tau^* \, \exp \, ( -\beta^* (w) \, \tau^* ) \,,
\end{equation}
and then the four-point function is written using Grassmann-odd variables
$\tau^*$, $\sigma^*_1$, $\sigma^*_2$, and $\sigma^*_3$ as
\begin{equation}
\begin{split}
	& \int \mathrm{d} \tau^* \mathrm{d} \sigma^*_1 \mathrm{d} \sigma^*_2 \mathrm{d} \sigma^*_3 \,
\langle \, \exp \, ( -\beta^* (w) \, \tau^* ) \,
\prod_{i=1}^3 \exp \, ( \sigma^*_i \gamma^* (u_i) ) \, \rangle \\
& = \int \mathcal{D} \beta^* \, \mathcal{D} \gamma^* \,
\mathrm{d} \tau^* \mathrm{d} \sigma^*_1 \mathrm{d} \sigma^*_2 \mathrm{d} \sigma^*_3 \,
\exp \, ( -I_{\beta^* \gamma^*}
-\beta^* (w) \, \tau^*
+\sum_{i=1}^3 \sigma^*_i \gamma^* (u_i) ) \,.
\end{split}
\label{beta^*-gamma^*^3}
\end{equation}
This takes the form of~\eqref{determinant-odd}
when we regard $( \beta^* (z), \sigma^*_1, \sigma^*_2, \sigma^*_3 )$
as variables of ghost number $-1$
and $( \gamma^* (z), \tau^* )$ as variables of ghost number $1$.
Let us now consider the corresponding integral
for Grassmann-even variables:
\begin{equation}
	\int \mathcal{D} \beta \, \mathcal{D} \gamma \, \mathrm{d} \tau \mathrm{d} \sigma_1 \mathrm{d} \sigma_2 \mathrm{d} \sigma_3 \,
\exp \, ( -I_{\beta \gamma}
-\beta (w) \, \tau +\sum_{i=1}^3 \sigma_i \gamma (u_i)  ) \,,
\end{equation}
where we regard $( \beta (z), \sigma_1, \sigma_2, \sigma_3 )$
as variables of ghost number $-1$
and $( \gamma (z), \tau )$ as variables of ghost number $1$.
This integral takes the form of~\eqref{determinant-even}.
It follows from the formula~\eqref{determinant-formula}
that this integral is related to~\eqref{beta^*-gamma^*^3} as
\begin{equation}
\begin{split}
	& \int \mathcal{D} \beta \, \mathcal{D} \gamma \, \mathrm{d} \tau \mathrm{d} \sigma_1 \mathrm{d} \sigma_2 \mathrm{d} \sigma_3 \,
\exp \, ( -I_{\beta \gamma}
-\beta (w) \, \tau +\sum_{i=1}^3 \sigma_i \gamma (u_i)  ) \\
& = \biggl[ \, \int \mathcal{D} \beta^* \, \mathcal{D} \gamma^* \,
\mathrm{d} \tau^* \mathrm{d} \sigma^*_1 \mathrm{d} \sigma^*_2 \mathrm{d} \sigma^*_3 \,
\exp \, ( -I_{\beta^* \gamma^*}
-\beta^* (w) \, \tau^*
+\sum_{i=1}^3 \sigma^*_i \gamma^* (u_i) ) \, \biggr]^{-1} \,.
\end{split}
\end{equation}
If we define $\delta ( \beta (w) )$ by
\begin{equation}
	\delta ( \beta (w) ) = \int \mathrm{d} \tau \, \exp \, ( -\beta (w) \, \tau ) \,,
\label{delta(beta)}
\end{equation}
we therefore have
\begin{equation}
\begin{split}
\langle \, \delta ( \beta (w) ) \,
\delta ( \gamma (u_1) ) \, \delta ( \gamma (u_2) ) \, \delta ( \gamma (u_3) ) \, \rangle
& = \frac{1}{\langle \, \beta^* (w) \, \gamma^* (u_1) \, \gamma^* (u_2) \, \gamma^* (u_3) \, \rangle} \,,
\end{split}
\end{equation}
where the correlation function on the right-hand side can be evaluated as
\begin{equation}
	\langle \, \beta^* (w) \, \gamma^* (u_1) \, \gamma^* (u_2) \, \gamma^* (u_3) \, \rangle
	= \frac{(u_1-u_2) \, (u_1-u_3) \, (u_2-u_3)}{(w-u_1) \, (w-u_2) \, (w-u_3)} \,.
	\label{betastar-gammastar}
\end{equation}
The integration of the Grassmann-even variable $\tau$ in~\eqref{delta(beta)}
should also be understood as an algebraic operation.
We treat $\mathrm{d} \tau$ as a Grassmann-odd object.

\subsection{Delta function operators for line integrals}
\label{sec:delta-function-operators}

We can also incorporate line integrals into correlation functions.
Consider, for example, the following correlation function:
\begin{equation}
\langle \, \beta^* [ \, f \, ] \, \gamma^* (u_1) \, \gamma^* (u_2) \, \gamma^* (u_3) \, \rangle
= \int \mathcal{D} \beta^* \, \mathcal{D} \gamma^* \, \exp \, ( -I_{\beta^* \gamma^*} ) \,
\beta^* [ \, f \, ] \, \prod_{i=1}^3 \gamma^* (u_i) \,,
\end{equation}
where $\beta^* [ \, f \, ]$ for $f(\xi)$ is defined by
\begin{equation}
	\beta^* [ \, f \, ] = \oint \frac{\mathrm{d}z}{2 \pi \mathrm{i}} \,
	\Bigl( \, \frac{\mathrm{d} f^{-1} (z)}{\mathrm{d}z} \, \Bigr)^{-1/2} \, \beta^* (z) \quad
	\text{with} \quad z = f(\xi) \,.
\end{equation}
This can also be written using Grassmann-odd variables
$\tau^*$, $\sigma^*_1$, $\sigma^*_2$, and $\sigma^*_3$ as
\begin{equation}
\begin{split}
	& \int \mathrm{d} \tau^* \mathrm{d} \sigma^*_1 \mathrm{d} \sigma^*_2 \mathrm{d} \sigma^*_3 \,
\langle \, \exp \, ( -\beta^* [ \, f \, ] \, \tau^* ) \,
\prod_{i=1}^3 \exp \, ( \sigma^*_i \gamma^* (u_i) ) \, \rangle \\
& = \int \mathcal{D} \beta^* \, \mathcal{D} \gamma^* \,
\mathrm{d} \tau^* \mathrm{d} \sigma^*_1 \mathrm{d} \sigma^*_2 \mathrm{d} \sigma^*_3 \,
\exp \, ( -I_{\beta^* \gamma^*}
-\beta^* [ \, f \, ] \, \tau^*
+\sum_{i=1}^3 \sigma^*_i \gamma^* (u_i) ) \,.
\end{split}
\label{beta[f]-gamma^*^3}
\end{equation}
This takes the form of~\eqref{determinant-odd}
when we regard $( \beta^* (z), \sigma^*_1, \sigma^*_2, \sigma^*_3 )$
as variables of ghost number $-1$
and $( \gamma^* (z), \tau^* )$ as variables of ghost number $1$.
Let us now consider the corresponding integral
for Grassmann-even variables:
\begin{equation}
	\int \mathcal{D} \beta \, \mathcal{D} \gamma \, \mathrm{d} \tau \mathrm{d} \sigma_1 \mathrm{d} \sigma_2 \mathrm{d} \sigma_3 \,
\exp \, ( -I_{\beta \gamma}
-\beta [ \, f \, ] \, \tau +\sum_{i=1}^3 \sigma_i \gamma (u_i)  ) \,,
\end{equation}
where we regard $( \beta (z), \sigma_1, \sigma_2, \sigma_3 )$
as variables of ghost number $-1$
and $( \gamma (z), \tau )$ as variables of ghost number $1$.
This integral takes the form of~\eqref{determinant-even}.
It follows from the formula~\eqref{determinant-formula}
that this integral is related to~\eqref{beta[f]-gamma^*^3} as
\begin{equation}
\begin{split}
	& \int \mathcal{D} \beta \, \mathcal{D} \gamma \, \mathrm{d} \tau \mathrm{d} \sigma_1 \mathrm{d} \sigma_2 \mathrm{d} \sigma_3 \,
\exp \, ( -I_{\beta \gamma}
-\beta [ \, f \, ] \, \tau +\sum_{i=1}^3 \sigma_i \gamma (u_i)  ) \\
& = \biggl[ \, \int \mathcal{D} \beta^* \, \mathcal{D} \gamma^* \,
\mathrm{d} \tau^* \mathrm{d} \sigma^*_1 \mathrm{d} \sigma^*_2 \mathrm{d} \sigma^*_3 \,
\exp \, ( -I_{\beta^* \gamma^*}
-\beta^* [ \, f \, ] \, \tau^*
+\sum_{i=1}^3 \sigma^*_i \gamma^* (u_i) ) \, \biggr]^{-1} \,.
\end{split}
\end{equation}
If we define $\delta ( \beta [ \, f \, ] )$ by
\begin{equation}
	\delta ( \beta [ \, f \, ] ) = \int \mathrm{d} \tau \, \exp \, ( -\beta [ \, f \, ] \, \tau ) \,,
\end{equation}
we therefore have
\begin{equation}
\begin{split}
\langle \, \delta ( \beta [ \, f \, ] ) \,
\delta ( \gamma (u_1) ) \, \delta ( \gamma (u_2) ) \, \delta ( \gamma (u_3) ) \, \rangle
& = \frac{1}{\langle \, \beta^* [ \, f \, ] \, \gamma^* (u_1) \, \gamma^* (u_2) \, \gamma^* (u_3) \, \rangle} \,.
\end{split}
\end{equation}

Let us prove the relation~\eqref{eq:delbetagamma} based on these definitions
of $\delta (\gamma (u))$ and $\delta ( \beta [ \, f \, ] )$.
The line integral $\beta_{-1/2}$ corresponds to $\beta [ \, f \, ]$ with $f(\xi) = \xi$:
\begin{equation}
	\beta [ \, f \, ] = \oint \frac{\mathrm{d}z}{2 \pi \mathrm{i}} \,
	\Bigl( \, \frac{\mathrm{d} f^{-1} (z)}{\mathrm{d}z} \, \Bigr)^{-1/2} \, \beta (z)
	= \oint \frac{\mathrm{d}z}{2 \pi \mathrm{i}} \, \beta (z) = \beta_{-1/2} \quad
\text{when} \quad z = f(\xi) = \xi \,.
\end{equation}
The action of $\delta (\beta_{-1/2})$ on $\delta (\gamma (0))$ is then
\begin{equation}
\delta (\beta_{-1/2}) \cdot \delta (\gamma (0))
= \int \mathrm{d} \tau \, \mathrm{e}^{-\beta_{-1/2} \, \tau }
\cdot \int \mathrm{d} \sigma \, \mathrm{e}^{\, \sigma \, \gamma (0)}
= \int \mathrm{d} \tau \mathrm{d} \sigma \, \mathrm{e}^{-\beta_{-1/2} \, \tau }
\cdot \mathrm{e}^{\, \sigma \, \gamma (0)} \,.
\label{beta_-1/2-gamma(0)}
\end{equation}
The action of $\beta_{-1/2}$ on an arbitrary function $f(\gamma(0))$ of $\gamma(0)$ is
\begin{align}
	\beta_{-1/2}\cdot f(\gamma(0)) = - f'(\gamma(0))
\end{align}
and thus
\begin{align}
	\mathrm{e}^{-\beta_{-1/2} \, \tau}\cdot f(\gamma(0)) =  f(\gamma(0)+\tau) \,.
\end{align}
We therefore have
\begin{equation}
\begin{split}
\delta (\beta_{-1/2}) \cdot \delta (\gamma (0))
& = \int \mathrm{d} \tau \mathrm{d} \sigma \, \mathrm{e}^{-\beta_{-1/2} \, \tau }
\cdot \mathrm{e}^{\, \sigma \, \gamma (0)}
= \int \mathrm{d} \tau \mathrm{d} \sigma \,
\mathrm{e}^{\, \sigma \, ( \gamma (0) +\tau)} \\
& = \int \mathrm{d} \tau \mathrm{d} \sigma \,
\mathrm{e}^{\, \sigma \tau}
= -\int \mathrm{d} \sigma \mathrm{d} \tau \,
\mathrm{e}^{\, \sigma \tau} = 1 \,,
\end{split}
\end{equation}
where we used the property~\eqref{x+a}.
This completes the proof of~\eqref{eq:delbetagamma}.
When we use the $\beta^* \gamma^*$ system,
the calculation corresponding to~\eqref{beta_-1/2-gamma(0)} is
\begin{equation}
\int \mathrm{d} \tau^* \, \mathrm{e}^{-\beta^*_{-1/2} \, \tau^* }
\cdot \int \mathrm{d} \sigma^* \, \mathrm{e}^{\, \sigma^* \, \gamma^* (0)}
= \beta^*_{-1/2} \cdot \gamma^* (0) = 1 \,,
\end{equation}
where
\begin{equation}
	\beta^*_{-1/2} = \oint \frac{\mathrm{d}z}{2 \pi \mathrm{i}} \, \beta^* (z) \,.
\end{equation}
The relation~\eqref{eq:delbetagamma} immediately follows from this.

Let us next prove the relation~\eqref{eq:delprimebetagamma}.
The action of $\delta' (\beta_{-1/2})$ on $\delta (\gamma (0))$ is
\begin{equation}
\delta' (\beta_{-1/2}) \cdot \delta (\gamma (0))
= -\int \mathrm{d} \tau \, \tau \, \mathrm{e}^{-\beta_{-1/2} \, \tau }
\cdot \int \mathrm{d} \sigma \, \mathrm{e}^{\, \sigma \, \gamma (0)}
= -\int \mathrm{d} \tau \mathrm{d} \sigma \, \tau \, \mathrm{e}^{-\beta_{-1/2} \, \tau }
\cdot \mathrm{e}^{\, \sigma \, \gamma (0)} \,.
\end{equation}
We have
\begin{equation}
\begin{split}
\delta' (\beta_{-1/2}) \cdot \delta (\gamma (0))
& = -\int \mathrm{d} \tau \mathrm{d} \sigma \, \tau \, \mathrm{e}^{-\beta_{-1/2} \, \tau }
\cdot \mathrm{e}^{\, \sigma \, \gamma (0)} \\
& = -\int \mathrm{d} \tau \mathrm{d} \sigma \, \tau \,
\mathrm{e}^{\, \sigma \, ( \gamma (0) +\tau)}
= -\int \mathrm{d} \tau \mathrm{d} \sigma \, ( \tau -\gamma (0) ) \,
\mathrm{e}^{\, \sigma \tau} \\
& = -\int \mathrm{d} \tau \mathrm{d} \sigma \, \tau \,
\mathrm{e}^{\, \sigma \tau}
+\gamma (0) \int \mathrm{d} \tau \mathrm{d} \sigma \,
\mathrm{e}^{\, \sigma \tau} \,,
\end{split}
\end{equation}
where we used the property~\eqref{x+a}.
Since
\begin{equation}
\int \mathrm{d} \tau \mathrm{d} \sigma \, \tau \, \mathrm{e}^{\, \sigma \tau}
= \int \mathrm{d} \tau \mathrm{d} \sigma \, \partial_\sigma \, \mathrm{e}^{\, \sigma \tau}  = 0\,,
\end{equation}
where we used the property~\eqref{integration-by-parts},
we find
\begin{equation}
\delta' ( \beta_{-1/2} ) \cdot \delta( \gamma(0) ) = \gamma (0) \,.
\end{equation}
This completes the proof of~\eqref{eq:delprimebetagamma}.

The relation~\eqref{eq:delprimebetagamma} can also be shown
by writing $\delta' ( \beta_{-1/2} )$ as
\begin{equation}
\delta' ( \beta_{-1/2} ) = [ \, \gamma_{1/2}, \delta( \beta_{-1/2} ) \, ] \,,
\end{equation}
where
\begin{equation}
	\gamma_{1/2} = \oint \frac{\mathrm{d}z}{2 \pi \mathrm{i}} \, \frac{1}{z} \, \gamma (z) \,.
\end{equation}
This follows from
\begin{equation}
[ \, \gamma_{1/2}, \beta_{-1/2} \, ] = 1 \,,
\end{equation}
and for an arbitrary function $f (\beta_{-1/2})$ of $\beta_{-1/2}$ we have
\begin{equation}
[ \, \gamma_{1/2}, f (\beta_{-1/2}) \, ] = f' (\beta_{-1/2}) \,.
\end{equation}
First, the operator $\delta( \gamma(0) )$ is annihilated by the action of $\gamma_{1/2}$:
\begin{equation}
\begin{split}
\gamma_{1/2} \cdot \delta( \gamma(0) )
& = \oint \frac{\mathrm{d}z}{2 \pi \mathrm{i}} \, \frac{1}{z} \, \gamma (z)
\int \mathrm{d} \sigma \, \mathrm{e}^{\, \sigma \, \gamma (0)}
= \int \mathrm{d} \sigma \, \gamma (0) \, \mathrm{e}^{\, \sigma \, \gamma (0)} \\
& = \int \mathrm{d} \sigma \, \partial_{\sigma} \, \mathrm{e}^{\, \sigma \, \gamma (0)}
= 0 \,.
\end{split}
\end{equation}
In fact, this is one of the conditions that the operator $\delta ( \gamma (0) )$
corresponds to the vacuum of the $\beta \gamma$ system
with picture number $-1$, where $\gamma_{1/2}$ plays a role of an annihilation operator.
The action of $\delta' ( \beta_{-1/2} )$ on $\delta( \gamma(0) )$
is then given by
\begin{equation}
\delta' ( \beta_{-1/2} ) \cdot \delta( \gamma(0) )
= \gamma_{1/2} \cdot ( \,\delta( \beta_{-1/2} ) \cdot \delta( \gamma(0) ) \, )
= \oint \frac{\mathrm{d}z}{2 \pi \mathrm{i}} \, \frac{1}{z} \, \gamma (z) = \gamma (0) \,,
\end{equation}
where we used~\eqref{eq:delbetagamma}.

\subsection{Mixed correlation functions}
\label{sec:mixed-correlation-functions}

Let us next consider mixed correlation functions
which involve delta function insertions such as $\delta (\gamma(u))$
and elementary fields $\beta (z)$ and $\gamma (z)$.
The simplest example is
\begin{equation}
\begin{split}
& \langle \, \delta ( \gamma (u_1) ) \, \delta ( \gamma (u_2) ) \, \gamma (u') \, \beta (w') \, \rangle \\
& = \int \mathcal{D} \beta \, \mathcal{D} \gamma \, \mathrm{d} \sigma_1 \mathrm{d} \sigma_2 \,
\exp \, ( -I_{\beta \gamma} +\sigma_1 \gamma (u_1) +\sigma_2 \gamma (u_2) ) \,
\gamma (u') \, \beta (w') \,.
\end{split}
\end{equation}
Based on the formula~\eqref{propagator-formula},
this can be expressed in terms of correlation functions of the $\beta^* \gamma^*$ system
as follows:
\begin{equation}
\begin{split}
& \langle \, \delta ( \gamma (u_1) ) \, \delta ( \gamma (u_2) ) \, \gamma (u') \, \beta (w') \, \rangle \\
& = \frac{\displaystyle{\int \mathcal{D} \beta^* \, \mathcal{D} \gamma^* \, \mathrm{d} \sigma^*_1 \mathrm{d} \sigma^*_2 \,
\exp \, ( -I_{\beta^* \gamma^*} +\sigma^*_1 \gamma^* (u_1) +\sigma^*_2 \gamma^* (u_2) ) \,
\gamma^* (u') \, \beta^* (w')}}
{\displaystyle{\biggl[ \, \int \mathcal{D} \beta^* \, \mathcal{D} \gamma^* \, \mathrm{d} \sigma^*_1 \mathrm{d} \sigma^*_2 \,
\exp \, ( -I_{\beta^* \gamma^*} +\sigma^*_1 \gamma^* (u_1) +\sigma^*_2 \gamma^* (u_2) ) \, \biggr]^2}} \\
& = \frac{\langle \, \gamma^* (u_1) \, \gamma^* (u_2) \, \gamma^* (u') \, \beta^* (w') \, \rangle}
{\langle \, \gamma^* (u_1) \, \gamma^* (u_2) \, \rangle^2} \,.
\end{split}
\label{eq:mixedcol}
\end{equation}
As described in subsection 10.3.2 of~\cite{Witten:2012bh},
this method can be extended to more general mixed correlation functions
of the $\beta \gamma$ system
and we can express them
in terms of correlation functions of the $\beta^* \gamma^*$ system.\footnote{
For example, when we evaluate the correlation function
$\langle \, \delta ( \gamma (u_1) ) \, \delta ( \gamma (u_2) ) \,
\gamma (u'_1) \, \beta (w'_1) \, \gamma (u'_2) \, \beta (w'_2) \, \rangle$,
we first express it in terms of
$\langle \, \delta ( \gamma (u_1) ) \, \delta ( \gamma (u_2) ) \,
\gamma (u') \, \beta (w') \, \rangle$
and $\langle \, \delta ( \gamma (u_1) ) \, \delta ( \gamma (u_2) ) \, \rangle$
by Wick's theorem
and then rewrite it in terms of
$\langle \, \gamma^* (u_1) \, \gamma^* (u_2) \,
\gamma^* (u') \, \beta^* (w') \, \rangle$
and $\langle \, \gamma^* (u_1) \, \gamma^* (u_2) \, \rangle$.
Note that the resulting expression is not related to
$\langle \, \gamma^* (u_1) \, \gamma^* (u_2) \,
\gamma^* (u'_1) \, \beta^* (w'_1) \, \gamma^* (u'_2) \, \beta^* (w'_2) \, \rangle$
in a simple way
because we use bosonic Wick contractions in the $\beta \gamma$ system
while we use fermionic Wick contractions in the $\beta^* \gamma^*$ system.
}

As an example, let us consider correlation functions
containing $\Xi [ \, f_1, f_2 \, ]$ given by
\begin{equation}
\begin{split}
\Xi [ \, f_1, f_2 \, ]
& = \int_0^1 \mathrm{d}t' \int \mathrm{d} \tilde{t'}
\int \mathrm{d} \zeta \mathrm{d} \tilde{\zeta} \,
\mathrm{e}^{-\{ \, Q',\, t' \zeta \beta [ \, f_1 \, ] \, \}} \,
\mathrm{e}^{-\{ \, Q',\, (1-t') \, \zeta \beta [ \, f_2 \, ] \, \}} \\
& = \int_0^1 \mathrm{d}t' \,
( \, \beta [ \, f_1 \, ] -\beta [ \, f_2 \, ] \, ) \,
\delta ( \, t' \beta [ \, f_1 \, ] +(1-t') \, \beta [ \, f_2 \, ] \, ) \,.
\end{split}
\end{equation}
Contributions to on-shell four-point amplitudes from Feynman diagrams with a quartic vertex
consist of terms with the following structure:
\begin{equation}
\langle \, \Xi [ \, f_1, f_2 \, ] \, X [ \, f_3 \, ] \,
\mathcal{V}_A (t_1) \, \mathcal{V}_B (t_2) \, \mathcal{V}_C (t_3) \, \mathcal{V}_D (t_4) \, \rangle \,,
\end{equation}
where $\mathcal{V}_A$, $\mathcal{V}_B$, $\mathcal{V}_C$, and $\mathcal{V}_D$
are on-shell unintegrated vertex operators in the $-1$ picture.
We choose them to be
\begin{equation}
\mathcal{V}_A = c \delta(\gamma) V_A \,, \quad
\mathcal{V}_B = c \delta(\gamma) V_B \,, \quad
\mathcal{V}_C = c \delta(\gamma) V_C \,, \quad
\mathcal{V}_D = c \delta(\gamma) V_D
\end{equation}
with $V_A$, $V_B$, $V_C$, and $V_D$ being superconformal primary fields
of weight $1/2$ in the matter sector.
Explicit forms of the conformal transformations $f_1 (\xi)$, $f_2 (\xi)$, and $f_3 (\xi)$
are complicated for the quartic vertex,
but the details do not make much difference
and we consider a simpler case where
\begin{equation}
f_1 (\xi) = \xi +t_1 \,, \quad
f_2 (\xi) = \xi +t_2 \,, \quad
f_3 (\xi) = \xi +t_3 \,.
\end{equation}
The action of $X [ \, f_3 \, ]$ changes
$\mathcal{V}_C$ in the $-1$ picture
to the following vertex operator in the $0$ picture:
\begin{equation}
X [ \, f_3 \, ] \, \mathcal{V}_C (t_3)
= c G_{-1/2} \cdot V_C (t_3) +\gamma V_C (t_3) \,.
\end{equation}
Because of the counting of the $bc$ ghost number
or the $\beta \gamma$ ghost number, only the second term gives
a nonvanishing contribution.
Then the $\beta \gamma$ ghost sector of the correlation function is given by
\begin{equation}
\langle \, ( \, \beta [ \, f_1 \, ] -\beta [ \, f_2 \, ] \, ) \,
\delta ( \, t' \beta [ \, f_1 \, ] +(1-t') \, \beta [ \, f_2 \, ] \, ) \,
\delta ( \gamma (t_1) ) \, \delta ( \gamma (t_2) ) \, \gamma (t_3) \, \delta ( \gamma (t_4) ) \, \rangle \,.
\label{beta-gamma-correlation-function}
\end{equation}
Using the formula~\eqref{propagator-formula},
this correlation function can be expressed
in terms of correlation functions of the $\beta^* \gamma^*$ system as follows:
\begin{equation}
\begin{split}
& \langle \, 
\delta ( \, t' \beta [ \, f_1 \, ] +(1-t') \, \beta [ \, f_2 \, ] \, ) \,
\delta ( \gamma (t_1) ) \, \delta ( \gamma (t_2) ) \, \delta ( \gamma (t_4) ) \,
\gamma (t_3) \, ( \, \beta [ \, f_1 \, ] -\beta [ \, f_2 \, ] \, ) \,
\, \rangle \\
& = \frac{\langle \,
( \, t' \beta^* [ \, f_1 \, ] +(1-t') \, \beta^* [ \, f_2 \, ] \, ) \,
\gamma^* (t_1) \, \gamma^* (t_2) \, \gamma^* (t_4) \,
\gamma^* (t_3) \, ( \, \beta^* [ \, f_1 \, ] -\beta^* [ \, f_2 \, ] \, ) \,
\rangle}
{\langle \, ( \, t' \beta^* [ \, f_1 \, ] +(1-t') \, \beta^* [ \, f_2 \, ] \, ) \,
\gamma^* (t_1) \, \gamma^* (t_2) \, \gamma^* (t_4) \, \rangle^2} \,.
\end{split}
\end{equation}
The correlation function in the denominator is
\begin{equation}
\begin{split}
& \langle \, ( \, t' \beta^* [ \, f_1 \, ] +(1-t') \, \beta^* [ \, f_2 \, ] \, ) \,
\gamma^* (t_1) \, \gamma^* (t_2) \, \gamma^* (t_4) \, \rangle \\
& = t' \, \langle \, \gamma^* (t_2) \, \gamma^* (t_4) \, \rangle
-(1-t') \, \langle \, \gamma^* (t_1) \, \gamma^* (t_4) \, \rangle \\
& = t' \, (t_2-t_4) -(1-t') \, (t_1-t_4) \,.
\end{split}
\end{equation}
The correlation function in the numerator is
\begin{equation}
\begin{split}
& \langle \,
( \, t' \beta^* [ \, f_1 \, ] +(1-t') \, \beta^* [ \, f_2 \, ] \, ) \,
\gamma^* (t_1) \, \gamma^* (t_2) \, \gamma^* (t_4) \,
\gamma^* (t_3) \, ( \, \beta^* [ \, f_1 \, ] -\beta^* [ \, f_2 \, ] \, ) \,
\rangle \\
& = \langle \, \beta^* [ \, f_1 \, ] \, \gamma^* (t_1) \,
\beta^* [ \, f_2 \, ] \, \gamma^* (t_2) \,
\gamma^* (t_4) \, \gamma^* (t_3) \, \rangle
= \langle \, \gamma^* (t_4) \, \gamma^* (t_3) \, \rangle = t_4 -t_3 \,,
\end{split}
\end{equation}
where we used
\begin{equation}
( \, t' \beta^* [ \, f_1 \, ] +(1-t') \, \beta^* [ \, f_2 \, ] \, ) \,
( \, \beta^* [ \, f_1 \, ] -\beta^* [ \, f_2 \, ] \, )
= \beta^* [ \, f_2 \, ] \, \beta^* [ \, f_1 \, ] \,.
\end{equation}
Note that the dependence on $t'$ disappeared
because $( \, \beta^* [ \, f_1 \, ] -\beta^* [ \, f_2 \, ] \, )^2 = 0 \,$.
The correlation function~\eqref{beta-gamma-correlation-function}
is thus given by
\begin{equation}
\begin{split}
& \langle \, ( \, \beta [ \, f_1 \, ] -\beta [ \, f_2 \, ] \, ) \,
\delta ( \, t' \beta [ \, f_1 \, ] +(1-t') \, \beta [ \, f_2 \, ] \, ) \,
\delta ( \gamma (t_1) ) \, \delta ( \gamma (t_2) ) \, \gamma (t_3) \, \delta ( \gamma (t_4) ) \, \rangle \\
& = \frac{t_4 -t_3}{( \, t' \, (t_2-t_4) -(1-t') \, (t_1-t_4) \, )^2} \,.
\label{beta-gamma-correlation-function-answer}
\end{split}
\end{equation}
Note that the correlation function~\eqref{beta-gamma-correlation-function}
diverges at
\begin{equation}
t' = \frac{t_4-t_1}{(t_4-t_1)+(t_4-t_2)} \,.
\end{equation}
If we take the limit $t_4 \to \infty$ as in subsection~\ref{sec:interpolation-super},
the correlation function~\eqref{beta-gamma-correlation-function}
diverges at $t' = 1/2$.
When we integrate the correlation function~\eqref{beta-gamma-correlation-function}
over $t'$ from $0$ to $1$, we have to deform the contour in the complex plane of $t'$
to avoid this pole.
The result does not depend on how we deform the contour
because the singularity consists of a double pole and does not contain a single pole.

In subsection~\ref{sec:delta-function-operators},
we considered the operator $\delta' ( \beta_{-1/2} )$.
If we carry out the integration over $\zeta$ of $X [ \, f \, ]$ in~\eqref{eq:defX},
the resulting form of $X [ \, f \, ]$ will involve $\delta' ( \beta [ \, f \, ] )$ defined by
\begin{equation}
	\delta' ( \beta [ \, f \, ] ) = -\int \mathrm{d} \tau \, \tau \, \exp \, ( -\beta [ \, f \, ] \, \tau ) \,.
\end{equation}
One way to calculate correlation functions containing $\delta' ( \beta [ \, f \, ] )$
is to write $\delta' ( \beta [ \, f \, ] )$ as
\begin{equation}
\delta' ( \beta [ \, f \, ] ) = [ \, \gamma [ \, f \, ], \delta( \beta [ \, f \, ] ) \, ] \,,
\end{equation}
where
\begin{equation}
	\gamma [ \, f \, ] = \oint \frac{\mathrm{d}z}{2 \pi \mathrm{i}} \,
	\Bigl( \, \frac{\mathrm{d} f^{-1} (z)}{\mathrm{d}z} \, \Bigr)^{3/2} \,
\frac{1}{f^{-1} (z)} \, \gamma (z) \quad
\text{with} \quad z = f(\xi) \,.
\end{equation}
The calculation then reduces to that of mixed correlation functions
containing $\delta( \beta [ \, f \, ] )$ and $\gamma (z)$.

\subsection{Correlation functions for general vertex operators}

We have seen in appendix~\ref{sec:mixed-correlation-functions}
that the correlation functions of the $\beta \gamma$ system
relevant for on-shell four-point scattering amplitudes
develop a singularity,
but it does not contain a single pole
so that the integral of the correlation functions
is defined without any ambiguity.
We show that correlation functions
relevant for the quartic vertex with general vertex operators
given by
\begin{equation}
\langle \, \delta ( \, t' \beta [ \, f_1 \, ] +(1-t') \, \beta [ \, f_2 \, ] \, ) \,
( \, \prod_{i=1}^{3} \delta ( \gamma (t_i) ) \, ) \, 
( \, \beta [ \, f_1 \, ] -\beta [ \, f_2 \, ] \, ) \,
( \, \prod_{j=1}^{N+1} \gamma (u_j) \, ) \,
( \, \prod_{k=1}^N \beta (w_k) \, ) \, \rangle
	\label{beta-gamma-correlation-function-2}
\end{equation}
do not contain a single pole as a function of $t'$
so that the quartic vertex is defined
for arbitrary string states without any ambiguity.

We have already calculated the correlation function for $N=0$.
The result can be written as
\begin{equation}
\begin{split}
& \langle \, \delta ( \, t' \beta [ \, f_1 \, ] +(1-t') \, \beta [ \, f_2 \, ] \, ) \,
( \, \prod_{i=1}^{3} \delta ( \gamma (t_i) ) \, ) \, 
( \, \beta [ \, f_1 \, ] -\beta [ \, f_2 \, ] \, ) \,
\gamma (u_1) \, \rangle \\
& = \frac{F_0 ( \, \{ t_i \}, u_1 ; t' \, )}
{\langle \, ( \, t' \beta^* [ \, f_1 \, ] +(1-t') \, \beta^* [ \, f_2 \, ] \, ) \,
\gamma^* (t_1) \, \gamma^* (t_2) \, \gamma^* (t_3) \, \rangle^2} \\
& = \frac{F_0 ( \, \{ t_i \}, u_1; t' \, )}
{( \, t' \, (t_2-t_3) -(1-t') \, (t_1-t_3) \, )^2} \,,
\end{split}
\end{equation}
where $F_0 ( \, \{ t_i \}, u_1; t' \, )$ is a polynomial
of $t_1$, $t_2$, $t_3$, $u_1$, and $t'$.
Its explicit form is
\begin{equation}
F_0 ( \, \{ t_i \}, u_1; t' \, )
= H ( t_3, u_1 ) \,,
\end{equation}
where
\begin{equation}
\begin{split}
H ( t_3, u )
& = \langle \,
( \, t' \beta^* [ \, f_1 \, ] +(1-t') \, \beta^* [ \, f_2 \, ] \, ) \,
( \, \prod_{i=1}^{3} \gamma^* (t_i) \, ) \,
\gamma^* (u) \, ( \, \beta^* [ \, f_1 \, ] -\beta^* [ \, f_2 \, ] \, ) \\
& = t_3 -u \,.
\end{split}
\end{equation}
Since $H ( t_3, u_1 )$ does not depend on $t'$,
the correlation function does not contain a single pole
as a function of $t'$.

For $N=1$, we use the formula~\eqref{four-point-formula} to obtain
\begin{equation}
\begin{split}
& \langle \, \delta ( \, t' \beta [ \, f_1 \, ] +(1-t') \, \beta [ \, f_2 \, ] \, ) \,
( \, \prod_{i=1}^{3} \delta ( \gamma (t_i) ) \, ) \, 
( \, \beta [ \, f_1 \, ] -\beta [ \, f_2 \, ] \, ) \,
\gamma (u_1) \, \gamma (u_2) \, \beta (w_1) \, \rangle \\
& = \frac{F_1 ( \, \{ t_i \}, u_1, u_2, w_1; t' \, )}
{\langle \, ( \, t' \beta^* [ \, f_1 \, ] +(1-t') \, \beta^* [ \, f_2 \, ] \, ) \,
\gamma^* (t_1) \, \gamma^* (t_2) \, \gamma^* (t_3) \, \rangle^3} \\
& = \frac{F_1 ( \, \{ t_i \}, u_1, u_2, w_1; t' \, )}
{( \, t' \, (t_2-t_3) -(1-t') \, (t_1-t_3) \, )^3} \,,
\label{beta-gamma-N1}
\end{split}
\end{equation}
where $F_1 ( \, \{ t_i \}, u_1, u_2, w_1; t' \, )$ is
a rational function of  $t_1$, $t_2$, $t_3$, $u_1$, $u_2$, $w_1$, and $t'$.
It takes the following form:
\begin{equation}
F_1 ( \, \{ t_i \}, u_1, u_2, w_1; t' \, )
= G ( \, \{ t_i \}, u_1, w_1; t' \, ) \,
H ( t_3, u_2 )
+G ( \, \{ t_i \}, u_2, w_1; t' \, ) \,
H ( t_3, u_1) \,,
\end{equation}
where
\begin{equation}
\begin{split}
G ( \, \{ t_i \}, u, w; t' \, )
& = \langle \,
( \, t' \beta^* [ \, f_1 \, ] +(1-t') \, \beta^* [ \, f_2 \, ] \, ) \,
( \, \prod_{i=1}^{3} \gamma^* (t_i) \, ) \,
\gamma^* (u) \, \beta^* (w) \, \rangle \,\\
&= t'\langle \,
\gamma^*(t_2) \, \gamma^*(t_3)\,
\gamma^* (u) \, \beta^* (w) \, \rangle
- (1-t')\langle \,
\gamma^*(t_1) \, \gamma^*(t_3)\,
\gamma^* (u) \, \beta^* (w) \, \rangle\,.
\end{split}
\end{equation}
The explicit expression for $G(\{t_i\},u,w;t')$ can be obtained using
the formula~\eqref{betastar-gammastar},
and we can see that $G(\{t_i\},u,w;t')$ is a linear function of $t'$
and a rational function of $t_1$, $t_2$, $t_3$, $u$, and $w$.
Therefore, $F_1 ( \, \{ t_i \}, u_1, u_2, w_1; t' \, )$ is a linear function of $t'$,
and the singularity of the correlation function~\eqref{beta-gamma-N1}
does not contain a single pole with respect to $t'$.

For general $N$, we have
\begin{equation}
\begin{split}
& \langle \, \delta ( \, t' \beta [ \, f_1 \, ] +(1-t') \, \beta [ \, f_2 \, ] \, ) \,
( \, \prod_{i=1}^{3} \delta ( \gamma (t_i) ) \, ) \, 
( \, \beta [ \, f_1 \, ] -\beta [ \, f_2 \, ] \, ) \,
( \, \prod_{j=1}^{N+1} \gamma (u_j) \, ) \,
( \, \prod_{k=1}^N \beta (w_k) \, ) \, \rangle \\
& = \frac{F_N ( \, \{ t_i \}, \{ u_j \}, \{ w_k \}; t' \, )}
{\langle \, ( \, t' \beta^* [ \, f_1 \, ] +(1-t') \, \beta^* [ \, f_2 \, ] \, ) \,
\gamma^* (t_1) \, \gamma^* (t_2) \, \gamma^* (t_3) \, \rangle^{N+2}} \\
& = \frac{F_N ( \, \{ t_i \}, \{ u_j \}, \{ w_k \}; t' \, )}
{( \, t' \, (t_2-t_3) -(1-t') \, (t_1-t_3) \, )^{N+2}} \,,
\label{beta-gamma-Ngen}
\end{split}
\end{equation}
where $F_N ( \, \{ t_i \}, \{ u_j \}, \{ w_k \}; t' \, )$ is
a rational function of the coordinates of the local operators and $t'$.
It is given by
\begin{equation}
	\begin{split}
	F_N ( \, \{ t_i \}, \{ u_j \}, \{ w_k \}; t' \, )
	=
	\sum_{j=1}^{N+1}\sum_{\sigma\in P_j}
	( \, \prod_{k=1}^N G( \, \{t_i\},u_{\sigma_k},w_{k};t' \, ) \, ) \, H(t_3,u_j) \,,
	\end{split}
\end{equation}
where $P_j$ denotes the set of permutations of the $N$ indices whose element $\sigma$ is described as
$\{1, 2, \cdots, \hat{j}, \cdots, N+1\} \to \{\sigma_1, \sigma_2, \cdots, \sigma_N \}$
with $j$ omitted, as indicated by $\hat{j}$.
The function $F_N ( \, \{ t_i \}, \{ u_j \}, \{ w_k \}; t' \, )$
is a polynomial of $t'$ and its degree is less than $N+1$.
Therefore, the singularity of the correlation function~\eqref{beta-gamma-Ngen}
does not contain a single pole with respect to $t'$.
Consequently, the integral of~\eqref{beta-gamma-Ngen} over $t'$ from $0$ to $1$
is defined without any ambiguity
and produces a rational function of
the coordinates of the local operators,
which does not have any branch cuts with respect to these variables.

\bibliographystyle{oophys}
{\small
\bibliography{refs2}
}

\end{document}